\documentclass[fedsstyle, oneside, onecolumn, Numbered]{jnl}

\usepackage{graphicx}
\usepackage{multirow}
\usepackage{amsmath,amssymb,amsfonts}
\usepackage[title]{appendix}
\usepackage{manyfoot}
\usepackage{booktabs}
\usepackage{accents}
\usepackage{bibunits}
\usepackage{titletoc}
\usepackage{lmodern}

\DeclareMathOperator{\Cov}{Cov}
\DeclareMathOperator{\Var}{Var}
\newcommand{\vect}[1]{\protect\accentset{\rightharpoonup}{#1}}

\defaultbibliographystyle{fedsstyle}

\hypersetup{urlcolor=black}

\newenvironment{aligned*}{
\begin{equation}
\begin{aligned}
}{%
\end{aligned}
\end{equation}%
}

\title{
Tunable spectral correlations of highly multimode visible light via broadband quantum frequency conversion
}

\author*[1]{\fnm{Federico} \sur{Presutti}}\email{fp267@cornell.edu}
\author*[1,2]{\fnm{Logan~G.} \sur{Wright}}\email{logan.wright@yale.edu}
\presentaddress{%
 \orgdiv{Department of Applied Physics},
 \orgname{Yale University},
 \orgaddress{%
   \state{CT}%
   , \country{USA}}}
\author[1]{\fnm{Shi-Yuan} \sur{Ma}}
\author[1]{\fnm{Tianyu} \sur{Wang}}
\author[1,3]{\fnm{Benjamin~K.} \sur{Malia}}
\author[1,2]{\fnm{Tatsuhiro} \sur{Onodera}}
\author*[1,4]{\fnm{Peter~L.} \sur{McMahon}}\email{pmcmahon@cornell.edu}
\affil[1]{%
 \orgdiv{School of Applied and Engineering Physics},
 \orgname{Cornell University},
 \orgaddress{%
   \state{NY}%
   , \country{USA}}%
}
\affil[2]{%
 \orgdiv{NTT Physics and Informatics Laboratories},
 \orgname{NTT Research, Inc.},
 \orgaddress{%
   \state{CA}%
   , \country{USA}}%
}
\affil[3]{%
 \orgdiv{Intelligence Community Postdoctoral Research Fellowship Program},
  \orgname{Cornell University},
 \orgaddress{%
   \state{NY}%
   , \country{USA}}%
}
\affil[4]{%
 \orgdiv{Kavli Institute at Cornell for Nanoscale Science},
 \orgname{Cornell University},
 \orgaddress{%
   \state{NY}%
   , \country{USA}}%
}

\begin{document}
\begin{bibunit}

\abstract{
Multimode squeezed states of light are a resource for achieving quantum advantage in computing and sensing, where spatial or temporal modes have been the experimental norm.
In our experiments, we generated highly frequency-multimode infrared quantum light, and show how adiabatic frequency conversion can be used to convert the quantum state to visible wavelengths, while concurrently manipulating the joint spectrum by realizing a configurable many-port frequency-domain-beamsplitter unitary transformation.
We report near-unity-efficiency quantum frequency conversion over a bandwidth \textgreater45\,THz, which allowed us to measure the state with an electron-multiplying CCD (EMCCD) camera-based spectrometer, at non-cryogenic temperatures.
The parametric amplification and conversion of \textgreater400 frequency modes yielded an overall mean of approximately 700 visible photons per shot, and photon statistics consistent with squeezing.
Our work shows how many-mode quantum states of light can be generated, manipulated, and measured with efficient use of hardware resources, motivating the use of frequency encoding in quantum optics.
}

\phantomsection
\addcontentsline{toc}{section}{Abstract}

\maketitle

\phantomsection
\addcontentsline{toc}{section}{Introduction}

\begin{figure}[t]
\centering
\includegraphics[width=0.75\normaltextwidth]{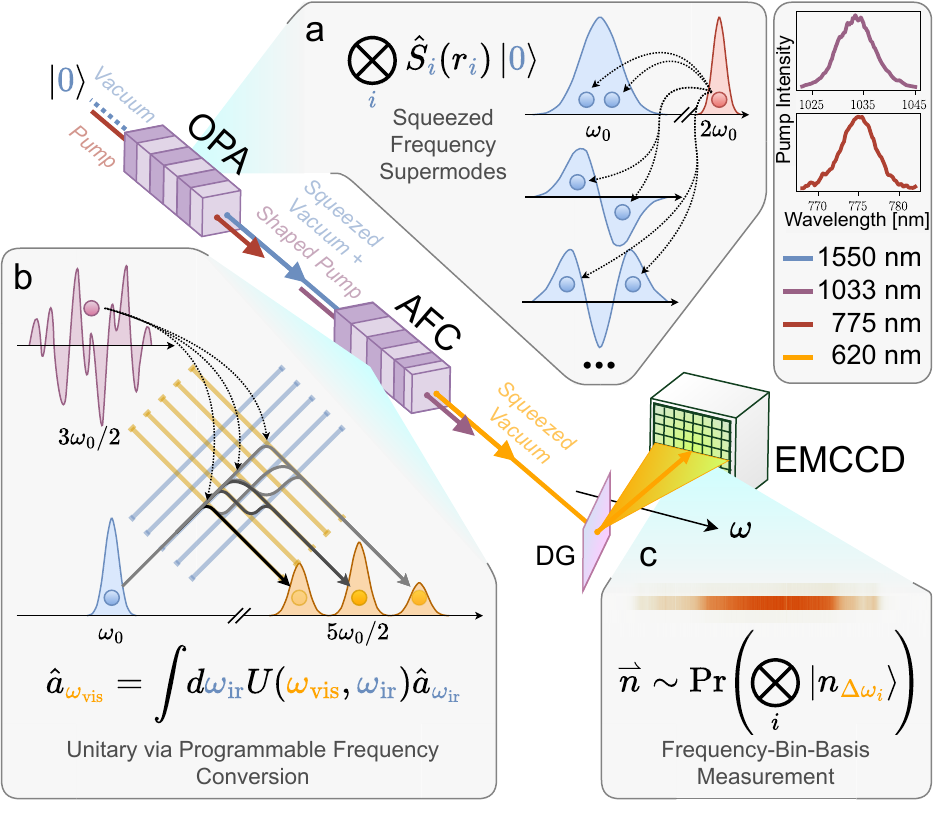}
\caption{
 \textbf{Frequency domain, multimode, visible squeezed state preparation and detection.}
 \textbf{a.} Highly multimode squeezed vacuum is generated in the degenerate optical parametric amplifier (OPA), at near infrared wavelengths.
 The squeezed modes occupy orthogonal frequency spectra.
 This may be represented by a set of squeezing operators $\hat{S}_i$ acting on the vacuum state $\lvert 0 \rangle$.
 Each operator squeezes a distinct frequency mode with some squeezing parameter $r_i$.
 The squeezing occurs around a central frequency $\omega_0$ equal to half the pump central frequency $2 \omega_0$.
 Concretely, these correspond to the wavelengths 1550\,nm and 775\,nm, respectively.
 \textbf{b.} Adiabatic frequency conversion (AFC) efficiently converts the squeezed light to visible wavelengths.
 The pump, a broadband pulse centered at 1033\,nm ($3\omega_0/2$), combines with the broadband 1550\,nm signal to yield 620\,nm ($5\omega_0/2$) light.
 This transformation between the two sets of frequencies can be changed by varying the spectro-temporal profile of the pump.
 This operation may be represented as a linear unitary transformation $U$ acting on the infrared and visible fields, represented by the operators $\hat{a}_{\omega_\text{ir}}$ and $\hat{a}_{\omega_\text{vis}}$.
 \textbf{c.} The final state, incident on a diffraction grating (DG), is split into frequency modes, and frequency-resolved photon counting is performed with an EMCCD camera.
 Each measurement yields some photon-number sequence, or vector, $\vect{n}$, whose probability distribution depends on the state (determined by $\hat{S}_i$ and $U$).
 The camera measures the spectrum in a discrete manner, as the pixels capture the photons within some frequency ``bins,'' denoted $\Delta \omega_i$.
 Hence, the state's overlap with these bin-basis Fock states, $\lvert n_{\Delta \omega_i} \rangle$, determines the probability distribution.
}
\label{fig:concept}
\end{figure}

The generation, control, and measurement of entangled multimode Gaussian states of light are crucial elements of continuous-variable (CV) quantum information processing \cite{ClusterStates, QuantumIllumination, WeedbrookLloyd, BourassaDhand}.
Most quantum technologies based on multimode quantum optics benefit from being able to use as many modes as possible.
As an example, a Gaussian boson sampler (GBS) \cite{RahimiLundRalph, HamiltonJex} is a special-purpose quantum computer that can -- at least in the ideal case -- perform certain calculations that are believed to be intractable on a classical computer when the number of modes and number of photons in the GBS are sufficiently large \cite{HamiltonJex, HangleiterEisert}.
The recent demonstration of Gaussian boson sampling in the regime of quantum computational supremacy, with tens to hundreds of squeezed modes and detected photons per shot \cite{ZhongPan, ZhongPan2021, DengPan, MadsenLavoie}, was a milestone in the development of CV-based quantum systems that was achieved because of the success in pushing to large numbers of modes and photons.
A GBS executes a sequence of three steps, which are common to many CV quantum-information-processing protocols:
(1) generate squeezed states, (2) apply a unitary transformation to entangle them, and (3) measure the final state (by photon counting).

Optics gives us the choice of several possible degrees of freedom within which to encode information -- specifically: space, time, frequency and polarization (or any combination thereof).
While large-scale GBS experiments have been realized using space \cite{ZhongPan, ZhongPan2021, DengPan} and time \cite{MadsenLavoie} encodings, the frequency domain remains to be explored.
Frequency encoding offers important potential advantages over space or time encoding for both the generation and the manipulation (unitary control) of multimode squeezed light: reduced hardware resources and complexity, and reduced loss.
The extremely broad bandwidth of light enables frequency-encoded systems to operate on many frequency modes in a compact system \cite{LukensLougovski, JoshiGaeta, JoshiGaeta_, HiemstraWalmsley, ReimerMorandotti, CabrejoSteinlechner, BasaniKrastanov}.
Many demonstrations of large-scale multimode squeezing, for example, use the frequency domain, e.g.,\ in broadband frequency combs \cite{ReimerMorandotti, CabrejoSteinlechner, LiuLi, OmiKannari, YamagishiKannari, RoslundTreps, CaiTreps, RomanParigi, YangYi, JahanbozorgiYi, PysherPfister, ChenPfister, XieWong, ChangWong, EldanPeer}.
Linear unitary operations in the frequency domain, i.e., acting on the frequency modes, can be implemented in a hardware-efficient way, operating on all frequency modes in parallel \cite{LuLougovski, LuLukens}.
One approach is to use one or more electro-optic modulators \cite{JavidLin, YuanFan, LuLougovski, LuLukens, ZhuPfister} (although there are limitations on the unitary from the driving microwave bandwidth being $\lesssim$100\,GHz); another is to use nonlinear wave mixing to convert photons in each frequency mode to photons in a combination of other frequency modes \cite{FrequencyBinReview}.
Unitaries based on nonlinear wave mixing mediated by a classical field, such as four-wave mixing \cite{JoshiGaeta2}, provide a route to realizing programmable unitaries that can operate over wide bandwidths, in compact hardware with low loss.
The programming of the unitary can be done by shaping the classical field(s) used to control the wave mixing of the modes containing quantum light, and the wave mixing can be implemented compactly in a single-spatial-mode waveguide.
However, a reconfigurable, frequency-domain unitary working over optical (\textgreater1\,THz) bandwidths for quantum light has yet to be realized at even moderate scale (more than 2 modes \cite{JoshiGaeta2}).

The final step, state measurement by photon counting, is a challenge for many multimode architectures.
Since the preferred nonlinear optical materials for generating squeezed vacuum work best at longer wavelengths,
squeezing at wavelengths centered near 1550\,nm is typical \cite{ZhongPan, ZhongPan2021, DengPan, MadsenLavoie, HiemstraWalmsley, ReimerMorandotti, CabrejoSteinlechner, BasaniKrastanov, LiuLi, OmiKannari, YamagishiKannari, RoslundTreps, CaiTreps, RomanParigi, YangYi, JahanbozorgiYi, EldanPeer}, also in part due to the convenience of being able to use optical components from telecommunications technologies.
However, good (high quantum-efficiency, low dark-count) single-photon detectors at 1550\,nm, namely superconducting nanowire detectors \cite{snspdreview2012, snspdreview2021}, are very expensive and require cryogenic cooling.
Silicon-based camera sensor technologies -- both charge-coupled device (CCD) and complementary metal--oxide--semiconductor (CMOS) detectors -- are well-established, comparatively inexpensive and compact,
and each camera comprises $10^5$--$10^6$ individual pixels, i.e., detectors.
Cameras capable of detecting single photons with low noise have recently become available, and
there is a growing literature of quantum-optics and sensing experiments that were performed with visible wavelengths and
these cameras \cite{DefienneFaccio, BolducLeach, HugoFaccio2, KumarMarino, LiAgarwal, SvihraNomerotski}.
Electron-multiplying CCD (EMCCD) cameras are arguably the current state-of-the-art, and
low-noise CMOS photon-number-resolving cameras are also a promising tool within this domain \cite{HamamatsuCMOS, GigajotCMOS}.

Here, we demonstrate how to use frequency conversion \cite{Kumar} to enable the use of these visible-light cameras
in combination with techniques for strong squeezing possible at longer wavelengths.
In addition, our demonstration is an improvement over existing methods of quantum frequency conversion:
previous demonstrations are limited to either modest bandwidths or efficiencies \cite{WangPan, VollmerSchnabel, BauneSchnabel, SamblowskiSchnabel, BonsmaMosley, AllgaierSilberhorn, SuchowskiSilberberg2}.
However, adiabatic frequency conversion (AFC) essentially eliminates this trade-off \cite{SuchowskiSilberberg, MosesKartner}.
We show how this method allows us to obtain robust, efficient and broadband conversion over \textgreater45\,THz (1390--1750 to 590--650\,nm) and near-unity efficiency.
Furthermore, it allows
unitary control of the multimode entanglement (with, in principle, no additional loss) by manipulation of the complex profile of the broadband pump used to drive the conversion.
This architecture provides the best of both worlds: squeezing at telecommunications wavelengths, and photon detection at visible wavelengths.
We will show that we are able to generate non-classically correlated light in over 400 frequency modes, resulting in states having a measured mean photon number of nearly 700; photon counting measurements of these optical modes follow the expected intensity and correlation statistics that are consistent with squeezing.
By using AFC to efficiently convert the light to visible wavelengths, and using the highly parallel photon counting made possible by a modern EMCCD, we can directly measure these states.
The non-classical statistics are preserved through frequency conversion.
We will also show that we can control the entanglement between different modes by using different spectrally shaped classical fields as the pump of the AFC process, resulting in different measured correlations between photon detections across the frequency modes.

\section*{Results}\label{sec:results}
\addcontentsline{toc}{section}{\nameref{sec:results}}

The experimental setup is illustrated in Fig.\,\ref{fig:concept};
an overview is as follows.
We use a waveguided degenerate optical parametric amplifier (OPA) pumped with a pulsed laser: this is known to provide squeezing over many frequency modes (known as the ``supermodes'' \cite{WasilewskiRadzewicz}), and in a single spatial mode by virtue of the waveguide.
An adiabatic frequency conversion (AFC) crystal subsequently converts this near-infrared light to the visible.
The temporal profile of the AFC pump pulse is shaped, which controls the conversion process -- the linear transformation between infrared and visible light frequencies.
Finally we detect the visible squeezed light with an electron-multiplying CCD (EMCCD) camera, serving as an array of high-quality photodetectors.

In the following section, we present characterizations of each of these stages.
We first measure the properties of the light in the infrared, directly out of the OPA.
Our characterizations enable us to measure the bandwidth, estimate the number of modes, and identify non-classical correlations.
We then demonstrate efficient conversion using AFC.
In addition, with the camera, we are able to observe spectral photon-number correlations throughout the whole bandwidth at high resolution.
With this ability, we generate qualitatively different joint spectra as a proof of concept of frequency-domain unitary transformations by pulse shaping.
Finally, we discuss detection and the photon numbers generated by this architecture.
Refer to the \nameref{sec:methods} section for details on the OPA, AFC, pulse shaper, the photon-counting spectrometer, as well as specific experiments.

In the theory of vacuum-seeded (unseeded), pulsed OPA, the light generated is well-understood to be squeezed light, where squeezing occurs to the quadrature components of the field in each spectral supermode.
Our experiments show sub-shot-noise photon-number correlations that are consistent with squeezed light.
However, because our experiments do not have direct access to the phase and thus can only infer quadrature squeezing from photon statistics, in this manuscript we use the term ``quantum light'' to emphasize that the experimentally produced light -- via OPA and AFC -- has not been proven to be squeezed, only non-classically correlated.
This limitation is due to the practical difficulty of creating a local oscillator that matches the spectral profile of the supermodes necessary for a homodyne (phase-sensitive) measurement.
We expect that because of the common-path nature of our experimental scheme, there are limited places where phase fluctuations could arise that would decohere squeezed light, and that a homodyne measurement with suitably shaped local oscillators would also reveal sub-shot-noise results consistent with squeezing.

\subsection*{Highly multimode quantum light in the frequency domain}\label{sec:results:dopa}
\addcontentsline{toc}{subsection}{\nameref{sec:results:dopa}}

\begin{figure}[!htbp]
\centering
\includegraphics[width=1\normaltextwidth]{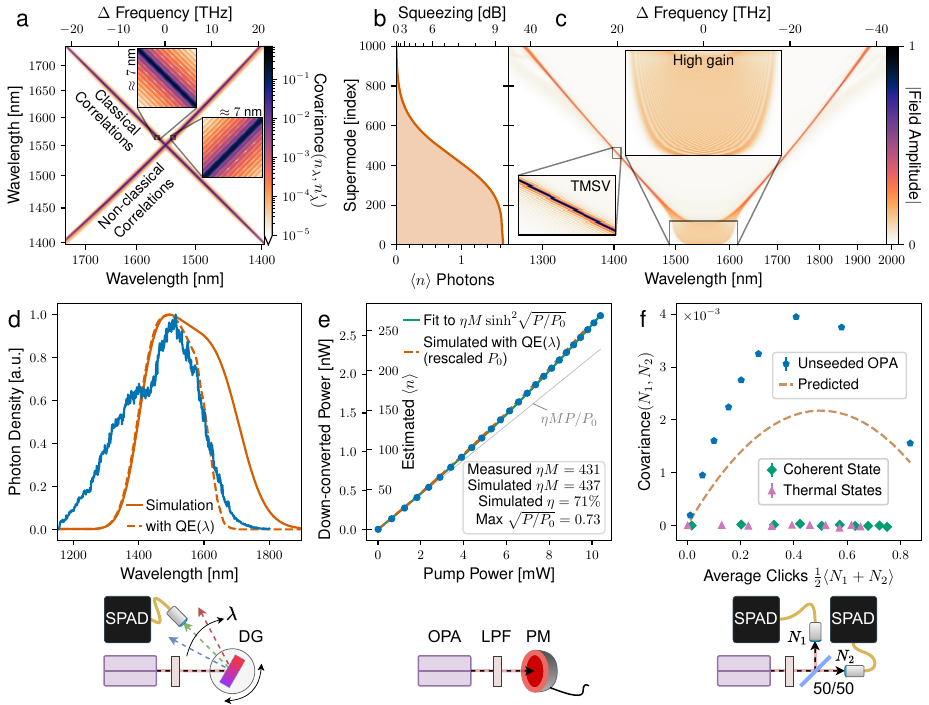}
\caption{
 \textbf{Frequency-multimode quantum light from a lithium niobate waveguide optical parametric amplifier.}
 \textbf{a.}
 Simulated photon-number covariance matrix of the state produced in the degenerate optical parametric amplifier (OPA).
 This is made up of classical correlations (due to photon-number variance) and non-classical correlations (due to entanglement).
 \textbf{b} and \textbf{c}
 show the decomposition of the simulated state into single-mode squeezed states (supermodes), and share a vertical axis.
 \textbf{b.}
 Simulated average photon number $\langle n \rangle$ and squeezing values of the first 1000 supermodes.
 \textbf{c.}
 Simulated supermode basis of the OPA.
 Each row represents a mode and its spectrum is represented by the colormap.
 The high gain supermodes are approximately Hermite--Gauss functions while weaker supermodes resemble two-mode squeezed vacuum (TMSV) states with two distinct frequency peaks.
 For \textbf{a--c},  the pump power is approximately equivalent to the highest used in the experiment.
 \textbf{d.}
 Measurement of the OPA output spectrum and comparison to simulation.
 As the detector quantum efficiency (QE) vanishes at longer wavelengths, the simulation spectrum is also plotted multiplied by the nominal relative QE for comparison.
 \textbf{e.}
 Parametric gain of the OPA, fit to the number of modes, and comparison to simulation.
 The photon number $\langle n \rangle$ is obtained by dividing the average power measurement by the repetition rate and the photon energy at the central wavelength.
 In the fit: $M$ represents the number of modes; $\eta$ the detection efficiency; $P$ the pump power; $P_0$ the fitted scaling factor.
 \textbf{f.}
 Coincidence detection, comparing unseeded OPA light, attenuated laser light (coherent) and broadband incandescent lamp (thermal).
 The prediction model is derived in Appendix~\ref{sec:app:c:qecoincidence}.
 Sketches of the experimental setups corresponding to \textbf{c--e} are shown below their respective plots.
 The OPA output is measured with either a single-photon avalanche diode (SPAD) or a power meter (PM), after isolating the signal with a long-pass filter (LPF).
}
\label{fig:dopa}
\end{figure}

We begin by studying the optical parametric amplification (OPA) process, and looking for behavior consistent with highly multimode squeezing.
Three experiments were performed to verify numerical predictions: a spectrum measurement, a parametric gain measurement and a coincidence detection measurement.
The purpose of the first is to verify that the bandwidth is indeed as broad as predicted.
The second is to infer the number of modes, based on how the output energy scales with pump energy.
The third is to verify the photon statistics, evidence that this light is made up of squeezed states.

The OPA was first simulated,
and the photon-number covariance is plotted in Fig.\,\ref{fig:dopa}a, which has a notable ``X'' shape
(as reported in Refs.~\cite{LandesRaymer,SpasibkoChekhova}).
The anti-diagonal is due to the non-classical frequency correlations of entangled photons (the joint spectral intensity) and energy conservation dictates its shape: down-converted photon pair energies must add up to the pump photon energy.
The diagonal represents the photon-number variance, and the correlations are classical (these statistics have the same properties as those of thermal states \cite{LvovskySqu}): this has a finite width, given by the phase-matching function of the OPA (typically a sinc function for periodic poling, as shown in the insets showing 7\,nm-wide enlarged view).
The Bloch--Messiah decomposition helps us interpret how the OPA generates squeezing \cite{WasilewskiRadzewicz, HosakaKannari},
by reducing the output to a set of supermodes and their corresponding squeezing or anti-squeezing values.
These are shown in Fig.\,\ref{fig:dopa}b and c.
For a 4\,cm LN waveguide, and 775\,nm 200\,fs pulse pump, we predict $\sim$600 supermodes with substantial average photon numbers, spanning over 50\,THz, or over 400\,nm.
The weakly squeezed supermodes (indices above 200) resemble two-mode squeezed vacuum (TMSV) states, occupying two sinc peaks on opposite sides of the central wavelength.
The highly squeezed supermodes (indices below 200) appear to approximately be the typical Hermite--Gauss functions  \cite{WasilewskiRadzewicz}.
The supermodes are not constant with respect to the pump power, and tend to transition from TMSV to Hermite--Gauss as the parametric gain increases.

The spectrum measurement, whose outcome is shown in Fig.\,\ref{fig:dopa}d, confirms that the spectrum is broad, spanning hundreds of nanometers.
The spectrum could not be measured past about 1700\,nm due to the detector response cut-off.
The spectrum may be thought of as the incoherent sum of the supermodes, scaled by average photon number.

The parametric gain experiment (Fig.\,\ref{fig:dopa}e) measures the output power of the OPA at varying pump powers.
The number of down-converted photons
scales linearly with the number of supermodes $M$ and the detection efficiency $\eta$, while each mode responds nonlinearly, containing on average $\langle n \rangle = \sinh^2 \sqrt{P / P_0}$ photons, for pump power $P$ and scaling factor $P_0$ (for a specific repetition rate and other experimental parameters).
We may estimate the number of modes by fitting to $\eta M \sinh^2 \sqrt{P / P_0}$, which assumes equally squeezed modes (not truly the case, as shown in Fig.\,\ref{fig:dopa}b, hence this is a lower bound).
From this measurement we infer the presence of at least $M \geq \eta M = 431$ frequency modes with substantial photon occupation -- on the order of 650 modes accounting for average detector responsivity, $\eta \approx 71\%$.
In order to achieve a sufficiently high -- measurable -- average output power, this experiment requires different operating conditions (a higher repetition rate pump laser), and we estimate that we would measure a slightly lower number of modes, 386 (465 corrected), under the usual conditions (Appendix~\ref{sec:app:tis}).

The coincidence experiment (inspired by Ref.~\cite{VaidyaVernon}) is shown in Fig.\,\ref{fig:dopa}f.
The OPA output beam is evenly split and the photon-number covariance between the two beams is measured.
The coincidences depend on the photon statistics of the state, and at low photon numbers, only the coincidence rate of highly multimode squeezed light is expected to scale linearly with respect to the average photon number.
We performed coincidence detection with the OPA output, attenuated 1550 nm laser light, and broadband incandescent lamp light; indeed only the former was nonzero and linear.

These three experiments are consistent with our predictions that the OPA is a source of highly multimode squeezed light, spanning hundreds of frequency modes.
A more thorough discussion of these results, including explanations of the experiment-theory discrepancies, is presented in the \hyperref[sec:methods:multimodesqueezing]{Methods}.

\subsection*{Adiabatic frequency conversion (AFC) of quantum light}\label{sec:results:afc}
\addcontentsline{toc}{subsection}{\nameref{sec:results:afc}}

\begin{figure}[!htbp]
\centering
\includegraphics[width=1\normaltextwidth]{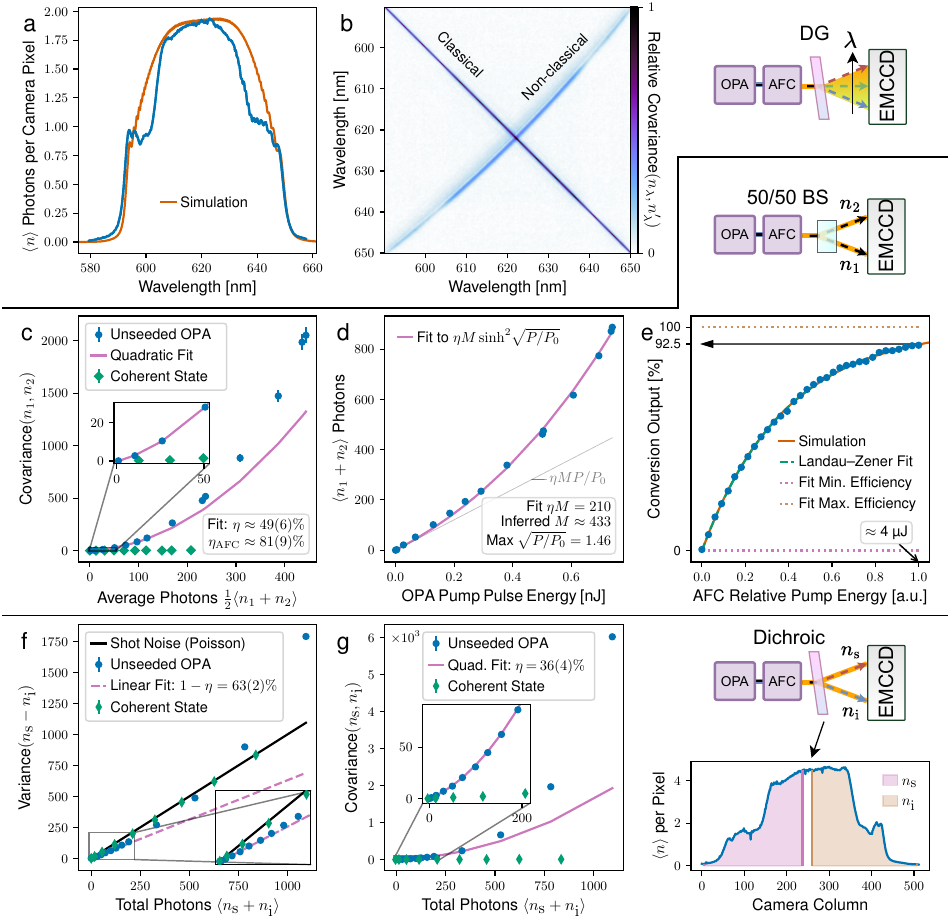}
\caption{
 \textbf{Broadband, adiabatic frequency conversion (AFC) of quantum light.}
 Experiments corresponding to plots \textbf{a} and \textbf{b} used a diffraction grating (DG) to use the EMCCD camera as a spectrometer.
 Experiments for plots \textbf{c--e} used a Wollaston prism configured as a 50:50 beamsplitter (BS).
 Experiments for plots \textbf{f} and \textbf{g} used the DG that simulates a dichroic mirror to partition the state into a signal and idler.
 ``Unseeded OPA'' and ``coherent'' states refer to the light generated by the OPA and to attenuated 1550\,nm laser light respectively, both converted by the AFC process.
 \textbf{a.} Converted spectrum measurement.
 This is compared to simulation matching 92\% conversion within the conversion bandwidth.
 \textbf{b.} Measured, wavelength-resolved, photon-number covariance matrix.
 The diagonal and anti-diagonal features can be recognized as the classical and non-classical correlations.
 \textbf{c.} Coincidence detection after conversion.
 This experiment measures the transmission of the state (effectively the purity) as $\eta=49\pm6$\%.
 \textbf{d.} Parametric gain measured after conversion.
 The fit to this curve estimates the number of frequency modes present in the converted state as 433, accounting for loss.
 \textbf{e.} Conversion as a function of pump power.
 A fit to the saturation curve estimates the conversion efficiency as 92.5\%.
 \textbf{f.} Sub-shot-noise correlations of partitioned state.
 The spectrum is split in half as shown in the panel on the right side.
 The variation in the difference of the shot-by-shot intensities were measured.
 Coherent light is found to vary like shot noise, as expected, while the multimode squeezed light is found to have sub-shot noise variance (effective transmission $\eta=37\pm2\%$), except at high average intensity (where the quadratic term dominates).
 \textbf{g.} Equivalence of covariance analysis.
 By performing a covariance analysis on the same data, we find consistent results: the effective transmission of the multimode squeezing is estimated as $\eta=36\pm4\%$.
 This reinforces the validity of the intensity-covariance technique.
}
\label{fig:afc}
\end{figure}

The next part of the experiment involves frequency conversion of the quantum light.
The infrared unseeded pulsed OPA output, centered around 1550\,nm, is up-converted with a pulsed 1033\,nm laser, to a central wavelength of 620\,nm, by sum frequency generation (SFG).
This ultra-broadband nonlinear process is possible through AFC \cite{SuchowskiSilberberg, MosesKartner},
which acts as a unitary transformation in the frequency domain,
and most importantly allows us to measure highly squeezed light with a sensitive, silicon-based camera.
Briefly, AFC involves a Landau--Zener transition, which avoids back-conversion and allows an arbitrarily broad phase-matching bandwidth.
This is unlike regular SFG, which involves a Rabi cycle (where each spatio-spectral mode may oscillate at different rates), and has a narrow phase-matching bandwidth -- typically on the order of 1\,nm for cm-long crystals.
(See Appendix~\ref{sec:app:afc:afcdesign} for the crystal quasi-phase-matching design used in this experiment.)

In this section we demonstrate frequency conversion of the pulsed OPA light, show that the non-classical correlations remain and measure the overall system fidelity, and provide estimates of the conversion efficiency and the number of converted modes.
To this end, two experiments are presented in this section: one based on our visible light, single photon spectrometer (Fig.\,\ref{fig:afc}a--b); and one based on two-mode coincidence detection (Fig.\,\ref{fig:afc}c--d).
The spectrometer uses a diffraction grating to separate the converted light into its wavelength components, which are imaged onto the EMCCD camera.
The coincidence experiment also uses the camera, but a with prism functioning as a 50:50 beamsplitter, replacing the grating.
The two beams are imaged onto the camera, which emulates two detectors via two small regions of interest on the CCD.

The spectrum measured with the EMCCD camera visible light spectrometer is shown in Fig.\,\ref{fig:afc}a.
The converted spectrum matches the designed conversion band.
The shoulders in the spectrum (roughly between 590--600 and 640--650\,nm), which differ from the simulation, are consistent with Fig.\,\ref{fig:dopa}c (around 1400\,nm): these features are from the original spectrum, not the conversion.
The intensity implies that, on average, the most brightly illuminated pixels receive almost 2 photons per pulse.

The photon-number covariance matrix is shown in Fig.\,\ref{fig:afc}b.
Shot-by-shot intensities are obtained by triggering the camera to capture individual pulses.
The camera is used in ``analog'' mode (not thresholded) due to the high photon numbers per pixel (see Appendix~\ref{sec:app:emccd}).
For this experiment, the AFC was pumped in such a way as to maximize the conversion and to perform a transformation as close as possible to the identity, although the pump bandwidth causes some broadening (see Appendix~\ref{sec:app:afc:afcdesign}; an identity would require a monochromatic pump).
The photon covariance matrix has the ``X'' shape we expect
from Fig.\,\ref{fig:dopa}a, with both classical and non-classical correlations.

The results of the coincidence experiment, similar to the one presented in the previous section, are shown in Fig.\,\ref{fig:afc}c.
We can now use this experiment to obtain the effective fidelity or purity of the state: where the photon statistics lie between those of pure squeezed vacuum and thermal noise, as parametrized by an effective loss or transmission.
For few photons in each supermode, $\langle n \rangle \ll 1$, we expect linear scaling; at intermediate photon numbers, where $\langle n \rangle = \sinh^2 \sqrt{P/P_0} \approx P/P_0$, we expect the covariance to scale quadratically; and at greater photon number we cannot predict the scaling without precise knowledge of the supermodes (see Appendix~\ref{sec:app:coincidence}).
In the first two cases, the coefficient of the linear component gives us an estimate of the overall transmission (a weighted average over all modes).
Hence, we perform a quadratic fit to the lowest photon number portion of the data.
The fit implies a $49\pm6\%$ overall effective transmission for all the modes.
This bounds the average vacuum squeezing to 3\,dB; by interpreting our experimental results as arising from squeezed states, we estimate that the experiment achieves 2.7 dB (Appendix~\ref{sec:app:squeezingestimate}).

The same data is used for Fig.\,\ref{fig:afc}d, to quantify the parametric gain.
By assuming the 49\% overall, effective transmission, the scaling implies the presence of over 430 populated frequency modes.
Some parasitic second harmonic generation is observed in the OPA at higher pump energies, which could reduce the pump energy available for parametric amplification and possibly cause this value to be slightly underestimated.

The conversion as a function of AFC pump power is used as an estimate of the conversion efficiency, which is shown in Fig.\,\ref{fig:afc}e.
Conversion efficiency (saturation) is described by the Landau--Zener formula: unconverted signal follows an exponential decay with respect to the pump power.
The fit implies 92.5\% conversion;
this is a reasonable upper-bound estimate on the conversion efficiency.
We have been careful to make the pump beam much larger than the signal, and have verified that it has minimal spatial chirp, as
a fit to the Landau--Zener formula may overestimate the efficiency if there is substantial mismatch in the spatio-temporal overlap in the pump and signal.
In addition, this is simply a photon conversion efficiency and cannot account for the impact of any other type of decoherence.
A lower bound estimate of the AFC efficiency can be obtained from the 49\% effective transmission.
By measuring the transmission of each optical element in the squeezed light beam path and taking into account the camera quantum efficiency (QE), we estimate the end-to-end passive optical transmission to be approximately 60\% (excluding AFC).
This bounds the conversion efficiency from below to about 81\%.
Note that the effective transmission figure here inherently includes non-loss decoherence (e.g. detector excess noise; weakly converted photons at wavelengths outside the conversion band).
We point out that the conversion efficiency is pump-power limited, meaning that achieving close to 100\% over these bandwidths is realistic without requiring phase-locking.

Finally, we performed a more conventional experiment \cite{VaidyaVernon} to demonstrate a sub-shot-noise intensity correlation between groups of entangled modes,
which corroborates the result that this state is lossy squeezed vacuum.
The results are shown in Fig.\,\ref{fig:afc}f.
After partitioning the state into two sets of wavelengths -- a ``signal'' and ``idler'' -- the variance of the difference in the intensities was measured.
Coherent light illumination clearly lies along the shot-noise line, while the unseeded OPA illumination lies below it.
Only at higher intensities does it surpass it:
this is expected when the signal and idler are not perfectly balanced (see Appendix~\ref{sec:app:c:vardeltan}), which is the case here due to their artificial division.
The partition shown was chosen as it minimized the variance: intuitively this requires capturing as many correlated signal and idler photons into the separate bins, while excluding photons from frequency-degenerate modes from ending up in the same bin, which will raise the variance of the number-difference.
The effective transmission in this experiment is significantly lower than the previous quantity, $\eta=37\pm2\%$.
This is expected as partitioning the state necessarily adds decoherence: the state is not bipartite, and the covariance matrix and supermodes are smooth functions; there is not a sharp boundary between signal and idler modes.
Hence it is not possible to divide the spectrum without either having excess variance in each half due to internal correlations (photons from the same mode), or without tracing out correlated wavelengths.
Fig.\,\ref{fig:afc}g shows an intensity covariance analysis of the same partitioned data.
The results are unsurprisingly consistent, yielding $\eta=36\pm4\%$, showing that this method (used previously for Fig.\,\ref{fig:afc}c and obtaining the $\eta=49\%$ figure) is equivalent to the analysis from the sub-shot-noise perspective.

For clarity, we summarize the three conversion efficiency estimation results.
Measuring the conversion saturation implied a 92\% efficiency: we consider this an optimistic upper bound.
The photon-statistics-based method (after accounting for linear losses) yielded a conversion efficiency of $81\pm9\%$;
we have the most confidence in this measurement.
A second, ``signal''--``idler'' intensity correlation experiment -- more conventional but less suitable for the states we measure -- yielded a $61\pm5\%$ efficiency;
we deem this a lower bound.

\subsection*{Upconversion as a frequency-domain unitary transformation}\label{sec:results:covs}
\addcontentsline{toc}{subsection}{\nameref{sec:results:covs}}

\begin{figure}[!htbp]
\centering
\vspace{-3\baselineskip}
\includegraphics[width=\normaltextwidth]{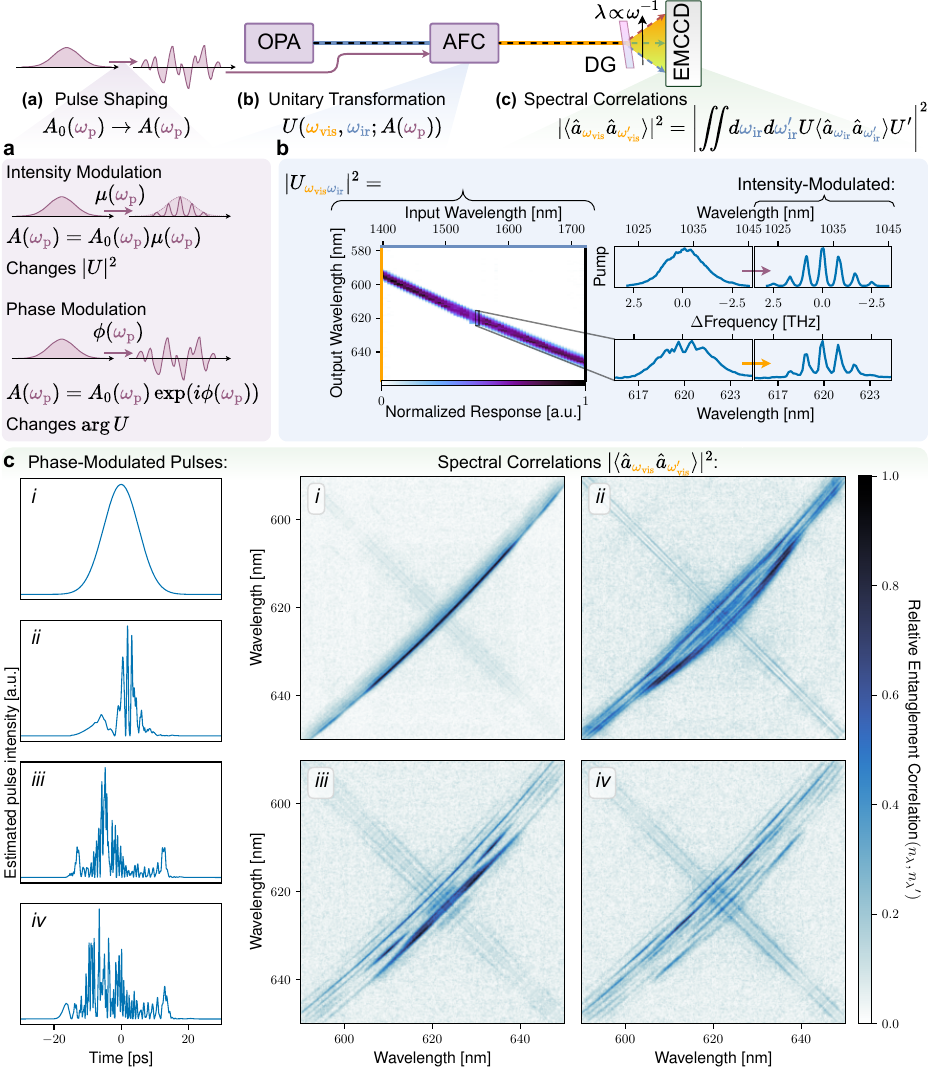}
\caption{
 \textbf{Preparing the joint spectrum via frequency conversion.}
 By pulse shaping the AFC pump $A(\omega_\text{p})$, we can modify the linear unitary transformation $U$ performed by the AFC process, hence the state and the measured spectral photon-number correlations.
 The pump spectral intensity and phase both play a role in the transformation.
 \textbf{a.}
 Pulse shaping involves two operations: intensity and phase modulation.
 The former ($\mu(\omega_\text{p})$) changes the spectral intensity of the pump, which in turn mainly affects the magnitude of elements of $U$, and the latter ($\phi(\omega_\text{p})$) changes the spectral phase of the pump, which mainly affects the complex phase of the elements of $U$.
 \textbf{b.}
 The phase-averaged linear transformation performed by the AFC.
 The rows of the unitary strongly resemble the pump spectrum, as shown in the subplots: the first pair shows how it compares to the original spectrum, and the second is an example of how it changes given an intensity-modulated pump.
 The frequency-difference axis is shared among the pump and unitary-row subplots.
 \textbf{c.}
 These plots show how a phase-modulated pump affects the conversion process and produces more complicated correlation structure.
 Example spectral intensity correlations are plotted on the right, and the corresponding (inferred) pump intensity profiles on the left.
 Specifically, we plot the entanglement contribution to photon correlation matrices (correlation matrices with the diagonal (the classical contribution) subtracted).
 The first shows the same data in Fig.\,\ref{fig:afc}b, for reference, with no pulse shaping.
 The next three are the result of random pulse shaping.
 The corresponding pump intensities, plotted on the right, are not measured directly, but estimated based on the phase modulation applied and the pulse shaper calibration.
 }
\label{fig:covs}
\end{figure}

Most quantum sensing or computing protocols in the optical domain require multiport-beamsplitter-like unitary operations to be performed on the overall state,
because each application requires some specific entanglement structure.
Here we demonstrate how the covariance matrix can be transformed by implementing linear-optical frequency unitaries through the broadband sum frequency generation process.
The unitary is applied during the conversion;
shaping the AFC pump pulse affects the nonlinear dynamics, changing which wavelengths convert to which.
This means that, within certain constraints, we can program the pulse to achieve a desired transformation.

Intuitively,
the relative phase of a pump frequency affects the phase of a signal it converts, thereby constructive and destructive interference may promote or suppress the conversion of a signal from one frequency to another, effectively forming a frequency-domain interferometer network.
This is illustrated in Fig.\,\ref{fig:covs}a--b.
We discuss this further in Appendix~\ref{sec:app:afclzgrid}.

Fig.\,\ref{fig:covs}b shows a measurement of the intensity transfer function, and how it changes with the spectral amplitude modulation of the AFC pump.
The effects of phase modulation are only observed in the covariance matrix, and some illustrative examples, which qualitatively demonstrate phase programmability, are shown in Fig.\,\ref{fig:covs}c.
The magnitude of the transfer matrix is kept roughly the same, as there is no amplitude modulation, yet new features appear in the photon covariance matrix, due entirely to a change in the phase profile.
This is explained by the interference effects between input modes during conversion, as expected under a beamsplitter-like unitary operation.
Phase modulations are applied with the pulse shaper, on top of a fixed quadratic phase (chirp) to guarantee a certain pump pulse duration.
The predicted pulse shapes (inferred from the applied phase modulation) are shown for each example.
For clarity, we show the correlation rather than covariance matrices,
and we subtract a fit to the diagonal (corresponding to the classical) part in order to show only the entanglement contribution, which can be thought of as the joint spectrum in two mode squeezing.
This is purely for visualization, as the diagonal is always the strongest feature, and does not significantly deviate from a Gaussian-like profile.
For reference, Fig.\,\ref{fig:covs}c(i) is derived from the same data as Fig.\,\ref{fig:afc}b.

In our experiment the bandwidth of the pump (\textless5\,THz) is much smaller than the bandwidth of the squeezed light ($\sim$47\,THz), which prevents all-to-all coupling
(the pump frequencies mediate the change in signal frequency).
Additionally, the AFC pump peak intensity is reduced by pulse-shaping, therefore reducing the efficiency.
Despite these limitations, it is possible to achieve qualitatively different joint spectra by simply changing the pulse shape.

\subsection*{Multimode quantum state sampling at visible wavelengths}\label{sec:results:sampling}
\addcontentsline{toc}{subsection}{\nameref{sec:results:sampling}}

\begin{figure}[htb]
\centering
\includegraphics[width=\normaltextwidth]{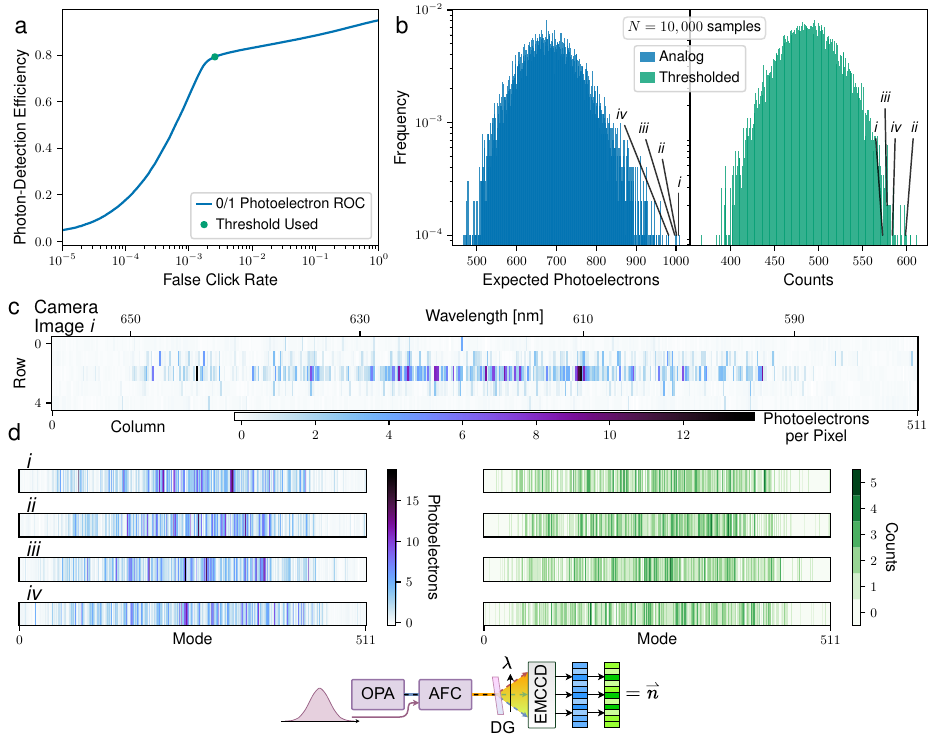}
\caption{
 \textbf{Parallel single-photon detection for multimode-quantum-state measurement: sampling the many-mode, many-photon distribution.}
 \textbf{a.} Receiver operating characteristic (ROC) curve for the EMCCD camera's detectors.
 This quantifies the trade-off between false click rate and photon-detection efficiency (PDE).
 \textbf{b.} Histogram of the total photon number per event: analog (noisy) and thresholded.
 We use the term photoelectrons to refer to the amplified charge on the CCD divided by the gain.
 Some example, high-photon-number events, are labeled on the histogram with callouts \textit{i--iv}; these are referenced in the subsequent subfigures.
 \textbf{c.} Example unprocessed camera image from one sampled event.
 \textbf{d.} Samples integrated vertically (left) and thresholded then summed vertically (right).
 The experimental schematic (bottom) refers to the experimental configuration for collecting the data shown in \textbf{b}--\textbf{d}.
 The data in \textbf{a} consists of dark frames collected with a closed shutter (see Appendix~\ref{sec:app:emccd:thresh} and \ref{sec:app:spectrometer:camchoice}.)
}
\label{fig:sampling}
\end{figure}

The final ingredient in any quantum-optical computing or sensing protocol is measurement, and in this work we focus on photon-counting measurements.
Any quantum advantage for computing or sensing typically scales with the number of photons detected
so high average photon numbers and high detection efficiency are crucial.
Here we demonstrate sampling of the photon-count distribution of our generated states using an EMCCD camera.

We used 5 rows and all 512 columns of a 512$\times$512 pixel EMCCD camera, whose CCD sensor has a QE of $\sim$95\% at 620\,nm \cite{nuvuhnuspec}.
The receiver operating characteristic (ROC) curve for thresholding for this camera is shown in Fig.\,\ref{fig:sampling}a:
this characterizes the false click rate against the photon-detection efficiency (PDE), parametrized by the threshold value (Appendix~\ref{sec:app:emccd:thresh}).
The green marker indicates the threshold used to generate the subsequent plots, resulting in a competitive photon-detection efficiency of $\sim$80\%.

Fig.\,\ref{fig:sampling}b shows a histogram of the number of photons detected per shot.
Blue (left) is prior to thresholding, green (right) is with thresholding.
The gain noise in EMCCDs is large, which makes it practically impossible to distinguish between photon numbers $\geq$1 incident on one pixel.
However, statistical averages tend to be accurate, as this is zero mean noise \cite{LantzDevaux} (see Appendix~\ref{sec:app:emccd}).
Henceforth, we use the term photoelectrons to refer to the amplified charge on the CCD divided by the gain: since the gain is a noisy process, this is not quantized.
The first histogram implies an average of almost 700 photons per shot; with thresholding the average number of clicks is just under 500 per shot.
The discrepancy is due to both the effective QE and the high rate of multiple photons incident on one pixel.
In Appendix~\ref{sec:app:afc:fluo} we analyze the contribution of dark counts and photons from parasitic processes:
these are not significant in number compared to the squeezed light photons.

For these frequency-resolving experiments we use more than one pixel-rows of the camera.
This is done in order to capture all the photons, due to the spectrometer's point spread function occupying a space larger than one pixel (Gaussian width of $\sigma_\text{PSF} \approx 0.6$~pixels).
An example of a raw sample is shown in Fig.\,\ref{fig:sampling}c, in units of photoelectrons.
With this configuration we could sample at a rate of just over 800\,Hz.
Using a single row would allow an average sampling rate of 15.7\,kHz, which is limited by data readout times.
Fig.\,\ref{fig:sampling}d shows samples integrated vertically, for each frequency bin mode.
The point spread function in the vertical direction can allow a thresholding camera to act as a pseudo-photon-number resolving detector, by illuminating more pixels (via astigmatic focus) (an established method \cite{FitchFranson, PaulJex, SperlingAgarwal} used e.g.\ in Refs.~\cite{DengPan}).
The point spread function in the horizontal dimension, however, acts as an effective decoherence and must be carefully engineered to fit within the dimensions of a pixel (see Appendix~\ref{sec:app:spectrometerpovm}).

\section*{Discussion}\label{sec:discussion}
\addcontentsline{toc}{section}{\nameref{sec:discussion}}

The creation, manipulation, and detection of highly multimode entangled squeezed states are important ingredients of many continuous-variable (CV) schemes for quantum computing, sensing, and communication using photonics.
When compared to space or time encoding, frequency encoding has significant advantages for integration and scalability, but it is not always practical to perform unitary transformations on frequency modes without substantial loss.
In this work, we have experimentally demonstrated efficient, broadband, quantum frequency conversion of highly multimode squeezed light generated in the near infrared into the visible using adiabatic frequency conversion (AFC), maintaining and manipulating underlying non-classical photon-number correlations.
This simultaneously allows efficient, parallel photon counting across over 400 squeezed and 500 detection modes using CCD-based photon-detector arrays,
and the programming of frequency correlations in the multimode squeezed light.
Specifically, we have shown that mode transformations of optical frequency modes can be performed during the conversion stage by pulse-shaping the pump to steer the conversion process.
The wavelength conversion enabled the use of an EMCCD camera as a high-quantum-efficiency, many-spatial-mode threshold detector on light that was originally in the near-infrared, which is well outside the detection range of the camera.
Our approach requires no active phase locking, uses a single optical beam path for the quantum light, and the number of modes and shot rates should be scalable well beyond what we have demonstrated.
The frequency-domain unitary is programmed by shaping a classical pump pulse before it interacts with the quantum light; the spectro-temporal shaping of the AFC pump pulse can be lossy and, so long as the pulse energy is sufficient to fully drive the AFC process, this loss will not affect the quantum light.
This is in contrast to the methods for realizing programmable unitaries with space or time encodings, which typically involve a tradeoff between programmability and loss.
Finally, our contributions also include the development and application of homodyne-free calibration methods for characterizing broadband frequency-multimode pulsed squeezed light, which we used to determine: the wavelength correlations, the effective transmission efficiency -- how well these correlations are preserved -- and the underlying number of supermodes that make up the state.

To the best of our knowledge, the quantum-optical states we produced are the largest partially configurable, photon-counted, multimode squeezed states produced by a moderate factor (about 2--4 times larger than previous results, albeit with different limitations and caveats), and by an order of magnitude the largest in the frequency domain
(see Appendix~\ref{sec:app:mmscomparison}).
Our work provides a path to constructing large-scale Gaussian boson samplers using frequency encoding.
Other possible applications may be found in quantum neural networks, variational quantum sensing and quantum networking.
Quantum reservoir computers \cite{GongPan, CiminiMcMahon}, can solve computational tasks via the variation of system parameters, and full control is not strictly necessary.
Just as for computing, many squeezed-light protocols for metrology also use photon detection (e.g. Refs.~\cite{BanchiPirandola, ZhuangZhang}, which show a quantum advantage in pattern recognition and object detection), so efficient measurements over hundreds of modes would present a practical advantage.
Furthermore, a variational optical beamsplitter unitary would bring about analogous benefits as those shown Ref.~\cite{MarciniakMonz} for magnetic-field sensing, but adapted, for example, to broadband optical phase or absorption sensing \cite{HermanDiddams, HaririZhang}.
The frequency domain unitary aspect of our work may also translate to quantum networking applications, for example, in conjunction with wavelength multiplexing of atomic or ionic systems \cite{ZhaoReza}.
By fully leveraging broadband nonlinear optics as we propose, multiplexing would offer not only high parallelism, but the entanglement-swapping operations could also be performed in the frequency domain.

Our experiment has limitations on the connectivity and programmability of the unitary transformation realized by AFC, as well as limited detection resolution and speed.
We explain these limitations and present prospects to overcome them in Appendix~\ref{sec:app:prospects}.
In addition, we have shown non-classical statistics using photodetection of light from a OPA followed by frequency conversion, but we did not show quadrature squeezing by performing homodyne measurement, due to the difficulty of engineering a local oscillator to match the spectro-temporal profile of the supermodes.
This would be possible in principle with an experimental setup adopting methods similar to those of Refs.~\cite{RoslundTreps,CaiTreps,RomanParigi}, scaled to a comparable number of modes and with the additional step of frequency conversion.
Similarly, quantitative benchmarking of the fidelity would be crucial in evaluating the system's programmability, and is currently limited by the ability to fully model -- thus program -- the system, due to a simulation--reality gap.

We will conclude with some thoughts on future research that could build on what we have already shown.
First, engineering the bandwidths of the squeezed light and of the pump to be the same will enable all-to-all correlations (i.e\ filling in the empty areas in Figs.\,\ref{fig:covs}b and c, as the pump would bridge the frequency gap between all the signal frequencies), and provide further evidence in support of the concept.
Second, there are many opportunities for improvements using integrated, on-chip, devices \cite{NehraMarandi, LedezmaMarandi}.
For example, one could implement the combined OPA and AFC source on a single chip, with single-spatial-mode waveguiding.
This could:
eliminate detrimental spatio-temporal effects in AFC;
allow for longer interaction lengths than in free space (limited by Rayleigh length) and allow for high intensities with less power, both resulting in more powerful unitary control;
and eliminate most interface losses of the quantum light prior to detection.
Using integrated photonics would also allow the engineering of dispersion to optimize the squeezing bandwidth, the number of modes in the OPA, and the interaction complexity imparted by AFC pump.
Frequency-encoded squeezed light is not restricted to using supermodes of a continuous basis: quantum states can be engineered to inhabit a more conventional basis of (ultrafast) frequency combs lines or discrete frequency bins \cite{OmiKannari, HurvitzArie, DragoAgata, MorrisonFedrizzi, FolgeSilberhorn}.
In Appendix~\ref{sec:app:simpleunitary} we show how tuning the pump's spectral amplitude and phase can be used to realize broadband two-mode unitary transformations with these kinds of input modes;
this procedure could be generalized to an arbitrary mode number.
Shaping the OPA pump (and possibly the domain poling) to engineer the OPA joint spectrum \cite{OmiKannari, HurvitzArie, DragoAgata, MorrisonFedrizzi, RozenbergArie} provides additional avenues for programmability, including the squeezing level of each individual supermode.
Finally, understanding how to program the AFC unitary, e.g.\ through the development of more quantitative models and better calibration procedures or variational feedback, is a valuable avenue of research;
it is also important to understand what parts of the state (Hilbert) space can be reached through OPA and AFC pulse shaping, as well as potentially incorporating additional frequency-domain operations
(Appendix~\ref{sec:app:afclzgrid} discusses the theory of AFC transformations).
Overall we hope our work will enable more widespread study of many-mode, many-photon entangled quantum states, motivate theory effort in modeling multimode adiabatic transformations, and provide a useful building block for large-scale frequency-encoded CV quantum technologies.


\vspace{\baselineskip}
{\centering \noindent\rule{0.5\normaltextwidth}{0.4pt}\par}
\vspace{\baselineskip}

\section*{Methods}\label{sec:methods}
\addcontentsline{toc}{section}{\nameref{sec:methods}}

\setcounter{figure}{0}
\renewcommand{\thefigure}{M\arabic{figure}}
\renewcommand{\theHfigure}{M\arabic{figure}}

\subsection*{Highly multimode quantum light in the frequency domain: details}\label{sec:methods:multimodesqueezing}
\addcontentsline{toc}{subsection}{\nameref{sec:methods:multimodesqueezing}}

To study the squeezing process, the OPA is first simulated with a split step-based partial differential equation solver, based on material properties of LN, and the pump laser (such as dispersion and nonlinearity; pulse duration and energy).
The Green's function, which describes how the OPA acts on the vacuum, is computed with this solver.
The Bloch--Messiah decomposition is then applied: this helps us interpret how the OPA generates squeezing \cite{WasilewskiRadzewicz, HosakaKannari},
by reducing the Green's function to a set of supermodes and corresponding squeezing or anti-squeezing values.
These are shown in Fig.\,\ref{fig:dopa}a and b.
For a 4\,cm LN waveguide, and 775\,nm 200\,fs pulse pump, we predict $\sim$600 supermodes with substantial average photon numbers, spanning over 50\,THz, or over 400\,nm.
Refer to Appendix~\ref{sec:app:multimodeQO:decomp} for the OPA equations of motion and the relationship between the Green's function and the Bloch--Messiah supermodes.

Three experiments were subsequently performed to verify the numerical predictions: a spectrum measurement, a parametric gain measurement and a coincidence detection measurement.
The purpose of the first is to verify that the bandwidth is indeed as broad as predicted.
The second is to infer the number of modes, based on how the output energy scales.
The third is to corroborate that this light is made up of squeezed states, through photon statistics.
The spectrum may be thought of as the incoherent sum of the supermodes, scaled by average photon number,
while for the parametric gain, we measure the total photon number as a function of pump power.

The spectrum is measured by monochromation of the OPA output beam to 1\,nm resolution and detecting with a single-photon avalanche diode (SPAD).
The spectrum could not be measured past about 1700 nm, due to the SPAD's InGaAs detector response cutoff.
As predicted by simulation, the output is broad, spanning hundreds of nanometers (see Fig.\,\ref{fig:dopa}c).
The spectrum is in fact broader than simulation, with a shoulder appearing and extending beyond the prediction around 1400\,nm. This additional phase matching may be due the presence of additional spatial frequencies in the poling.
The 775\,nm pump, unless otherwise specified, is generated from a 200\,fs, 1033\,nm laser by nonlinear frequency conversion.

The parametric gain experiment involves measuring the output power of the OPA with a power meter at varying pump powers (see Fig.\,\ref{fig:dopa}d).
A high repetition rate laser is necessary to obtain a sufficiently high average output power, hence an 80\,MHz Ti:S laser is used to pump the OPA.
The data is fit to the function $\eta M \sinh^2 \sqrt{P / P_0}$, where $\eta$ represents overall transmission, $M$ number of modes, $P$ pump power, and $P_0$ power scaling factor (a single squeezed state has $\langle n \rangle = \sinh^2 r$, with squeezing parameter $r \propto \sqrt{P}$).
This fit assumes $M$ equally squeezed modes each contributing the same number of photons (thus, a lower bound).
Remarkably, the simulation, which includes the responsivity of the detector as a function of wavelength, scales ostensibly the same (albeit with a different nonlinear efficiency $P_0$).
From this measurement we can infer the presence of at least 430 modes -- on the order of 650 modes if accounting for detector responsivity.
The Ti:S pulses, 100\,fs, have more bandwidth than the 200\,fs upconverted 775\,nm.
This is accounted for in simulation, which predicts that the parametric gain bandwidth will be moderately broader, hence a marginally larger $M$ than in the 200\,fs case, but the central modes are not significantly affected.
If a 200\,fs experiment were possible at a lower repetition rate, we predict a slightly lower measurement of $\eta M$ (see Appendix~\ref{sec:app:tis} for comparisons).

The coincidence experiment (inspired by Ref.~\cite{VaidyaVernon}) involves splitting the OPA output beam with a broadband 50:50 beamsplitter.
Each beam is detected by a gated SPAD, with the goal of measuring the photon-number covariance between the two beams.
The coincidences depend on the photon statistics of the state (namely the bunching, which may also be thought of in terms of g$^{(2)}$ or Mandel Q).
Fig.\,\ref{fig:dopa}e shows the covariance between SPAD clicks against the average click rate (as these are threshold detectors).
At low photon numbers, only the coincidence rate of highly multimode squeezed light is expected to scale linearly with respect to the average photon number (see Appendix~\ref{sec:app:coincidence} for details).
To verify this, we performed coincidence detection with the OPA light, attenuated 1550\,nm laser light, and broadband incandescent lamp light.
Indeed, only the unseeded OPA light has nonzero, linear covariance where $\langle n \rangle \ll 1$.
As the thresholding detectors yield Bernoulli random variables, the covariance between the two is restricted to a parabola, hence the covariance curves back down around $\langle n \rangle = 1/2$.
The slope of the function depends on the transmission (loss).
In the \hyperref[sec:results:afc]{Section} that followed we use this method as one way to estimate the purity of the detected visible light, but due to the significantly inhomogeneous responsivity of the detectors, this is not possible in the infrared.
We can, however, compare to theory (``Predicted'' in Fig.\,\ref{fig:dopa}e):
a simple model (see Appendix~\ref{sec:app:c:threshold}) can approximately predict the output, based on the spectrum in Fig.\,\ref{fig:dopa}c and the detector QE, assuming biphotons centered around 1550\,nm.
It predicts a similar scaling to the one measured.
As this coincidence measurement is quite sensitive to wavelength-dependent loss,
the prediction estimates a decoherence amounting to 98\% effective loss while we measured 96\%, hence the factor of two between measurement and prediction.
This deviation is due to inaccuracies in the QE and spectrum measurement.
We emphasize that this decoherence is due to the limited spectral range and QE of the infrared detector, and not due to the purity of the states.

These three experiments support our predictions that the OPA is a source of highly multimode squeezed light, spanning hundreds of frequency modes.

\subsection*{OPA Waveguide}\label{sec:methods:dopawg}
\addcontentsline{toc}{subsection}{\nameref{sec:methods:dopawg}}

The OPA waveguide consists of 4\,cm MgO-doped LN ridge waveguide (Covesion WG-1550-40 WGCK40), poled for second harmonic generation of 1550 to 775\,nm and designed to be single-mode at near-IR wavelengths.
The LN device is mounted in an oven (Covesion PV40).
The oven is mounted with a V-shaped mount (Thorlabs VC3C) and a 3D-printed sleeve to ensure horizontal orientation.
The oven and objective lenses (Newport 5726-B-H, 5726-C-H) are mounted on linear axes stages (Newport 562-XYZ, 561D-XYZ-LH, 561D-XYZ respectively).
Prior to the waveguide input, a dichroic mirror is used to merge the 775/1550\,nm pump and alignment beams (Layertec 109068).
A custom mirror (Layertec) is used following the output to separate the pump and signal.
The output pump beam is imaged onto a camera (Basler acA800-510um) to view and optimize the pump's spatial profile in the waveguide.

\subsection*{AFC}\label{sec:methods:afc}
\addcontentsline{toc}{subsection}{\nameref{sec:methods:afc}}

The crystal used for AFC is a 3 cm poled KTP crystal (Raicol Crystals).
The poling profile is designed such that the spatial frequency (inverse of the poling period) varies linearly over the length of the crystal, except at the facets of the crystal.
The poling spatial frequency is designed to phase match the conversion of all the target wavelengths.
The front and back 0.25\,mm of the crystal are poled with a rapidly varying tanh profile.
This improves the rapid passage of the wavelengths that convert near the ends of the crystal.
See Appendix~\ref{sec:app:afc:afcdesign} for more details.
The crystal is designed for use at 48\textdegree C and held in an oven (Eksma Optics HP30), mounted on a 7-axis mount (Thorlabs K6XS, SM1P1).

To optimize the conversion efficiency, bright CW laser light is sent through the OPA: it thus has same spatial mode as the down-converted light generated by the OPA.
This alignment beam is used to optimize the spatial overlap of the pump with the signal.
The conversion efficiency of the pulsed OPA light is then optimized by maximizing the temporal overlap.
This consists of a sweep of the time delay, followed by a sweep of the pump pulse chirp (i.e.\ duration).
As AFC is phase-insensitive, there is no need for phase stabilization.
The delay line, a motorized stage with a retro-reflecting right-angle mirror, adjusts the timing of 775\,nm pump prior to the OPA waveguide.
The chirp is programmed by adjusting the spectral phase profile of the pulse shaper.
See the spectrometer section below
for more details.

To mitigate spatio-temporal effects in the AFC, we designed the pump to have a beam waist of 300\,{\textmu}m, and the signal beam to have a waist of approximately 70\,{\textmu}m when focused in the crystal.

To measure the phase-averaged response of the AFC process, monochromated super-continuum (NKT Origami, Thorlabs HN1550, and the infrared monochromator described below) is sent through the AFC, one wavelength at a time, and the resulting visible spectrum measured on the EMCCD camera spectrometer (described below).

\subsection*{775~nm generation from 1033~nm}\label{sec:methods:lbo}
\addcontentsline{toc}{subsection}{\nameref{sec:methods:lbo}}

Pulsed 775\,nm light is generated by cascading a second harmonic generation (SHG) and OPA process.
Both processes use lithium triborate (LBO; Newlight Photonics).
Both crystals are temperature tuned and non-critically phase matched (the optical polarization are aligned to the crystal axes, to avoid spatial walk-off).
The lengths are designed to match the temporal walk-off of the 200\,fs pulses.
The first crystal, 6\,mm long, held at 182.6\textdegree C, generates 517\,nm ($y$-axis polarized) from 1033\,nm ($z$-axis).
The second, 10\,mm long, held at 145.4\textdegree C, carries out the 517\,nm~($y$)~$-$~1548.5\,nm~($z$)~$\rightarrow$~775\,nm~($z$) process.
The 1548.5\,nm seed is generated by a diode laser (ILX Lightwave 79800C), amplitude-modulated by a stabilized (Oz Optics MBC-SUPER-PD-3A) electro-optic modulator (EOM) (Eospace AZ-DS5-10-PFA-PFA-LV-LR), to generate 10\,ns square pulses.
These seed pulses are optically amplified (Pritel PMFA-20), following an isolator.
The EOM pulsing is triggered by the Amplitude Satsuma 1033\,nm laser, with an appropriate time delay (IDQuantique ID900).
The EOM is driven by a pulse/function generator (HP-8116A).

\subsection*{Parametric gain measurements}\label{sec:methods:parametricgain}
\addcontentsline{toc}{subsection}{\nameref{sec:methods:parametricgain}}
For this measurement, the OPA is pumped with a 775\,nm, 80\,MHz Ti:S laser (Spectra-Physics Tsunami).
(All other experiments use the 775\,nm pump generated from a 1033\,nm laser.)
The high repetition rate ensures a sufficiently high average power of down-converted light.
The 1550\,nm output beam power is measured with a pW-sensitive power meter (Thorlabs S154C), after coupling into multimode fiber (Thorlabs FG200LEA), which is about 95\% efficient.
The OPA pump power is removed with a dichroic mirror (Layertec) and measured concurrently (Thorlabs S130VC/S130C).
The amplified vacuum light is filtered further with long-pass filters (Thorlabs FELH1150).

At visible wavelengths, the 620\,nm power is measured on the camera by integrating the signal on all illuminated pixels, after appropriate filtering of the pump (Thorlabs DMLP900; Semrock FF01-632/148).

\subsection*{Single photon spectrometers}\label{sec:methods:spectrometers}
\addcontentsline{toc}{subsection}{\nameref{sec:methods:spectrometers}}

The infrared wavelength spectrometer is based on a diffraction grating on motorized rotation stage (Thorlabs GR25-0616; K10CR1).
The first order reflection couples into SMF-28 single mode fiber (Thorlabs F260APC-1550), which monochromates the input.
The fiber coupling efficiency is approximately 35\%.
For single photon detection, this coupled into a InGaAs SPAD (IDQuantique ID Qube NIR Gated), set to the nominal 15\% QE.
The wavelength-angle correspondence is calibrated using tunable lasers (JDSU mTLG-C1C1L1) between 1527\,nm and 1609\,nm and fitting to the grating equation, $m \lambda = d (\sin(\theta_{i0} + \Delta) - \sin(\theta_{ir} - \Delta) = 2 d \cos[(\theta_{i0} + \theta_{r0}) / 2] \sin[(\theta_{i0} - \theta_{r0}) /2 + \Delta]$, where $d$ is the grating constant, $\theta_{i0}, \theta_{r0}$ are some reference incidence and reflection angles, and $\Delta$ is the rotation angle of the grating.
The extrapolation to uncalibrated wavelengths is deemed correct as the spectrum stops sharply at 1150\,nm, which matches the long pass filter cutoff (Thorlabs FELH1150).
However, the coupling efficiency as a function of wavelength uncalibrated.

The visible light spectrometer is based on a diffraction grating (Ibsen PCG-1908/675-972) imaged by an objective lens (Olympus UPLFLN4x) onto the N\"uV\"u HN\"u 512 IS EMCCD camera.
See Appendix~\ref{sec:app:spectrometer:design} for more details.
The wavelength-pixel correspondence is calibrated by monochromating supercontinuum (NKT Origami, Thorlabs HN1550) with the infrared monochromator, and converting this light through AFC.
The AFC pump is amplitude-modulated to a narrow bandwidth, thus also effectively monochromated.
The monochromated supercontinuum wavelength-to-angle is calibrated with an optical spectrum analyzer (Ando AQ6317B).
An EM gain of 3000 is used on the camera.

The SPAD and camera are triggered by the Amplitude Satsuma laser, with an appropriate time delay (IDQuantique ID900).

\subsection*{Coincidence detection}\label{sec:methods:coincidence}
\addcontentsline{toc}{subsection}{\nameref{sec:methods:coincidence}}

In the infrared, the beam out of the OPA is split with a broadband beamsplitter and achromatic waveplate (Thorlabs UFBS50502, AHWP10M-1600).
The two beam paths are coupled into two SPADs (IDQuantique ID Qube NIR Gated) using fixed collimators (Thorlabs F260APC-1550).
The SPADs are set to the nominal 15\% QE.
The fiber coupling efficiency is approximately 85\%.
Data acquisition and triggering is configured with an IDQuantique ID900 time controller.
This includes feedback for enforcing a global dead time after either of the SPADs fired, and reducing the triggering rate to 2 MHz (maximum SPAD response rate) when using the 80\,MHz Spectra-Physics Tsunami Ti:S.

The coincidence detection of visible light is performed with the N\"uV\"u EMCCD camera.
The diffraction grating is replaced with a Wollaston prism (Thorlabs WP10-A) rotated to achieve 50:50 in the two output beams, at around 45\textdegree.
The beams are focused with a 30\,mm lens (Thorlabs AC254-030-A-ML).
The camera is configured to detect in two separate regions of interest (opposite corners of the CCD).
The signal from these two 16$\times$16 pixel regions are integrated to obtain the overall power.
An EM gain of 500 is used on the camera, which is used in analog mode.

The SPADs and camera are triggered by the Amplitude Satsuma 1033\,nm laser, with an appropriate time delay (IDQuantique ID900).

Refer to Appendix~\ref{sec:app:coincidence} for a theoretical motivation for these experiment.

\subsection*{Temporal pulse shaper}\label{sec:methods:pulseshaper}
\addcontentsline{toc}{subsection}{\nameref{sec:methods:pulseshaper}}

Light diffracted by a transmission grating (Ibsen Photonics PCG-1765-808-981) is focused by a 150\,mm cylindrical lens (Thorlabs LJ1629L2-B) onto an spatial light modulator (SLM; Meadowlark P1920-0600-1300-PCIe).
A vertically oriented blazed grating is written on the SLM (orthogonal to the direction that frequency components are dispersed).
The SLM is oriented such that the light reflected in the 1st order travels back through the cylindrical lens and grating, following the same path except for a slight downward angle, such that backward-traveling beam is separated by a pick-off mirror.
The spatial phases (vertical translations) of the blazed grating along the wavelength axis impart spectral phases to the pulses.
Amplitude modulation is also possible by reducing the grating amplitude at a given wavelength, thereby reducing the diffraction efficiency.

The pixel-column-to-wavelength calibration is performed by writing narrow Gaussian-shaped amplitude modulation patterns along the wavelength axis of the SLM.
The beam is then measured with an optical spectrum analyzer (Ando AQ6317B) to determine the transmitted wavelength.

The distance between the three optics is optimized by rounding the ellipticity in the beam shape, as viewed on a camera, and minimizing the pulse duration, measured with an autocorrelator (APE Pulsecheck).
The orientations of the SLM, grating, and input mirrors are tuned to eliminate the spatial chirp after the initial alignment.
The spatial chirp as a function of the transverse directions of the beam is measured by translating the end of a multimode fiber connected to an optical spectrum analyzer (Ando AQ6317B).

The pulse shaper is about 60\% efficient without any phase modulations.
Phase modulations can reduce the efficiency if the features are smaller than the imaging resolution of one wavelength onto the SLM.

This pulse shaper is based on the design in Ref.~\cite{FrumkerSilberberg}.

\subsection*{EMCCD camera operation}\label{sec:methods:cameraanalog}
\addcontentsline{toc}{subsection}{\nameref{sec:methods:cameraanalog}}

The N\"uV\"u HN\"u EMCCD camera is water-cooled by a thermo-electric chiller (Solid State TCube Edge), and the camera operates with a CCD temperature of -60\textdegree C, cooled by the built-in TEC.
The camera is mounted on one rotation and two linear stages (Newport RS65; Thorlabs PT1A).
One linear stage is used to center the camera with respect to the beam.
The other two are actuated by motorized micrometers (Thorlabs Z812B; Newport TRA25CC), in order to align the camera to the objective's focal plane, such that the entire spectrum is in focus when used as a spectrometer.
These two degrees of freedom are optimized by minimizing the point spread function of two different wavelengths, alternating until convergence.

The laser repetition rate of the Satsuma is chosen to match the maximum frame-rate of the camera (under a given experimental configuration), and the camera is triggered by the laser (IDQuantique ID900 is used to adjust the delay).
The blanking and exposure times are generally set to 0.1\,ms, as these values do not adversely affect the frame-rates within the experimental configurations reported.

The camera digital readout (``pixel value'', $p$) is converted to photo-electrons, $\langle n_e \rangle$, by subtracting the bias $b$ and dividing by the total gain.
The latter is comprised of the electron multiplication (EM) gain and the analog-to-digital conversion factor $k$ (here 21.43 photoelectrons per pixel unit).
Hence $\langle n_e \rangle = k p / g - b$.
Bias subtraction is calibrated per pixel where possible, as these exhibited small variations.
See Appendix~\ref{sec:app:emccd} for an explanation of the measurements performed with an EMCCD.\\

\begin{figure}[htb]
\centering
\includegraphics[width=0.72\normaltextwidth]{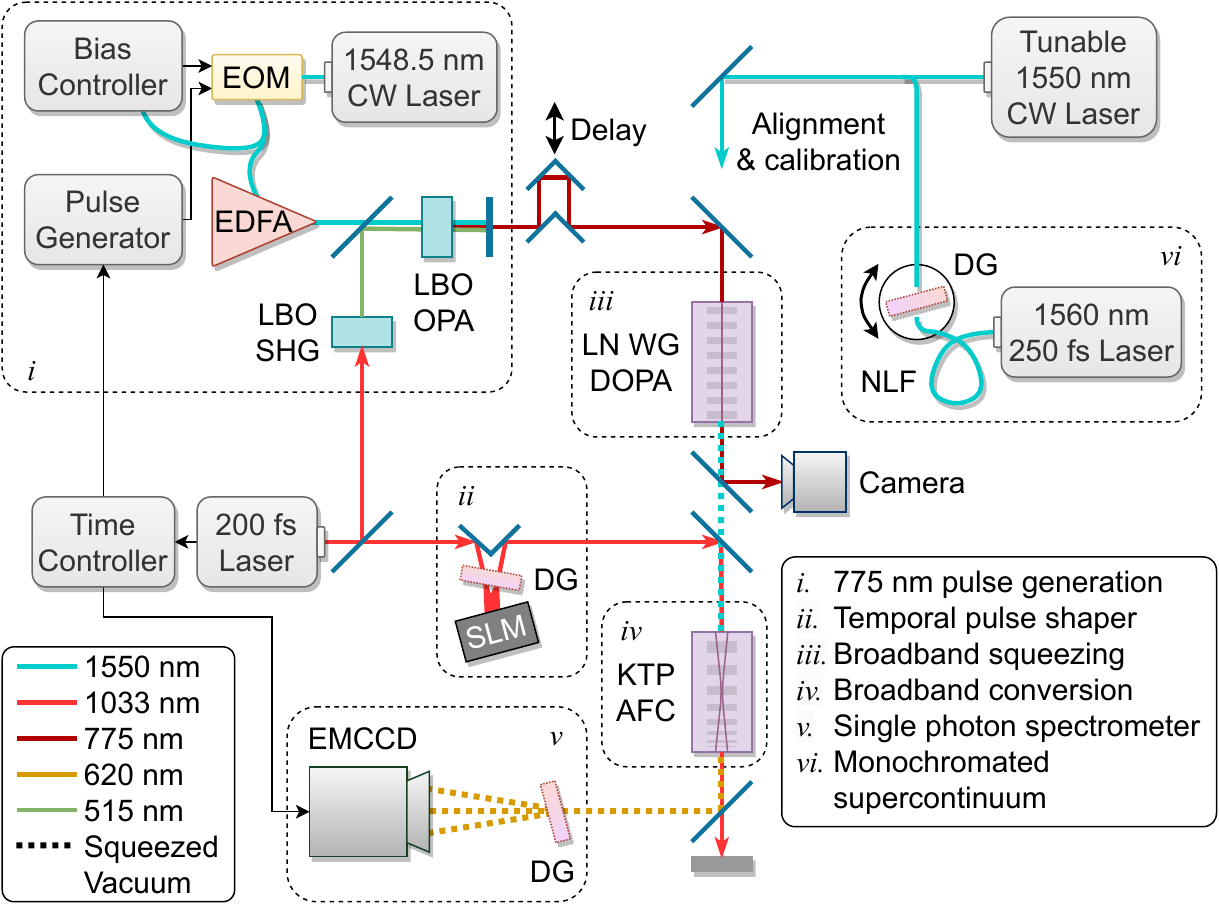}
\caption{
 \textbf{Simplified experimental diagram.}
 EDFA: erbium doped fiber amplifier.
 WG: waveguide.
 NLF: nonlinear fiber.
 CW: continuous wave.
}
\label{fig:experimentdiagram}
\end{figure}

\noindent%
See Appendix~\ref{sec:app:photographs} for photographs.
Fig.\,\ref{fig:experimentdiagram} shows an experimental diagram of the experiment.

\vspace{\baselineskip}
{\centering \noindent\rule{0.5\normaltextwidth}{0.4pt}\par}
\vspace{\baselineskip}

\phantomsection
\addcontentsline{toc}{section}{Declarations}

\bmhead{Acknowledgments}

We acknowledge helpful discussions with Noah Flemens, Alexey Gorshkov and Mandar Sohoni.
We thank Valeria Cimini, Alen Senanian, James Williams and Ryotatsu Yanagimoto for helpful comments regarding the manuscript.

\bmhead{Funding}
We thank NTT Research for their financial and technical support.
Portions of this work were supported by the National Science Foundation (award CCF-1918549) and a David and Lucile Packard Foundation Fellowship.
P.L.M.\ acknowledges membership of the CIFAR Quantum Information Science Program as an Azrieli Global Scholar.
This research was supported in part by an appointment to the Intelligence Community Postdoctoral Research Fellowship Program at Cornell University, administered by Oak Ridge Institute for Science and Education through an interagency agreement between the U.S.\ Department of Energy and the Office of the Director of National Intelligence.

\bmhead{Data availability}
Experimental data and scripts to replicate the figures in this paper based on this data, are available at
\url{http://doi.org/10.5281/zenodo.10492956}.

\bmhead{Code availability}
All simulations were performed using our software library for nonlinear-optical design, simulation, and Gaussian quantum optics, which is found at \href{https://github.com/mcmahon-lab/MultimodeNonlinearOptics}{github.com/mcmahon-lab/MultimodeNonlinearOptics}.

\bmhead{Author contributions}
F.P.\ and L.G.W.\ designed the experimental setup.
F.P.\ performed the simulations; designed the nonlinear optics; and built and ran the experiment with assistance from L.G.W.\, B.K.M.\ and S.-Y.M.
S.-Y.M.\ evaluated and calibrated photon-counting cameras for multimode squeezed light detection.
T.W.\ designed and tested the EMCCD-camera spectrometer.
F.P.,\ L.G.W.,\ T.O.\ and S.-Y.M.\ worked on the quantum-optical modeling of the experiment.
P.L.M.\ and L.G.W.\ conceived and supervised the project.
All authors contributed to preparing the manuscript.


\pagebreak
\newgeometry{
    top={26mm},
    headheight={12pt},
    headsep={5.15mm},
    text={\dimexpr8.5in-40mm,\dimexpr11in-50mm},
    marginparsep=5mm,
    marginparwidth=12mm,
    footskip=10.13mm
}

\phantomsection
\addcontentsline{toc}{section}{References}

\putbib[%
 references%
]
\restoregeometry
\end{bibunit}
\pagebreak


\pagebreak
\begin{appendices}
\begin{bibunit}

\setcounter{figure}{0}
\setcounter{table}{0}
\renewcommand{\thefigure}{A\arabic{figure}}
\renewcommand{\theHfigure}{A\arabic{figure}}
\renewcommand{\thetable}{A\arabic{table}}

\startcontents[sections] 
{
\hypersetup{linkcolor=black}
\section*{Appendix Contents}
\printcontents[sections]{}{0}{}
}

\vspace{\baselineskip}
{\centering \noindent\rule{0.5\normaltextwidth}{0.4pt}\par}
\vspace{\baselineskip}

\addtocontents{toc}{\protect\nohyperlinkcontentsline{section}{Theory}{}\%}

\section{Photon statistics in multimode Gaussian quantum optics}\label{sec:app:multimodeQO}

The aim of this section is to provide a working knowledge of the physics of multimode Gaussian quantum optics, especially as it relates to our experiments.
We provide an explanation of the underlying theory in our preferred formalism, show how to apply the nonlinear optics equations, and how to predict the supermodes and squeezing values.
Finally, we describe the photon number properties of Gaussian states.

We refer to as ``Gaussian'' the class of states
that are fully described by the first two moments of its field operators, or, equivalently,
whose phase space distribution is a Gaussian.
These states are the most typical in experimental quantum optics
(this is a consequence of weak optical nonlinearities and the large -- classical -- driving fields that are required to achieve them).
Furthermore, while multimode or multivariable quantum-mechanical states have generally exponentially large representations,
Gaussian states have efficient phase-space representations.
This makes working with large multimode Gaussian states tractable.

\subsection*{The bosonic covariance matrix}
\label{sec:app:multimodeQO:covmat}
\addcontentsline{toc}{subsection}{\nameref{sec:app:multimodeQO:covmat}}

To fully characterize a zero-mean Gaussian state, we must know the squeezing, the thermal noise, the loss and the correlations between all modes; all this information is encoded in the covariance matrix.
In typical convention \cite{Olivares, FerraroOlivaresParis}:
\begin{equation}
\sigma = \frac12 \left\langle\left\lbrace \hat{\xi}, \hat{\xi}^\dagger \right\rbrace\right\rangle - \left\langle \hat{\xi} \right\rangle \left\langle \hat{\xi}^\dagger \right\rangle,
~\hat{\xi}^\intercal = \left(\hat{a}_1, \ldots, \hat{a}_M, \hat{a}_1^\dagger, \ldots, \hat{a}_M^\dagger \right) ,
\end{equation}
where $\hat{a}_i^\dagger$ and $\hat{a}_i$ are the bosonic creation and annihilation operators for some mode $i$, defined in the usual manner.
$\sigma$ is Hermitian and, for later convenience, we can write it in terms of the submatrices:
\begin{equation}
\sigma
=
\left(\begin{matrix}
V + I_M / 2 & U \\ U^* & V^\intercal + I_M / 2
\end{matrix} \right)
\end{equation}
such that $V$ is Hermitian and $U$ is symmetric; $I_M$ is the $M\times M$ identity.
As we shall see, $V \sim \langle \hat a_i^\dagger \hat a_j + h.c.\rangle$ contains the information pertaining to the state's classical properties and thermal correlations, and $U \sim \langle \hat a_i^\dagger \hat a_j^\dagger + h.c.\rangle$ encodes the entanglement and higher-order-correlation physics.

Note that it is more common in the quantum optics and information literature to see the quadrature covariance matrix, which use the real-valued canonical variables $\hat x \propto \hat a + \hat a^\dagger$ and $\hat p \propto \hat a - \hat a^\dagger$.
The quadrature basis has the advantage of being a real-valued symplectic space.
However, the $\hat a$-basis is a more natural convention for photon-number properties of the field.

Overall, the Gaussian states defined by $\sigma$ have $2M^2$ free parameters (real numbers), plus an additional $M$ local phase degree of freedom that may be ignored, as they have no effect on the photon number statistics.

\subsection*{Constructing and decomposing the covariance matrix}
\label{sec:app:multimodeQO:decomp}
\addcontentsline{toc}{subsection}{\nameref{sec:app:multimodeQO:decomp}}

Here we summarize some of the body of work pertaining to the matrix representations of multimode Gaussian systems.
For more depth, see Refs.~\cite{WasilewskiRadzewicz, OpatrnyLeuchs, HosakaKannari, Olivares, AdessoLee, WeedbrookLloyd, FerraroOlivaresParis}.

In this formalism, the space of Gaussian states is closed under Gaussian operations, some of which
are represented as linear transformations.
We call the transformation a Green's function if it represents the outcome of a process described by a linear differential equation.
In a discrete basis of $\hat a$-operators,
any lossless Gaussian operation can be represented as:
\begin{aligned*}
\hat a_\text{out} &= C \hat a_\text{in} + S \hat a_\text{in}^\dagger \\
\hat \xi_\text{out} &= \left(
\begin{matrix}
C & S \\
S^* & C^*
\end{matrix} \right)
\hat \xi_\text{in}
= G \hat \xi_\text{in}
\end{aligned*}
where $C, S$ are linear transformations,
or in a continuous basis parametrized by $\omega$:
\begin{equation}
\hat a_\text{out} = \int C(\omega, \omega') \hat a_\text{in}(\omega') + S(\omega, \omega') \hat a_\text{in}(\omega')^\dagger d\omega' .
\end{equation}
$S$ may only be nonzero only if there is squeezing (if the Hamiltonian contains an $\hat a^\dagger \hat a^\dagger + c.c.$ term).

In our experiment, specifically, we are interested in the Green's function of the OPA process, a combination of parametric amplification and dispersion described by the differential equation:
\begin{equation}
\frac{d \hat a}{dz}(\Delta \omega) = i D(\Delta \omega) \hat a + i A(z, \Delta \omega) * \hat a^\dagger
\end{equation}
where $D$ represents the dispersion (phase matching, group velocity difference, and higher order dispersion), and the coupling term $A$ represents the classical pump field, convolved with the quantum field.
This equation is the same as you would find in classical nonlinear optics, but with the operator $\hat a$ replacing the classical field term.
Classically, we could interpret this as the expectation value of the field operator and its evolution; quantum mechanically, we should think of this as the evolution of $C, S$ that act on the operator.
In our experiment, we expect $S$ to be predominantly anti-diagonal due to energy conservation, and $C$ to depend on the phase-matching function.

\begin{figure}[htb]
\centering
\includegraphics[width=0.666\normaltextwidth]{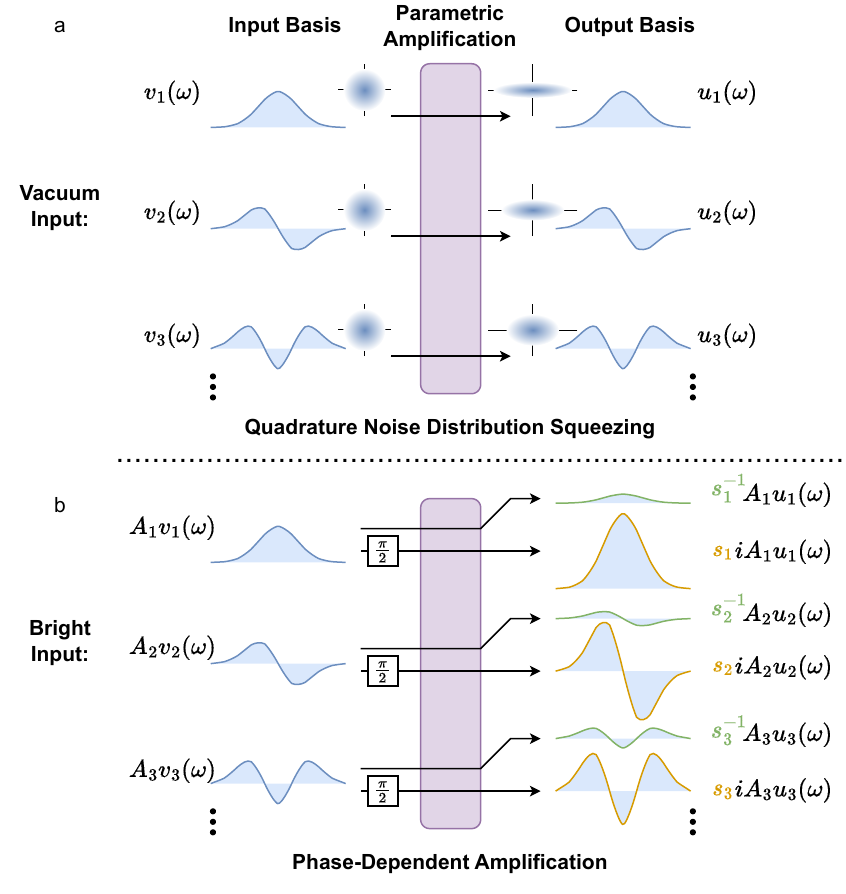}
\caption{
 \textbf{Parametric amplification and the Bloch--Messiah decomposition.}
 The Bloch--Messiah decomposition reduces the parametric amplification to a phase-dependent amplification of an orthogonal set of modes.
 Each input mode is transformed to an output mode during this process: these modes are generally not the equal due to other effects in the OPA that may alter the signal (e.g.\ dispersion).
 \textbf{a.} In the case of vacuum input, squeezed vacuum is generated in each output mode independently.
 \textbf{b.} In the case of bright input, a signal may be decomposed into input modes, mapped to output modes, and each component is (de)amplified depending on the phase.
}
\label{fig:bmdecomposition}
\end{figure}

Such a Green's function can be decomposed by the Bloch--Messiah decomposition, which informs us of how much squeezing there is and over what modes.
To take advantage of standard computational methods, this is performed in the quadrature basis of $\hat x, \hat p$ (denote this Green's function by $G'$).
This matrix decomposition returns canonically conjugate squeezed and anti-squeezed ``supermodes.''
Concretely:
\footnote{Note: this assumes the convention $\hat x = \hat a^\dagger + \hat a$ and $\hat p = i (\hat a^\dagger - \hat a)$.}
\begin{aligned*}
G' &= O_\text{out} \Sigma O_\text{in}^\intercal \\
\Sigma &= \tfrac12 \text{diag}(\{ s_i \}_{i=1}^M, \{ s_i^{-1} \}_{i=1}^M)
 = \tfrac12 \text{diag}(\{ \mathrm{e}^{r_i} \}_{i=1}^M, \{ \mathrm{e}^{-r_i} \}_{i=1}^M)
\end{aligned*}
where $s_i$ are the symplectic eigenvalues, related to the squeezing parameters $r_i$, and orthogonal matrices $O$ contain the ``input'' and ``output'' symplectic eigenvectors or supermodes, which come in pairs (squeezed and anti-squeezed).
The input supermodes are not important if the initial state is vacuum, as the squeezed vacuum only depends on the output supermodes.
Intuitively, Bloch--Messiah may be thought of as a singular value decomposition that preserves commutation relations.
These bases diagonalize the Green's function into single mode squeezing operations.
Fig.\,\ref{fig:bmdecomposition} illustrates the process.
Each $s_i$ represents an independent squeezed mode, and a source of photons with $\langle n \rangle = \sinh^2 r_i$ distributed over the mode $O_{\text{out},i}$,
or from a quantum noise perspective, a $20\log_{10} s_i$ dB noise reduction in the mode $O_{\text{out},i}$.
A change of basis from $\hat x, \hat p$ back to $\hat a$ transforms the orthogonal matrices into unitary, and the diagonal matrix into a matrix where the quadrants are diagonal:
\begin{equation}
\sigma = \left(
\begin{matrix}
\text{diag}(\{ \cosh r_i \}_{i=1}^M) & \text{diag}(\{ \sinh r_i \}_{i=1}^M) \\
\text{diag}(\{ \sinh r_i \}_{i=1}^M) & \text{diag}(\{ \cosh r_i \}_{i=1}^M)
\end{matrix}
\right) .
\end{equation}

With vacuum input, a covariance matrix can be generated as follows:
\begin{equation}
\sigma_{ij} = \tfrac12 \{ G_{ik} \hat \xi_k, G_{jk} \hat \xi_k \} \Rightarrow \sigma = \tfrac12 G G^\dagger .
\end{equation}
Additional operations can be similarly applied to the covariance matrix:
\begin{equation}
\sigma \rightarrow G \sigma G^\dagger .
\end{equation}
Note that vacuum $\sigma = I_{2M} / 2$.

As previously mentioned, (going back to the $\hat x, \hat p$ basis) the input supermodes cancel out in the vacuum case:
\begin{equation}
\tfrac12 G' G'^\intercal
= \tfrac12 O_\text{out} \Sigma O_\text{in}^\intercal O_\text{in} \Sigma O_\text{out}^\intercal
= \tfrac12 O_\text{out} \Sigma^2 O_\text{out}^\intercal
\end{equation}
and we see that the Bloch--Messiah decomposition is also valid on the the covariance matrix, although one must account for squared diagonal matrix
(however, we note that the Williamson decomposition must be used when the state is not pure: in the case of loss, thermal noise, or if the state has been partially traced out).

We can observe that:
\begin{aligned*}
\sigma &= \tfrac12 G G^\dagger =
\frac12 \left(
\begin{matrix}
S S^\dagger + C C^\dagger & C S^\intercal + S C^\intercal  \\
C^* S^\dagger + S^* C^\dagger & S^* S^\intercal + C^* C^\intercal
\end{matrix} \right)
\notag \\
&= \left(
\begin{matrix}
S S^\dagger + I/2 & (C S^\intercal + S C^\intercal)/2 \\
( C^* S^\dagger + S^* C^\dagger )/2 & S^* S^\intercal + I/2
\end{matrix} \right)
=
\left(\begin{matrix}
V + I/2 & U \\ U^* & V^\intercal + I/2
\end{matrix} \right) .
\end{aligned*}
Note that for the $U$ quadrants to be nonzero, a squeezing operation must be involved through a nonzero $S$.
No unitary $C$ applied to any thermal state can populate $U$.
On the other hand, the squeezing contributes to ``thermal'' components $V$, as it adds photons to the field, thus increasing the photon-number variance.

The formalism introduced so far -- where all the operations are unitary or symplectic and preserve commutation relations -- cannot account for losses or inefficiencies that must be considered during the frequency conversion and detection steps of our experiment.
In both cases, the state occupies unobserved modes, that are traced out.
Tracing out modes in a covariance matrix is simply equivalent to removing the corresponding rows and columns,
in other words, taking a principal submatrix that omits the traced out modes.
For example, this is used to derive the action of a lump loss or noise on a covariance matrix (modeled as passing through a fictitious beamsplitter and tracing out the second port):
\begin{equation}
\sigma \rightarrow \sqrt{\eta \eta^\intercal} \circ \sigma + (1-\eta) \circ \nu \circ I
\end{equation}
where $\eta$ is the transmission, $\nu = {\bar n} + 1/2$ represents any thermal noise added ($1/2$ for vacuum).
Both are vectors in the general case.
The $\circ$ operator is the element-wise or Hadamard product.

We now use these results to describe the physics relevant to our experiment:
we convert the infrared squeezed light to visible, but do not detect the remaining infrared light.
For notational simplicity let $\hat \xi = [ \hat {\vect a}_\text{vis}, \hat {\vect a}_\text{vis}^\dagger, \hat {\vect a}_\text{ir}, \hat {\vect a}_\text{ir}^\dagger ]$.
The initial state is:
\begin{equation}
\sigma_\text{tot}(z=0) =
\left( \begin{matrix}
\sigma_\text{vis} & \sigma_\text{vis,ir} \\
\sigma_\text{vis,ir}^\dagger & \sigma_\text{ir}
\end{matrix} \right)
=
\left( \begin{matrix}
I_{2M} / 2 & 0 \\
0 & \sigma_\text{ir}(0)
\end{matrix} \right) .
\end{equation}
The AFC (over crystal length $L$) acts as a unitary on the entire state:
\begin{equation}
\sigma_\text{tot}(z=L)
=
\left( \begin{matrix}
G_\text{vis,vis} & G_\text{vis,ir} \\
G_\text{ir,vis} & G_\text{ir,ir}
\end{matrix} \right)
\left( \begin{matrix}
I_{2M} / 2 & 0 \\
0 & \sigma_\text{ir}
\end{matrix} \right)
\left( \begin{matrix}
G_\text{vis,vis} & G_\text{vis,ir} \\
G_\text{ir,vis} & G_\text{ir,ir}
\end{matrix} \right)^\dagger .
\end{equation}
Expanding and tracing out the infrared modes we obtain the covariance matrix of the observed visible modes:
\begin{equation}
\sigma_\text{vis}(L)
= G_\text{vis,ir} \sigma_\text{ir}(0) G_\text{vis,ir}^\dagger + G_\text{vis,vis} G_\text{vis,vis}^\dagger / 2 .
\end{equation}
As the conversion tends to unity, $G_\text{vis,ir}$ becomes unitary and $G_\text{vis,vis}$ vanishes.

The sum frequency generation equations that yield the Green's function in this case are:
\begin{aligned*}
\frac{d \hat a_\text{vis}}{dz}(\Delta \omega)
&= i D_\text{vis}(z, \Delta \omega) \hat a_\text{vis} + i A(z, \Delta \omega) * \hat a_\text{ir} \\
\frac{d \hat a_\text{ir}}{dz}(\Delta \omega)
&= i D_\text{ir}(z, \Delta \omega) \hat a_\text{ir} + i A^*(z, \Delta \omega) * \hat a_\text{vis}
\end{aligned*}
where, again, $A$ is the pump field and $D$ is the dispersion, which is notably a function of $z$, due to aperiodic poling.

\subsection*{Photon number properties of zero-mean Gaussian states}
\label{sec:app:multimodeQO:photons}
\addcontentsline{toc}{subsection}{\nameref{sec:app:multimodeQO:photons}}

\newcommand{\normal}[1]{:\mathrel{#1}:}
\newcommand{\weyl}[1]{:\mathrel{#1}:_W}

For convenience, in the context of photon number statistics, we can make the covariance matrix symmetric and complex-valued, by defining the transformation $X$, as first introduced in \cite{HamiltonJex}:
\begin{equation*}
\begin{gathered}
\sigma' = \sigma - I_{2M} / 2, \quad X \sigma'
=
\left(\begin{matrix}
U & V^\intercal \\ V & U^*
\end{matrix} \right)
, \\
X = \left(\begin{matrix}
0 & I_M \\ I_M & 0
\end{matrix} \right) .
\end{gathered}
\end{equation*}
$\sigma'$ defined in this manner is useful, because we can use it to find any of the following photon number expectation values:
\begin{equation}
\left\langle \prod_{i} \hat n_i^{m_i} \right\rangle = \left\langle \prod_{i} \left( \hat a_i^\dagger \hat a_i \right)^{m_i} \right\rangle = \text{Haf}(X \sigma_{\vect m}'), ~\vect m \in \lbrace{ 0, 1 \rbrace}^M .
\end{equation}
using the conventional definition of the Hafnian (Haf).
$\sigma_{\vect m}'$ indicates the principal submatrix with indices given by $\vect m$.
Suffice to say, the Hafnian function is central to the (Gaussian or perturbative) multimode physics of bosons, since it can be thought of as an implementation of Wick's theorem for Gaussian integrals;
for zero-mean Gaussian variables $x_{i_j}$:
\begin{equation}
\langle x_{i_1} \ldots x_{i_{2m}} \rangle = \sum_{\mathcal{P}} \langle x_{k_1} x_{k_2} \rangle \ldots \langle x_{k_{2m-1}} x_{k_{2m}} \rangle
\end{equation}
where the sum is over all pairings $\mathcal{P}$ -- all possible ways to group the $i$-indices into $m$ pairs of $k$-indices --
hence the Hafnian, a function defined for this very purpose \cite{Caianiello}.
And of course the operators $\hat a_i$ are jointly Gaussian (in their quasi-probability distributions).

The general case with $\vect m \in \mathbb{N}^M$ is more complicated, as it requires more careful consideration of operator ordering.
Indeed, we were being hasty: $\hat a$ is not a random variable, it is an operator.
However, Wick's theorem tells us that this was allowed as long as the state and operator have compatible representations.
For example, one way to ``treat our operators as though random variables'' is by converting our expression to Weyl ordering (denoted by $\weyl{\ldots}$).
This requires a prescribed order of $\hat{a}$'s and $\hat{a}^\dagger$'s, and once we have this expression, we may move the operators within without incurring a commutation relation, e.g.\ $\weyl{\hat{a}^\dagger \hat{a}}\ =\ \weyl{\hat{a} \hat{a}^\dagger}$.
The correct procedure is to expand out the operator expression into Weyl-ordered expressions, and replace each one with the corresponding Hafnian,
i.e.:
\begin{aligned*}
\langle \hat n_1^{m_1} \hat n_2^{m_2} \ldots \rangle
&= \langle (\hat a_1^\dagger \hat a_1)^{m_1} (\hat a_2^\dagger \hat a_2)^{m_2} \ldots \rangle \\
&= \sum_\text{orderings} c_\text{order} \langle \weyl{ {\hat a_1^\dagger}^{m_{\text{order} 1}} \hat a_1^{m_{\text{order} 2}} \ldots } \rangle \\
&= \sum_\text{orderings} c_\text{order} \left[ \sum_{\mathcal{P}(\text{order})} \prod_{(k_1, k_2)} \langle \weyl{ \hat a_{k_1} \hat a_{k_2} } \rangle \right] \\
&= \sum_\text{orderings} c_\text{order} \text{Haf}(X \sigma_{\bar m_\text{order}}) .
\end{aligned*}
It is well known how to relate normal (operator expressions where all the $\hat{a}^\dagger$ precede the $\hat{a}$, denoted by $\normal{\ldots}$) and Weyl forms \cite{CahillGlauber, FerraroOlivaresParis}, in our case:
\begin{equation}
\normal{{\hat a^\dagger}^m \hat a^m}\ = \sum_{k=0}^{m} k! \binom{m}{k}^2 \left(-\frac12 \right)^k \weyl{ {\hat a^\dagger}^{m-k} \hat a^{m-k} }
\end{equation}
and all modes are treated separately.
Additionally, the normal form of $(\hat a^\dagger \hat a)^m$ can be obtained by repeatedly applying commutation relations.
This can be generalized to a sum represented by Stirling numbers $S$ \cite{Katriel}:
\begin{equation}
(\hat a^\dagger \hat a)^m = \sum_{i=1}^m S(m, i) {\hat a^\dagger}^i \hat a^i .
\end{equation}
In the case of $\vect m \in \lbrace{ 0, 1\rbrace}^M$, as above, $\hat a^\dagger \hat a =\ \weyl{\hat a^\dagger \hat a} - 1/2$, and we can subtract the constant directly from the covariance matrix as we did for $\sigma'$ ($\sigma$ is conventionally defined as above, with the anticommutator, such that it is conveniently in Weyl form).
\footnote{Note, however, that because in this case $\hat{n}^1 = \hat{a}^\dagger \hat{a}$ is already normal ordered, we end up implicitly converting $\sigma$ to normal ordering by subtracting $I/2$ (i.e.\ $\sigma'_{ii} = \langle \normal{\hat{a}_i^\dagger \hat{a}_i} \rangle = \langle \hat{a}_i^\dagger \hat{a}_i \rangle$ from $\sigma_{ii} = \langle \weyl{\hat{a}_i^\dagger \hat{a}_i} \rangle = \langle \{\hat{a}_i^\dagger, \hat{a}_i\} \rangle$).}
Similarly, for the second order case,
\begin{equation}
\langle \hat n^2 \rangle = \langle (\hat a^\dagger \hat a)^2 \rangle
= \langle \normal{{\hat a^\dagger}^2 \hat a^2 + \hat a^\dagger \hat a} \rangle
= \langle \weyl{ {\hat a^\dagger}^2 \hat a^2 - \hat a^\dagger \hat a } \rangle
\Rightarrow \text{Haf}(X \sigma^{(2)}) - \text{Haf}(X \sigma) .
\end{equation}
In this example we use $\sigma^{(2)}$ to denote a larger covariance matrix where we repeat the rows and columns of $\sigma$ for this mode.

These equations involving Hafnians will look familiar to the reader familiar with the recent Gaussian Boson Sampling literature.
Indeed, for multimode Gaussian states, calculating the probabilities is largely the dual of calculating the expectation values.
The photon number distribution is essentially a generalized Bose--Einstein distribution that incorporates modes, interference and entanglement.
To quote the result proven in \cite{RahimiLundRalph, HamiltonJex}, the probability of a given measurement, $\vect n \in \mathbb{N}^M$, has the form:
\begin{aligned*}
P(\vect n) &= \frac{1}{\vect n! \lvert \sigma + I_{2M}/2 \rvert^{1/2}} \text{Haf} \left(X A_{\vect n} \right) \\
A &= I_{2M} - (\sigma + I_{2M}/2)^{-1},
\quad
\vect{n}! = {\textstyle\prod_i} n_i !
\end{aligned*}
where the Hafnian's argument refers to the principal submatrix with indices given by $\vect n$.
To gain some intuition,
one may notice the Bose--Einstein resemblance via $\sigma_{ii} = \langle \hat a_i^\dagger \hat a_i + h.c.\rangle/2 = \langle n_i \rangle + 1/2$, so $(\sigma + I/2)_{ii} = \langle n_i \rangle + 1$, thus in some limiting cases:
\begin{equation}
\lvert \sigma + I/2 \rvert^{-1/2} \rightarrow \frac{1}{\langle n \rangle + 1}
, \quad
I - (\sigma + I/2)^{-1}
\rightarrow
\frac{\langle n \rangle}{\langle n \rangle + 1}
\end{equation}
More precisely, the Bose--Einstein distribution, which is simply geometric, can be thought of as stemming from the recurrence relation:
\begin{equation}
P_\textrm{BE}(n) = P_\textrm{BE}(n-1) a = P_\textrm{BE}(0) a^n = \frac{1}{\langle n \rangle + 1} \left( \frac{\langle n \rangle}{\langle n \rangle + 1} \right)^n
\end{equation}
while this general version follows the similar, yet more complex, Hafnian (Wick's theorem) recursive property:
\begin{equation}
\forall i \text{ s.t.\ } n_i > 0, P(n_1, \ldots, n_M) = \frac{P(0, \ldots, 0)}{\vect n !} \sum_{\substack{j=1\\j\neq i}}^{2N} a_{ij} \text{Haf}(X A_{\vect n - \{i,j\}}),
\quad N = \sum_i {\vect n}_i
\end{equation}
where the notation $A_{\vect n - \{i,j\}}$ denotes the subtraction of rows and columns $i,j$ from the matrix $A_{\vect n}$
(note e.g.\ with $j = i + N$, $\text{Haf}(A_{\vect n - \{ij\}}) \propto P(n_1, \ldots, n_i-1, \ldots, n_M)$, $a_{ij} = P(0, \ldots, n_i = 1, \ldots, 0)$).

From the quantum computing and complexity theory perspective, the Hafnian is of interest as it belongs to \#P, a class of classically intractable functions \cite{ArkhipovAaronson, HamiltonJex, HangleiterEisert}.

A brief sketch of a derivation of the probabilities is as follows;
it is perhaps a bit more cumbersome than using phase space formalism (which abstracts away the use of operators and orders) but it is hopefully more transparent.
We use the normal form, as this allows us to use a convenient representation of the photon-number Fock state projection operator:
the vacuum projection is known to be
\begin{equation}
\lvert 0 \rangle \langle 0 \rvert =\ \normal{\exp(-\hat{a}^\dagger \hat{a})}
\end{equation}
(see, for example, \cite{Fan} Eqns.~17--19 for a short proof), and it follows that \cite{FerraroOlivaresParis}:
\begin{aligned*}
\lvert \vect{n} \rangle \langle \vect{n} \rvert &=
\lvert n_1, \ldots, n_m \rangle \langle n_1, \ldots, n_m \rvert =
\prod_i (\hat{a}_i^\dagger)^{n_i} \lvert 0 \rangle \langle 0 \rvert \prod_i \hat{a}_i^{n_i} / n_i! =\
\normal{ \prod_{i=1}^M \exp(-\hat{a}_i^\dagger \hat{a}_i) \frac{\left( \hat{a}_i^\dagger \hat{a}_i \right)^{n_i}}{n_i!}}
\\
P(\vect{n}) &=
\left\langle \lvert \vect{n} \rangle \langle \vect{n} \rvert \right\rangle =
\frac{1}{\vect{n}!} \left\langle \normal{ \prod_{i=1}^M \exp(-\hat{a}_i^\dagger \hat{a}_i) \left( \hat{a}_i^\dagger \hat{a}_i \right)^{n_i}} \right\rangle 
= \left\langle \normal{ \prod_{i=1}^M \frac{\mathrm{e}^{-\hat{n}_i} \hat{n}_i^{n_i}}{n_i!}} \right\rangle .
\end{aligned*}
Since the operator is normal-ordered, we must also use the normal-ordered covariance matrix,
which is $\sigma'$ as defined above
(since, e.g.\ for the first matrix quadrant, $\sigma'_{ij} = \langle \normal{\hat{a}_i^\dagger \hat{a}_j} \rangle = \langle \hat{a}_i^\dagger \hat{a}_j \rangle = \langle \{ \hat{a}_i^\dagger, \hat{a}_j \} \rangle - \delta_{ij} = \langle \weyl{\hat{a}_i^\dagger \hat{a}_j} \rangle - \delta_{ij} = \sigma_{ij} - \delta_{ij}$).
Lastly, notice that we can set
\begin{equation}
\lvert 0 \rangle \langle 0 \rvert =\
\normal{\exp \left( -\tfrac12 {\hat\xi}^\dagger \hat{\xi} \right)}
\end{equation}
so that the notation is compatible with that of the state's covariance matrix.

Hence, the problem is reduced once more to calculus; multivariate Gaussian integrals.
We first find $P(0)$, and the rest of the probabilities follow from Wick's theorem, with a modified covariance matrix due to the above $\lvert 0 \rangle \langle 0 \rvert$ Gaussian.
The former can be solved with the usual tricks.
Let $Z\sim\mathcal{N}(\vect 0, I)$ be a vector of standard normal random variables.
By affine transformation, we can transform it to a vector of arbitrary jointly Gaussian variables: $\sigma'^{1/2} Z + \mu$.
Since we are only interested in the case of zero displacement, let the means $\mu=0$.
We thus simplify our calculation by computing the expectation over the multivariate standard normal distribution:
\begin{equation}
\left\langle \exp\left( -\tfrac12 (\sigma'^{1/2} Z)^\dagger \sigma'^{1/2} Z \right) \right\rangle
= (2\pi)^{-M} \int_{\mathbb{R}^{2M}} dZ \exp\left( -\tfrac12 Z^\dagger (\sigma' + I) Z \right)
= |\sigma' + I|^{-1/2}.
\end{equation}
This is the zero-photon probability.
We note that $\sigma' + I = \sigma + I/2$ corresponds to the anti-normal covariance matrix (all the $\hat{a}$ precede the $\hat{a}^\dagger$).

Finally, we can include monomial terms in the integral to calculate other probabilities, using Wick's theorem once more, after solving for the effective covariance matrix, $A$ -- since the distribution and operator Gaussians combine:
\begin{equation}
\exp\left( -\tfrac12 Z^\dagger Z  -\tfrac12 Z^\dagger \sigma'^{-1} Z \right) =
\exp\left( -\tfrac12 Z^\dagger A^{-1} Z \right) .
\end{equation}
$A$ can be found by invoking the matrix inversion lemma:
\begin{aligned*}
A^{-1} &= \sigma'^{-1} + I \\
A &= (\sigma'^{-1} + I)^{-1} \\
&= I - \sigma'^{-1} (\sigma^{-1} + I)^{-1} \\
&= I - (\sigma' + I)^{-1} ,
\end{aligned*}
as per convention.
We have not been completely rigorous, however, since $\sigma'$ may not be invertible
(if, in a diagonal basis, sigma has some $\langle \normal{ \hat{a}_i^\dagger \hat{a}_i }\rangle = 0$,%
\footnote{The field can never be a pure vacuum at finite temperature, although for all intents and purposes it is at optical frequencies, since the photon energy $\hbar \omega \gg k_B T$ (the Boltzmann constant temperature product) at room temperature.}
in which case the Gaussian tends towards a delta function);
hence, besides for the last line, the above equalities may not be strictly valid.
Nonetheless, the moments exist, which is what we need for the moment expansion in Wick's theorem.

\subsection*{Photon number expectation values}
\label{sec:app:multimodeQO:expectations}
\addcontentsline{toc}{subsection}{\nameref{sec:app:multimodeQO:expectations}}

We can solve for some lower order photon number moments in terms of $\sigma$'s elements.
To eliminate local phase degrees of freedom, we set all diagonal $U_{ii}$ terms to be real and positive.
Here, wherever relevant, we have $i < j < k$ (switching order requires taking complex conjugates of some terms and can therefore introduce inconsistencies).
\begin{aligned*}
\langle n_i \rangle &= \text{Haf}(X \sigma_i') = V_{ii} \\
\langle n_i n_j \rangle &= \text{Haf}(X \sigma_{ij}') = |U_{ij}|^2 + |V_{ij}|^2 + \langle n_i \rangle \langle n_j \rangle \\
\langle n_i^2 \rangle &= \text{Haf}(X \sigma_{ii}) - \text{Haf}(X \sigma_{i}) = |U_{ii}|^2 + 2 \langle n_i \rangle^2 + \langle n_i \rangle \\[3pt]
\langle n_i n_j n_k \rangle &= \text{Haf}(X \sigma_{ijk}') \\
&= \langle n_i \rangle \langle n_j n_k \rangle + \langle n_j \rangle \langle n_i n_k \rangle + \langle n_k \rangle \langle n_i n_j \rangle - 2 \langle n_i \rangle \langle n_j \rangle \langle n_k \rangle \\
&\quad + 2 \mathcal{R} \left(U_{ij}^* (V_{ik} U_{jk} + V_{jk} U_{ik}) + V_{ij} (U_{ik}^* U_{jk} + V_{ik}^* V_{jk}) \right)  \\[3pt]
\langle n_i^2 n_j \rangle &= \text{Haf}(X (\sigma_{iij} - \tfrac12 \delta_{jj}) - \text{Haf}(X (\sigma_{ij} - \tfrac12 \delta_{jj})) \\
&= \langle n_i n_j \rangle (4 \langle n_i \rangle + 1) + \langle n_j \rangle (|U_{ii}|^2 - 2 \langle n_i \rangle^2) + 2 U_{ii} (U_{ij}^* V_{ij} + U_{ij} V_{ij}^*)
\\
\langle n_i n_j^2 \rangle &= \text{Haf}(X (\sigma_{ijj} - \tfrac12 \delta_{ii}) - \text{Haf}(X (\sigma_{ij} - \tfrac12 \delta_{ii})) \\
&= \langle n_i n_j \rangle (4 \langle n_j \rangle + 1) + \langle n_i \rangle (|U_{jj}|^2 - 2 \langle n_j \rangle^2) + 2 U_{jj} (U_{ij} V_{ij} + U_{ij}^* V_{ij}^*)
 \\[3pt]
\langle n_i^2 n_j^2 \rangle &= \text{Haf}(X \sigma_{iijj}) + \text{Haf}(X \sigma_{ij}) - \text{Haf}(X \sigma_{iij}) - \text{Haf}(X \sigma_{ijj}) \\
&= \langle n_i^2 n_j \rangle (4 \langle n_j \rangle + 1) + \langle n_i n_j^2 \rangle (4 \langle n_i \rangle + 1) - \langle n_i n_j \rangle (4\langle n_i \rangle +1)(4\langle n_j \rangle+1) \\
&\quad + 4 (\langle n_i n_j \rangle - \langle n_i \rangle \langle n_j \rangle)^2 + (2 \langle n_i \rangle^2 - |U_{ii}|^2) (2 \langle n_j \rangle^2 - |U_{jj}|^2) \\
&\quad + 2 U_{ii} U_{jj} (V_{ij}^2 + U_{ij}^2 + {V_{ij}^*}^2 + {U_{ij}^*}^2)
+ 8 |U_{ij}|^2 |V_{ij}|^2 .
\end{aligned*}

A few important observations.
The equations for the statistics can be separated into a trivial component (composed of lower order statistics) and an interference term, due to complex-valued $U$ and $V$ terms.
If the $U$ terms are zero, there are no nontrivial higher order correlations: everything can be described by second- or first-order statistics.
Hence why thermal states do not have interesting higher order correlations.

Relatedly, where $U=0$, there are a vast number of states that have the same photon number distribution, as there is no information beyond photon number mean and covariance.
Measurements carry no information about the underlying supermodes.

\begin{figure}[htb]
\centering
\includegraphics[width=1\normaltextwidth]{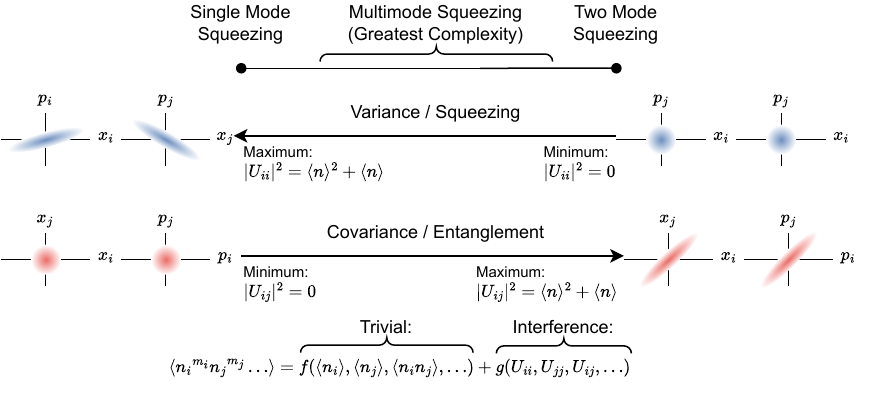}
\caption{
 \textbf{Multimode squeezing within the continuum between single and two-mode squeezing.}
}
\label{fig:stats}
\end{figure}

The value of $U_{ii} \in [0, \langle n_i \rangle (\langle n_i \rangle + 1)]$, and the two extremes correspond to two-mode and single-mode squeezing respectively.
States that have intermediate values
can be thought of as a multimode generalization of these two concepts, where the value of $U_{ii}$ represents the locality of the entanglement.
When $U_{ii}=0$, certain higher order correlations also lose interference terms.
Therefore, measurements of two-mode squeezed vacuum also lack certain information about the underlying state.
These points are illustrated in Fig.\,\ref{fig:stats}.

Finally,
for a state with sufficient information encoded in $U$, the inverse problem of retrieving the state and modes from observations is possible, within some limitations.
This is within the realm of possibility and future experiments should seek to achieve this.

Returning to the experiment, the photon covariance matrix in Fig.\,\ref{fig:afc}b (and the simulation in Fig.\,\ref{fig:dopa}a) consists of contributions from both the $V$ (diagonal) and $U$ (anti-diagonal) components.

\section{EMCCD camera statistics}\label{sec:app:emccd}

The electron-multiplying (EM) gain process used in EMCCD cameras is what makes these instruments highly sensitive: by magnifying the charge of a few photoelectrons to macroscopic levels, the camera is capable of detecting single photons.
However, the nature of the gain process is stochastic, which introduces noise.
Here we introduce the model for the gain process, and discuss how it influences the measurements we are interested in.

\subsection*{Photon statistics after EM gain}\label{sec:app:emccd:average}
\addcontentsline{toc}{subsection}{\nameref{sec:app:emccd:average}}

The EM gain process adds a high amount of noise, which typically precludes the possibility of resolving the exact photon numbers in EMCCD cameras.
When $n$ photons are captured as photoelectrons on the CCD sensor and amplified with EM gain $g$, the conditional probability of yielding $x$ amplified electrons follows a gamma (Erlang) distribution \cite{LantzDevaux}:
\begin{equation}
P(x; n) = \frac{x^{n-1} \exp(-x / g)}{g^n (n-1)!}.
\end{equation}
To estimate the statistics of the EM gain output $x$, let $P_n$ be the photon-number distribution incident on a pixel.
We are first able to calculate the conditional moments of $x$ with respect to $n$, use these to calculate (unconditional) moments of $x$, and finally calculate the photon number moments:
\begin{aligned*}
\langle x^k \rangle_n &= \int_{0}^{\infty} x^k P(x; n) dx = g^k \frac{(n+k-1)!}{(n-1)!}, \\
\langle x^k \rangle &= \sum_{n \geq 0} P_n \langle x^k \rangle_n, \\
g^{-k} \langle x^k \rangle &= \sum_{n \geq 0} P_n \prod_{i=0}^{k-1} n+i
= \left\langle \prod_{i=0}^{k-1} n+i \right\rangle.
\end{aligned*}
This can be extended to correlations between multiple variables, e.g.:
\begin{equation*}
\langle x_i^k x_j^l \rangle = \sum_{n_i, n_j \geq 0} P_{n_i n_j} \langle x_i^k \rangle_{n_i} \langle x_{j}^l \rangle_{n_j}
= g^{k+l} \left\langle \left(\prod_{m=0}^{k-1} n_i+m \right) \left( \prod_{m=0}^{l-1} n_j+m \right) \right\rangle.
\end{equation*}
We may then solve for the photon number moments by inverting the above equations.
Hence we can write down the formulae for our photon statistics of interest (omitting the gain term $g$):
\begin{aligned*}
\langle n \rangle &= \langle x \rangle \\
\langle n_i n_j \rangle &= \langle x_i x_j \rangle \\
\langle n_i n_j n_k \rangle &= \langle x_i x_j x_k \rangle \\
\langle n^2 \rangle &= \langle x^2 \rangle - \langle x \rangle \\
\langle n_i^2 n_j \rangle &= \langle x_i^2 x_j \rangle - \langle x_i x_j \rangle \\
\langle n_i^2 n_j^2 \rangle &= \langle x_i^2 x_j^2 \rangle - \langle x_i^2 x_j \rangle - \langle x_i x_j^2 \rangle + \langle x_i x_j \rangle.
\end{aligned*}

Finally, the electronic stages between the EM gain process and a digitized pixel value will introduce additional noise, such as readout noise.
However this noise is independent of the signal $x$, hence it does not affect the above our ability to estimate the photon number.

\begin{figure}[htb]
\centering
\includegraphics[width=1\normaltextwidth]{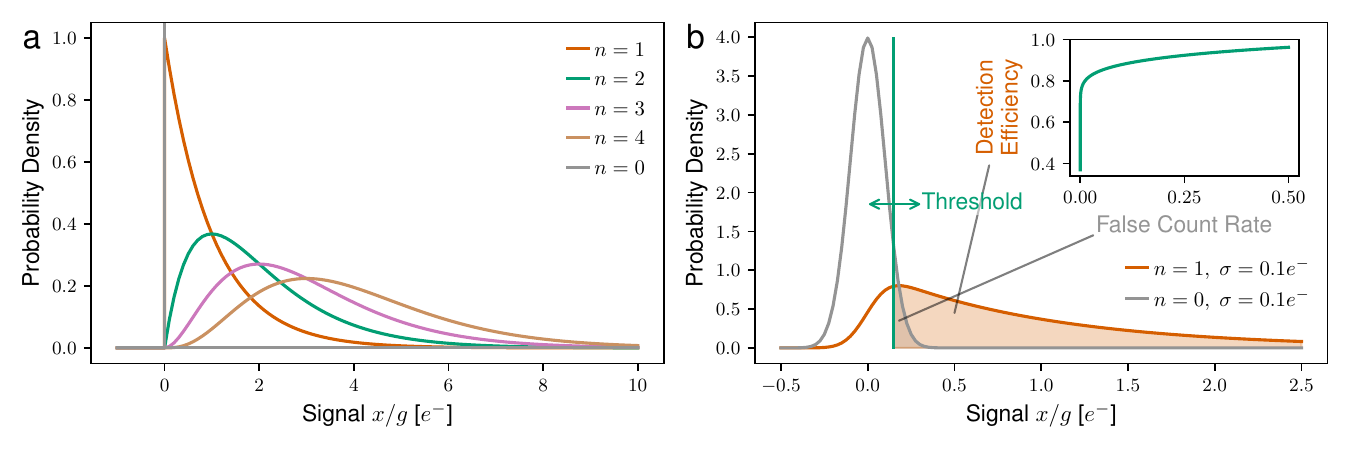}
\caption{
 \textbf{Electron multiplying gain model and thresholding.}
 \textbf{a.} The probability distribution of an amplified signal for a given number of photoelectrons, prior to additional noise.
 We model this with the Erlang distribution.
 The signal $x$ is measured in electrons ($e^{-}$) divided by the gain $g$ (the expected value of this is the original number of photelectrons).
 \textbf{b.} The probability distribution of the signal with Gaussian readout noise.
 Although the two probability distributions begin to overlap, in this example, thresholding can be used to distinguish between 0 and 1 (or more) photons with some degree confidence, incurring a tradeoff between the photon-detection efficiency and false count rate.
 The readout noise has standard deviation $\sigma$.
 The inset shows the analytical (ideal) ROC curve for this example.
}
\label{fig:emgain}
\end{figure}

\subsection*{Thresholded operation}\label{sec:app:emccd:thresh}
\addcontentsline{toc}{subsection}{\nameref{sec:app:emccd:thresh}}

As discussed above, the electronic signal $x$, ensuing an EM process with a gain of $g$, follows a random distribution parametrized by the number of photons captured $n$ -- this is shown in Fig.\,\ref{fig:emgain}.
Due to the stochastic nature of the EM gain process, it is impossible to resolve the original number of photoelectrons $n$ on the CCD sensor from the number of amplified electrons $x$.
However, if there are no photons absorbed by the CCD sensor during the detection window, there is no signal (no electrons) to amplify in the gain process.
Hence, there is no excess noise from the EM process, and the variance of the output signal depends solely on the readout noise $\sigma$.
With a high EM gain, the effective readout noise, $\sigma \ll 1\,e^-$, we can set a threshold on the output signal to discriminate between the absence or presence of photoelectrons with a high degree of confidence.

We evaluate the EMCCD camera in the single-photon detection mode using a threshold to detect the absence or presence of a photon.
The threshold value will determine the photon-detection efficiency (PDE) and the false click rate.
In general, a lower threshold increases the effective PDE but increases the false click rate, and vice versa.
See Fig.\,\ref{fig:emgain}b.
By varying the threshold value, we obtain a receiver operating characteristic (ROC) curve to characterize the performance of the photon counter (Fig.\,\ref{fig:emgain}b inset).
The ROC curve quantifies the trade-off in the effective quantum efficiency and dark count rate.
In practice, we obtain the false click rate as a function of the threshold by obtaining dark frames -- images from the sensor with no illumination -- and count how many times a given threshold is exceeded.
We obtain the PDE by determining the readout noise from this data, and computing the PDE through the model by comparing $g$ and $e$, multiplied by the quantum efficiency (QE) of the CCD sensor.

\section{Coincidence detection}\label{sec:app:coincidence}

In this section we explain how coincidence detection may be used as a tool to distinguish between different pulsed photonic states, as well as measure the purity of the state.

\subsection*{Derivation}\label{sec:app:coincidence:derivation}
\addcontentsline{toc}{subsection}{\nameref{sec:app:coincidence:derivation}}

\begin{figure}[htb]
\centering
\includegraphics[width=0.22\normaltextwidth]{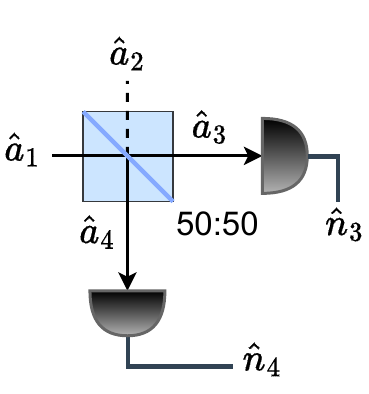}
\caption{
 \textbf{Toy model for the coincidence detection experiment.}
}
\label{fig:coincidencemodel}
\end{figure}

Consider a beam of squeezed light incident on a balanced beamsplitter, which is followed by two detectors on each side of the output (Fig.\,\ref{fig:coincidencemodel}).
The first arm of the beamsplitter, with operator $\hat a_1$, has some $\langle \hat n \rangle$ and $\Var(\hat n) = 2 \langle \hat n \rangle (\langle \hat n \rangle + 1)$, in the lossless case ($\eta=1$).
The second input port, with operator $\hat a_2$ has vacuum input.

The third and fourth ports, defined by operators $\hat a_3, \hat a_4$, have $\langle \hat n_3 \rangle = \langle \hat n_4 \rangle = \langle \hat n \rangle / 2$.
Evaluating their statistics:
\begin{aligned*}
\left\langle \hat n_3 \hat n_4 \right\rangle &= \left\langle \hat a_3^\dagger \hat a_3 \hat a_4^\dagger \hat a_4 \right\rangle \\
&= \frac14 \left\langle (\hat a_1 + \hat a_2)^\dagger (\hat a_1 + \hat a_2) (\hat a_1 - \hat a_2)^\dagger (\hat a_1 - \hat a_2) \right\rangle \\
&= \frac14 \left\langle \hat n_1^2 - \hat a_2 \hat a_2^\dagger \hat n_1 \right\rangle \\
&\Rightarrow \frac14 \left( \left\langle \hat n_1^2 \right\rangle - \langle \hat n_1 \rangle \right), \eta=1
\end{aligned*}

We can now compute the covariance of the number measurements on each detector.
\begin{aligned*}
\Cov(\hat n_3, \hat n_4) &= \langle \hat n_3 \hat n_4 \rangle - \langle \hat n_3 \rangle \langle \hat n_4 \rangle \\
&= \frac14 \left( \left\langle \hat n_1^2 \right\rangle - \langle \hat n_1 \rangle \right) - \frac14 \langle \hat n_1 \rangle^2 \\
&= \frac14 \left( \Var(\hat n_1) - \langle \hat n_1 \rangle \right) \\
&\Rightarrow \frac14 \langle \hat n \rangle (2\langle \hat n \rangle + 1 \rangle),~ \eta=1
\end{aligned*}

Note that for coherent states, with $\Var(\hat n) = \langle \hat n \rangle$, and for thermal states, with $\Var(\hat n) = \langle \hat n \rangle (\langle \hat n \rangle + 1)$, the covariances are 0 and $\langle \hat n \rangle^2 / 4$, respectively.
Similarly, Fock states with $\Var(\hat n) = 0$, have negative covariance; this is the well-known anti-bunching behavior.
These different scaling behaviors allow us to experimentally distinguish these different photonic states.

To account for loss, we introduce a fictitious unbalanced beamsplitter operation such that $\langle n_1' \rangle = \eta \langle n_1 \rangle$:
\begin{aligned*}
\Var(\hat n_1')
&= \left\langle \left(\sqrt{\eta} \hat a_1 + \sqrt{1-\eta} \hat a_0 \right)^\dagger \left(\sqrt{\eta} \hat a_1 + \sqrt{1-\eta} \hat a_0 \right) \left(\sqrt{\eta} \hat a_1 + \sqrt{1-\eta} \hat a_0 \right)^\dagger \left(\sqrt{\eta} \hat a_1 + \sqrt{1-\eta} \hat a_0 \right) \right\rangle - \eta^2 \langle \hat n_1 \rangle^2 \\
&= \eta^2 \left\langle \hat n_1^2 \right\rangle + \eta (1-\eta) \langle \hat n_1 \rangle - \eta^2 \langle n_1 \rangle^2 \\
&= \eta^2 \Var(\hat n_1) + \eta (1-\eta) \langle \hat n_1 \rangle
\end{aligned*}

So the covariance will simply change by a factor $\eta^2$:
\begin{aligned*}
\Cov(\hat n_3, \hat n_4) &= \frac14 \left( \Var(\hat n_1') - \langle \hat n_1' \rangle \right) \\
&= \frac{\eta^2}{4} \left( \Var(\hat n_1) - \langle \hat n_1 \rangle \right) \\
&\Rightarrow \frac{\eta^2}{4} \left( 2 \langle \hat n_1 \rangle^2 + \langle \hat n_1 \rangle \right)
\end{aligned*}

If we consider the case of asymmetric loss, it can be shown that the transmission efficiency can be replaced by the individual efficiencies of the left and right arms, $\eta^2 \rightarrow \eta_L \eta_R$.

Finally, in the multimode case, the total covariance is simply the sum of the covariances of all the individual modes (as a consequence of mode independence/commutation).
Therefore, a lossy multimode squeezed state has:
\begin{aligned*}
\Cov(\hat n_3, \hat n_4) &= \sum_i \frac{\eta_{Li} \eta_{Ri}}{4} \left( \Var( \hat n_i) - \langle \hat n_i \rangle \right) \\
&\Rightarrow \sum_i \frac{\eta_{Li} \eta_{Ri}}{4} \left( 2 \langle \hat n_i \rangle^2 + \langle \hat n_i \rangle \right)
\end{aligned*}

In the case where all modes have $\langle \hat n_i \rangle \ll 1$, the covariance is a linear function of the photon number
\begin{equation*}
\Cov(\hat n_3, \hat n_4) \approx \sum_i \frac{\eta_{Li} \eta_{Ri}}{4} \langle \hat n_i \rangle
\end{equation*}
Notably, the slope is determined by the overall transmission efficiency, $\eta / 2$ (assuming a constant transmission), since the average number of photons per detector is
\begin{equation}
\langle N \rangle = \sum_i \frac{\eta_i}{2} \langle \hat n_i \rangle
\end{equation}
Therefore, in addition to being a test for squeezing, this experiment may also serve as a measurement of loss.

To measure loss at slightly larger photon numbers, a quadratic fit may be used:
\begin{equation}
\Cov(\hat n_3, \hat n_4)
\approx \frac{\eta}{2} \langle N \rangle + \frac{\eta}{2} C \langle N \rangle^2,
\quad C \approx \sum_i \left( \frac{\eta_i \langle \hat n_i \rangle}{\langle N \rangle} \right)^2
\end{equation}
so long as $\langle \hat n_i \rangle / \langle N \rangle \approx \textrm{const}$.
In other words, the photon numbers in the state must scale linearly.
This requires that the squeezed modes remain in the region where $\langle \hat n \rangle = \sinh^2 r_i \approx r_i^2$.
Once again, the transmission is given by the linear coefficient.

\subsection*{Threshold detection}
\label{sec:app:c:threshold}
\addcontentsline{toc}{subsection}{\nameref{sec:app:c:threshold}}

Finally, we must consider the case of threshold detectors, where the detector only provides ``clicks,'' if it registers any number of photons.
This reduces the two observables to Bernoulli variables.
While the total number of photons $\langle N \rangle \ll 1$, the physics remains the same.
However, the covariance of Bernoulli variables is bound by a parabola, therefore the photon covariance will behave as such:
\begin{aligned*}
\Cov(n_3, n_4) =&~ \langle n_3 n_4 \rangle - \langle n_3 \rangle \langle n_4 \rangle \\
=&~ P(n_3=1 | n_4=1) P(n_4=1) \\
& - \left( P(n_3=1| n_4=1) P(n_4=1) + P(n_3=1| n_4=0) P(n_4=0) \right) \langle n_4 \rangle \\
=&~ \left( P(n_3=1 | n_4=1) + P(n_3=1| n_4=0) \right) (1-\langle n_4 \rangle) \langle n_4 \rangle
\end{aligned*}
where we expect the probabilities to be constant in the many mode, $\langle \hat n_i \rangle \ll 1$, scenario, where coincidences are predominantly due to biphotons.

This explains the parabolic curvature in the measurements in Fig.\,\ref{fig:dopa}e.

\subsection*{Predicting coincidence experiment outcomes with wavelength-dependent detector quantum efficiency}
\label{sec:app:c:qecoincidence}
\addcontentsline{toc}{subsection}{\nameref{sec:app:c:qecoincidence}}

The above treatment of coincidence detection assumes a uniform quantum efficiency.
However, in our infrared-wavelength experiment, we have that $\eta_{Li}, \eta_{Ri} \rightarrow \eta_{Li}(\lambda), \eta_{Ri}(\lambda)$.
This thermalizes the state further than uniform loss.
Because the QE cutoff is 1700\,nm, any photon above this wavelength will never be detected, it is traced out and its sister photon therefore effectively becomes a thermal photon.
In addition, the spectrum measured in experiment is far broader than in simulation.
Therefore, while it is straightforward to derive the covariance expected for a given QE for the simulated state, the simulation is not representative of experiment in this case.

To account for this, we derive a simple model to estimate the photon number vs covariance scaling we expect in the experiment (Fig.\,\ref{fig:dopa}e), using the spectrum measured experimentally.
In addition, we use the specified detector quantum efficiency.

Assume that the number of modes $M \gg \langle N \rangle$, the total number of photons, and therefore the probability of any given mode yielding photon pairs, $P_m \ll 1$, where $m$ indexes the mode.
We truncate the wavefunction of each mode in the photon number basis, as per the biphoton approximation:
\begin{equation}
\vert \psi \rangle_m \approx \sqrt{1 - P_m} \vert 0 \rangle_m + \sqrt{P_m} \vert 2 \rangle_m + \mathcal{O}(P_m)
\end{equation}
Furthermore assume that the probability of producing more than two photons per event is negligible.
Then we can define the probability $\mathcal{P}_m$ that detected photons arise from any given mode:
\begin{aligned*}
\langle N \rangle &= 2 \sum_m P_m \\
\sum_m \mathcal{P}_m = 1 &\Rightarrow \mathcal{P}_m = \frac{2}{\langle N \rangle} P_M
\end{aligned*}
The covariance is a sum over all covariances, which depend on the coincidence of two photons on the two different detectors:
\begin{aligned*}
\Cov(\hat n_3, \hat n_4) = \sum_m \Cov_m(\hat n_3, \hat n_4) &\approx \sum_m \frac12 P_m \eta_L \eta_R \text{QE}(\omega_{m,1}) \text{QE}(\omega_{m,2}) + \mathcal{O}(P_m^2) \\
&= \sum_m \langle N \rangle \mathcal{P}_m \eta_L \eta_R \text{QE}(\omega_{m,1}) \text{QE}(\omega_{m,2}) + \mathcal{O}(P_m^2)
\end{aligned*}
By switching to the continuous frequency basis and parametrizing by $\Delta \omega$ about the central wavelength $\omega_0$, we can obtain the expression:
\begin{aligned*}
\Cov(\hat n_3, \hat n_4) &\approx \langle N \rangle \eta^2 \int_0^{\Delta \omega_\text{max}} d \Delta \omega \mathcal{P}(\Delta \omega) \text{QE}(\omega_0 - \Delta \omega) \text{QE}(\omega_0 + \Delta \omega) \\
&= \left. \langle N \rangle \eta^2 \int_{\lambda_0}^{\lambda_\text{min}} \frac{d \lambda}{\lambda^2} \left[ \mathcal{P}(\lambda) \text{QE}(\lambda) \text{QE} \left( \frac{2}{\lambda_0} - \frac{1}{\lambda} \right) \right] \middle / \int_{\lambda_0}^{\lambda_\text{min}} \frac{d \lambda}{\lambda^2} \mathcal{P}(\lambda) \right.
\end{aligned*}
where $\mathcal{P}(\lambda)$ is now the spectral density (corresponding to Fig.\,\ref{fig:dopa}c in the experiment).
As expected, the covariance is linear in $\langle N \rangle$.

For simplicity, it is assumed above that the biphotons are perfectly correlated in frequency, which for a finite bandwidth pump and finite number of supermodes is not physically correct, but a reasonable approximation for a narrow phase-matching bandwidth (with respect to $d\text{QE}/d\omega$).

\subsection*{Intensity Difference Variance and Sub-shot-noise Intensity Correlations}
\label{sec:app:c:vardeltan}
\addcontentsline{toc}{subsection}{\nameref{sec:app:c:vardeltan}}

An equivalent coincidence method to verify intensity correlations is to consider the photon number difference variance,
\begin{equation}
\Var(\hat n_1 - \hat n_2) = \Var(\hat n_1) + \Var(\hat n_2) - 2 \Cov(\hat n_1, \hat n_2) ,
\end{equation}
which of course relates to the covariance.
The scope of this method is more limited -- it is only informative for bipartite entangled states, i.e.\ two-mode squeezed vacuum -- but when this is the case, it has the advantage of an intuitive and unambiguous interpretation:
vanishing variance implies that $n_1, n_2$ are completely entangled;
when the variance is at least Poissonian (i.e.\ with equal mean and variance, as shot noise) it is an indication that the state may be a classical one (although not necessarily).
Hence this is a good verification for even a moderately bipartite state.

Here we rederive the relation between intensity difference variance, squeezing and loss \cite{VaidyaVernon,ChristSilberhorn} in a slightly more general manner than usual to better predict our experiments.
Consider a set of two-mode squeezed states (such that the variance is minimized and the covariance is maximized), where all the signal modes and all the idler modes, indexed by $i$, are measured together by a pair of detectors.
As we have computed earlier, for each mode we have that:
\begin{aligned*}
\Var(\hat n_i) &= \eta_i^2 \left(\langle \hat n_{i}^2 \rangle - \langle \hat n_i \rangle^2 \right) + \eta_i (1-\eta_i) \langle \hat n_i \rangle \\
&= \eta_i^2 \langle \hat n_{i} \rangle^2 + \eta_i \langle \hat n_i \rangle \\
\Cov(\hat n_{i,1}, \hat n_{i,2}) &= \eta_{i,1} \eta_{i,2} \left(\langle \hat n_{i,1} \hat n_{i,2} \rangle - \langle \hat n_{i,1} \rangle \langle \hat n_{i,2} \rangle \right) \\
&= \eta_{i,1} \eta_{i,2} \left( \langle \hat n_{i} \rangle^2 + \langle \hat n_i \rangle \right) .
\end{aligned*}
As these statistics are additive, we find that:
\begin{aligned*}
\Var(\hat n_1 - \hat n_2) &=& \sum_i &\,
\eta_{i,1}^2 \langle \hat n_i \rangle^2 + \eta_{i,1} \langle \hat n_i \rangle \\
&&+&\,
\eta_{i,2}^2 \langle \hat n_i \rangle^2 + \eta_{i,2} \langle \hat n_i \rangle \\
&&-2&\,
\eta_{i,1} \eta_{i,2} \left( \langle \hat n_{i} \rangle^2 + \langle \hat n_i \rangle \right) \\
&=& \sum_i &\,
(\eta_{i,1} + \eta_{i,2} - 2\eta_{i,1} \eta_{i,2}) \langle \hat n_i \rangle \\
&&+&\,
(\eta_{i,1}^2 + \eta_{i,2}^2 - 2\eta_{i,1} \eta_{i,2}) \langle \hat n_i \rangle^2 .
\end{aligned*}

If the signal and idler losses are perfectly balanced, $\eta_1 = \eta_2$, then the quadratic term vanishes and we have that
\begin{equation}
\Var(\hat n_1 - \hat n_2) = 2\sum_i (1 - \eta) \eta \langle \hat n_i \rangle,
\end{equation}
hence by plotting this variance against the total photon number $2\sum_i \eta \langle \hat n_i \rangle$, the average transmission or loss can be obtained via the slope.
If the two transmissions are not perfectly balanced for all modes, however, the quadratic term remains, and will dominate at higher intensities.
Finally, as discussed in Appendix~\ref{sec:app:multimodeQO:expectations},
when the modes involved are not two-mode squeezed states, the intensity variance of each arm will increase and their covariance will decrease, which will spuriously present itself as excess loss by increasing the difference variance and hence the slope.
The limit of this is single mode squeezed states on a balanced beam splitter, as was discussed earlier in this section.

\section{Multimode AFC} \label{sec:app:afclzgrid}

The aim of this section is to provide some theoretical background and insight into the adiabatic frequency conversion process.
Mainly, we are interested in the frequency linear transformations realized by the AFC.

\subsection*{The Landau--Zener grid problem}\label{sec:app:afclzgrid:lz}
\addcontentsline{toc}{subsection}{\nameref{sec:app:afclzgrid:lz}}

\begin{figure}[htb]
\centering
\includegraphics[width=0.36\normaltextwidth]{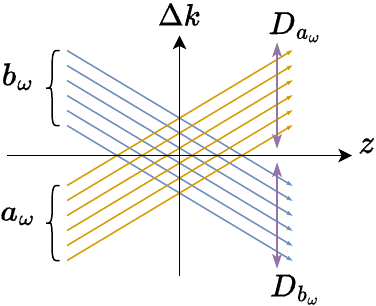}
\caption{
 \textbf{Multistate Landau--Zener grid in AFC.}
 The diagonal lines give the diabatic states.
 There is some probability of a transition when these coupled states cross.
}
\label{fig:lzgrid}
\end{figure}

AFC is often modeled as a two-level system, to which the Landau--Zener formula applies.
Energy and time in the canonical system correspond to momentum and space in this nonlinear-optical analog:
instead of the energy separation of the states varying with time, the wavenumber difference between the two frequencies varies with propagation distance.

However, when there are many frequency components at play, the equations for broadband AFC can be generalized to a bipartite system of modes:
\begin{equation}
\frac{d}{dz}
\left( \begin{matrix}
a_\omega \\
b_\omega
\end{matrix} \right)
= i \left( \begin{matrix}
D_{a_\omega} + \beta_0 z & K(A_{\omega}) \\
K(A_{\omega})^\dagger & D_{b_\omega} - \beta_0 z
\end{matrix} \right)
\left( \begin{matrix}
a_\omega \\
b_\omega
\end{matrix} \right)
\end{equation}
where, in the rotating frame,
$2\beta_0$ is the rate of change in the quasi-phase matching (QPM) wavenumber (e.g.\ such that the central wavelengths are phase matched at $z=0$),
the matrices $D = \text{diag}(\Delta \beta_1 \Delta\omega + \tfrac12 \beta_2 \Delta\omega^2 + \tfrac16 \beta_3 \Delta\omega^3 + \ldots)$ represent the dispersion,
and $K(A_\omega)$ represents the coupling between modes.
$K(A_\omega)$ is the pump convolution matrix, i.e.\ Toeplitz matrix of the pump spectrum $A$.
States in $a$ couple to states in $b$ through the pump $A$, but there is no intra-band coupling (unless unrelated nonlinear-optical processes such as cross-phase modulation come into play).

This kind of bipartite system is sometimes referred to as the Landau--Zener grid, due to the system's graphical representation, shown in Fig.\,\ref{fig:lzgrid}.
This general problem is mostly unsolved \cite{BrundoblerElser, YuDemkov, Usuki, Shytov, MallaSinitsyn, ShevchenkoNori}, but the concept is straightforward:
each level crossing has some transition probability, and the state can effectively perform a random walk, where amplitudes through different paths may interfere \cite{Harmin, BouwmeesterWoerdman}.
Only certain scattering probabilities, or special cases of the system, have been solved analytically.

The following is an explanation of Fig.\,\ref{fig:lzgrid}.
The horizontal axis represents the propagation distance, and the vertical axis the momentum.
The diagonal lines, the diabatic states, represent frequency modes.
When two lines cross they are phase-matched (equal momentum) and it is possible to have conversion between the modes.
The entire process is coherent, hence paths may interfere.
This is why applying a phase profile to the pump results in different transformations.

\subsection*{Linear transformations on bipartite systems}\label{sec:app:afclzgrid:transforms}
\addcontentsline{toc}{subsection}{\nameref{sec:app:afclzgrid:transforms}}

Consider a bipartite system of discrete modes -- without loss of generality: frequency (e.g.\ a comb) -- such as the one illustrated above in Fig.\,\ref{fig:lzgrid}.
An infinite number of modes, and a constant spacing in $D$ -- no higher order dispersion -- will result in translational invariance in frequency:
that is, any shift in the input frequencies would result in the same trajectory and a shifted, but otherwise identical, output.
Therefore, up to finite size effects due to a finite length in $z$, this symmetry implies that the linear transformations must be Toeplitz.

However, dispersion -- uneven spacing between modes represented by $D$ -- can break that symmetry.
With appropriately different spacings, the diabatic levels of the modes will cross at unique points along $z$, so in principle each crossing may be addressed individually.
This could potentially allow arbitrary transformations, if we extend the model to allow $A$ to not be fixed, but a function of propagation, $A(z)$.
Slowly changing $D$ -- as a function of $z$ -- may also help to achieve this.

In the context of ultrafast AFC the pump is fixed: once it is prepared and released, it cannot be changed as it propagates.
This limits the control over the dynamics, which impacts the programmability.
Nonetheless, dispersion at the pump wavelengths will impart relative spectral phases, and self-phase modulation may occur, hence $A$ may still be as a function of $z$, even if it is not controllable.

\subsection*{Programming the AFC pump to alter the transformation}\label{sec:app:afclzgrid:afcmodulation}
\addcontentsline{toc}{subsection}{\nameref{sec:app:afclzgrid:afcmodulation}}

In order to alter the AFC dynamics and produce a variety of linear transformations, we are able to perform intensity and phase modulations.
We can represent the shaped pump as:
\begin{equation}
A(\omega) = A_0(\omega) \sqrt{\mu(\omega)} \mathrm{e}^{i \phi(\omega)} ,
\end{equation}
where $A_0(\omega)$ is the original pump spectrum, $0 \leq \mu(\omega) \leq 1$ is the spectral intensity modulation, and $\phi(\omega)$ is the phase modulation.
With a discrete representation, control of $\mu(\omega)$ and $\phi(\omega)$ allows us to modify the Toeplitz coupling matrix as follows:
\begin{equation}
K(A_\omega) =
\left(
\begin{matrix}
\ddots & \vdots & \vdots & \\
\cdots & A_j & A_{j-1} & \cdots \\
\cdots & A_{j+1} & A_j & \cdots \\
& \vdots & \vdots & \ddots
\end{matrix}
\right)
\rightarrow
\left(
\begin{matrix}
\ddots & \vdots & \vdots & \\
\cdots & A_j \sqrt{\mu_j} \mathrm{e}^{i \phi_j} & A_{j-1} \sqrt{\mu_{j-1}} \mathrm{e}^{i \phi_{j-1}} & \cdots \\
\cdots & A_{j+1} \sqrt{\mu_{j+1}} \mathrm{e}^{i \phi_{j+1}} & A_j \sqrt{\mu_j} \mathrm{e}^{i \phi_j} & \cdots \\
& \vdots & \vdots & \ddots
\end{matrix}
\right) .
\end{equation}

\section{Spectral linear-optical unitary transformations by conversion engineering} \label{sec:app:simpleunitary}

\begin{figure}[htbp!]
\centering
\vspace{-1.5\baselineskip}
\includegraphics[width=0.95\normaltextwidth]{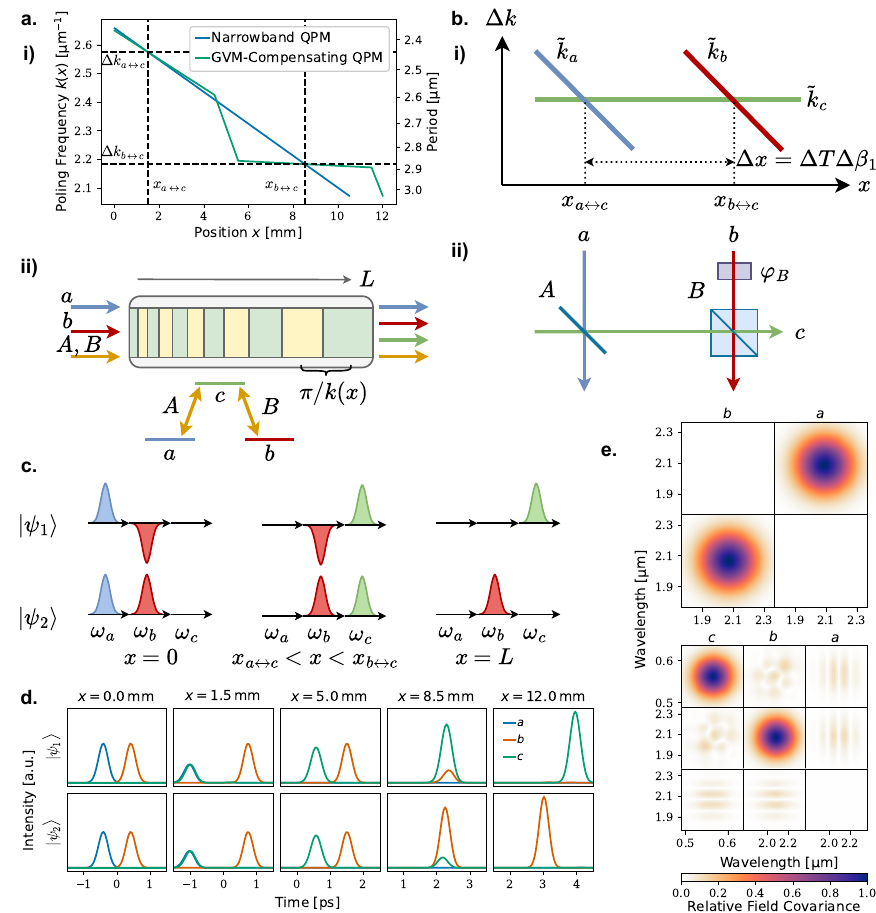} \\
\caption{
 \textbf{AFC-based interferometer of two mode squeezed states.}
 \textbf{a.} Delay compensating quasi-phase matching profile.
 Over the length of the crystal $L$, we sweep the spatial frequency of the poling $k$, as illustrated in \textbf{i}.
 The sweep passes through $\Delta k_{a \leftrightarrow c}$ and $\Delta k_{b \leftrightarrow c}$, the phase matching frequencies necessary for conversion between the pair of modes $a,c$ and $b,c$ respectively.
 These conversions occur at locations $x_{a \leftrightarrow c}$, $x_{b \leftrightarrow c}$ along the length of the crystal.
 Two pump modes $A,B$ couple $a,b$ to $c$.
 \textbf{b. i} Eigenvalue diagram of the system.
 The change in $k$ translates to changes in the effective phase mismatch between the $a,c$ and $b,c$ conversion processes.
 In the rotating frame of $c$ (where its spatial frequency $\tilde k_c$ [eigenvalue] is constant), $a$ and $b$ rotate with $\tilde k_a$ and $\tilde k_b$.
 Coherent conversion occurs as the eigenvalues cross.
 The state can be interpreted to travel along these trajectories, analogous to the spatial beamsplitter and phase-shifter configuration illustrated in \textbf{ii}.
 The magnitudes and relative phases of pumps $A,B$ determine the transmissions, reflection, and phase ($\varphi_B$) of the beamsplitter interaction.
 \textbf{c.} Two-mode squeezed state conversion in the frequency-domain.
 The initial state is composed of two supermodes: an odd and even superposition of $a,b$; $|\psi_1 \rangle, |\psi_2\rangle$.
 Following the first conversion, the $a$ component of both supermodes is converted to $c$.
 The second conversion is tuned to a balanced beamsplitter interaction, via appropriate intensity and phase modulations of the pumps,
 resulting in the full conversion of one state to $b$, and the other to $c$, disentangling the modes.
 \textbf{d.} Simulation of the states in the time domain.
 Two-mode squeezing requires that mode $a$ have greater (lesser) group velocity than $b$.
 In order for the states to interact with each other during conversion they must be coincide in time.
 We pick $c$ to have the least (greatest) group velocity, such that $b,c$ return to each other.
 At $x_{a \leftrightarrow c}$, $b,c$ will have a separation of $\Delta T$.
 \textbf{e.} Simulated photon covariance matrices (joint spectral intensities) of the two dominant supermodes, at the input and output: before and after conversion.
 Fully entangled, two-mode squeezed states are converted into effectively fully independent single-mode states, up to the conversion efficiency.
 This process can, in principle, be made arbitrarily broadband by appropriately tuning the slope and curvature of $k(x)$.
}
\label{fig:simpleafc}
\end{figure}

In the main text we have shown the ability to measure and manipulate photonic spectral correlations by frequency conversion, and experimentally validated that the physics of multimodal quantum frequency conversion closely adheres to our theoretical predictions.
To complement these proofs of concept, we briefly explore how broadband quasi-phase-matched frequency conversion methods can be extended beyond our current experimental capabilities.
We thus present a toy model, with the aim of demonstrating how bespoke AFC processes can be conceived to perform linear transformations, especially on frequency bin or frequency comb modes.
Namely, we describe a $3\times3$ unitary for converting between two-mode and single-mode squeezed vacuum.
This illustrative example provides intuition on how the quasi-phase matching profile, pump phase modulation, and pump spectral intensity play a role in the coherent conversion process.

A prevalent approach for preparing squeezed vacuum involves two-mode squeezing of group-velocity mismatched (GVM) propagation modes \cite{UrenRaymer};
typically entangling different wavelength or polarization modes.
In order to satisfy the group velocity condition, (which also determines the spectral bandwidths of the two modes -- adding an additional constraint if these are to be equal), a non-degenerate scheme is often necessary \cite{XinLoncar, BabelSilberhorn}.
In addition, in all cases the two modes are temporally separated, due to the GVM requirement, contributing delay distinguishability in addition to spectral.

Distinguishability in wavelength, temporal, polarization and bandwidth differences, however, can be overcome by broadband nonlinear optical conversion:
an AFC process can be designed to realize spectro-temporal beamsplitter-like interactions to coherently interfere these modes, as we demonstrate in this example.
This example can be generalized for use in conjunction with frequency-multimode squeezing schemes proposed, for example, in Refs.~\cite{HurvitzArie, DragoAgata, MorrisonFedrizzi, RozenbergArie}.

Fig.\,\ref{fig:simpleafc} illustrates this broadband-conversion-based unitary scheme over three modes, which could be implemented, for example, via Type 0 and II conversion in lithium niobate.
Call the OPA modes $a,b$, and let $c$ be a third mode at a different frequency.
Fig.\,\ref{fig:simpleafc}a represents the quasi-phase matching (QPM) design that can perform the frequency beamsplitting to mode sort a two-mode squeezed state (in $a, b$) into two independent states (in $a, c$).
The instantaneous QPM function $k(x)$ must cover the poling periods necessary to convert $a\leftrightarrow c$ and $b\leftrightarrow c$, as pumped by modes $A, B$ (Fig.\,\ref{fig:simpleafc}b).
Crucially, the second conversion point must coincide with the temporal overlap of $b, c$, as discussed later.
Furthermore, the AFC QPM profile can be designed to compensate for phase due to dispersion, which enables broadband beamsplitting: a linear profile is sufficient for narrowband conversion; piecewise linear compensates GVM for broadband conversion; nonlinearity may account for group velocity dispersion.

Fig.\,\ref{fig:simpleafc}c represents the eigenvalue diagram, introduced in Appendix~\ref{sec:app:afclzgrid}.
The eigenvalues represent the spatial rotation frequency $\Delta k$ of the modes as a function of position $x$ in the crystal (in this case in the rotating frame of mode $c$).
The rotation frequencies of $a,b$ ($\tilde k_a$, $\tilde k_b$) depend on their phase mismatch ($\Delta k_{a\leftrightarrow c}, \Delta k_{b\leftrightarrow c}$) and change solely based on the QPM $k(x)$.
When the diabatic trajectories cross, any state in one mode will undergo an avoided crossing, converting into the other mode.
The efficiency depends on the pump and the $k(x)$ slope, as given by the Landau--Zener formula.
Fig.\,\ref{fig:simpleafc}d gives an analogy in terms of physical beamsplitters, with a given transmission and reflection values, phase plates imparting a given phase, and distance imparting a given group delay.

Fig.\,\ref{fig:simpleafc}e illustrates how a two-mode squeezed vacuum state evolves throughout the AFC process, in the frequency domain.
Two-mode squeezed vacuum is made up of a symmetric and antisymmetric supermode of the frequencies $\omega_a$ and $\omega_b$.
The fastest mode $a$ is converted into the slowest mode $c$, via $A$, with near unit efficiency.
The next conversion, between $b\leftrightarrow c$ is driven to 50\% conversion (a balanced beamsplitter), via appropriately amplitude- and phase-modulated $B$.
Due to the symmetric and antisymmetric supermodes' phase difference, one state converts completely to $c$ and one to $b$.
Fig.\,\ref{fig:simpleafc}e simulates the dynamics of the process in the time domain.
Modes $a,b$ begin separated in time.
After mode $a$ converts to $c$, mode $b$ catches up, at which point the $b\leftrightarrow c$ conversion occurs, separating the underlying states.
Fig.\,\ref{fig:simpleafc}g is a simulation of the joint spectral intensity, or photon number covariance, of a fully entangled, two mode-squeezed state given as the input to AFC.
The output, shown below, demonstrates the off-diagonal correlations moving into the matrix diagonal, indicating independence between the states.

Although for simplicity here we have shown the conversion of only two states, this method is compatible with the parallel conversion of an arbitrarily wide frequency comb, as long as the QPM appropriately compensates for any dispersion.
In addition, by adjusting the pump parameters, any effectively two-mode unitary can be realized.
Generalizing to multimode unitary transformations is ostensibly a matter of realizing an optimal network of parallel conversions between all frequency bin modes, equivalent to adding more intersecting lines (beamsplitters) to Fig.\,\ref{fig:simpleafc}b.

When performing the simulations for this example we have used parameters consistent with two-mode squeezing of both polarization modes at 2066\,nm in lithium niobate via a 1033\,nm pump, in a 3\,mm Gaussian-apodized QPM OPA.
The group indices are given as 2.27, 2.375 and 2.48 for the $\hat y$-polarized signal ($a$), pump and $\hat z$-polarized signal ($b$) modes -- values achievable via dispersion engineering in thin-film waveguides.
The respective effective indices are given as 1.835, 2.054 and 1.769.
We assume no group velocity dispersion.
The AFC process is simulated over a 12\,mm QPM, as shown in the Figure, via 775\,nm pump.
The following section discusses how to arrive at the QPM profile.
The group and effective indices used for the $\hat z$-polarized 560\,nm mode ($c$) are 2.56 and 2.266.
For completeness we include the $\hat y$-polarized 560\,nm mode, but it has little effect as its conversion is highly phase mismatched.
The 775\,nm AFC pump has a $\hat y$ and a $\hat z$ component ($A, B$), with 2.110, 2.183 effective indices, responsible for the Type II and 0 conversions respectively.
To create the desired unitary operation, the intensity in the $\hat z$ component is first optimized via a sweep and interpolation to achieve 50\% conversion, and the phase is similarly optimized to interferometrically separate the two supermodes; these comprise the necessary amplitude and phase modulations.
In principle, different pump bandwidths may be used to compensate for pulsed two-mode squeezed states with different bandwidth modes.

\subsection*{Designing the AFC to compensate for time delay}\label{sec:app:simpleunitary:delay}
\addcontentsline{toc}{subsection}{\nameref{sec:app:simpleunitary:delay}}

Consider three frequency modes $a, b, c$.
Their group velocity are assumed to be $v_{a} > v_{b} > v_{c}$ (where the group velocity is the inverse of the first-order propagation constant $v=\beta_1^{-1}$)
such that the pulse in mode $a$ first converts to $c$, and experiences a time delay with respect to $b$.
Once the pulses in $b$ and $c$ overlap, the second conversion occurs.
Let $\Delta T$ be the original delay between the two pulses in $a$ and $b$ at the first conversion point, which is due to the propagation length within the OPA ($L_\text{OPA}$) and within the AFC before the conversion point ($L_\text{AFC}^{(1)}$):
\begin{equation}
\Delta T \approx \left(L_\text{OPA} + L_\text{AFC}^{(1)} \right) \left( \beta_1^{(b)} - \beta_1^{(a)} \right),
\end{equation}
then the distance between the two conversion points, $a\rightarrow c$ and $b\leftrightarrow c$, must be
\begin{equation}
L_\text{AFC}^{(2)} \left( \beta_{1}^{(c)} - \beta_{1}^{(b)} \right) = \Delta T .
\end{equation}

Within this distance the quasi-phase matching for the AFC must change from $\Delta k_{(ac)} = k_p + k_a - k_c$ to $\Delta k_{(bc)} = k_p + k_b - k_c$
The simplest way to do this is by varying the QPM linearly to connect the two points.
This works sufficiently well, however, as discussed in the next section, for broadband processes it is preferable to vary the slope according to the dispersion.

Finally, consider some error in the design, such that there is a deviation from the optimal length
$L_\text{AFC}^{(2)} \rightarrow L_\text{AFC}^{(2)} + \delta L$.
Varying the pump wavelengths (hence the central wavelength of mode $c$) can compensate for this error.
The change in frequency affects $k_p$ and $k_c$, hence $L_\text{AFC}^{(1)}$ and $\Delta T$ for the altered mode $c$, and consequently $L_\text{AFC}^{(2)}$.

\subsection*{Designing the slope for broadband conversion}\label{sec:app:simpleunitary:slope}
\addcontentsline{toc}{subsection}{\nameref{sec:app:simpleunitary:slope}}

We want to design the slope of the AFC such that the different wavelengths all experience the same phase and time delay.
To achieve this, the conversion points for different wavelengths, occurring in different locations along the propagation axis of the crystal, must be spaced according to the dispersion.

The slope is defined as the rate of change in the instantaneous quasi-phase matching frequency $k$ over the propagation axis:
\begin{equation}
S(x) = \frac{dk}{dx} .
\end{equation}
Let $x_1$ and $x_2$ be the two conversion points for the central frequencies, as discussed previously.
Consider a frequency $\delta\omega$ away from the initial central frequency $\omega_0$.
Due to dispersion, it will experience a different wavenumber, which will result in a phase and time delay; this can be compensated for by propagation over a longer or shorter distance.
Let $\delta k_1 = \Delta k(\delta \omega + \omega_0) - \Delta k(\omega_0)$, and similarly $\delta k_2$, be the difference in the wavenumbers involved in the first and second conversion.
The overall phase delay $\phi$ (with respect to the central frequency, and which we wish to eliminate) will be the spatial integral of the wavenumber between the two conversion points, and its derivative is the time delay $\Delta T$:
\begin{aligned*}
\phi(\delta \omega) &= \int_{0}^{x_1(\delta\omega)} \delta k_1(\delta\omega) dx - \int_{0}^{x_2(\delta\omega)} \delta k_2(\delta\omega) dx \\
-\Delta T(\delta \omega) &= \frac{d}{d\omega} \left[ \int_{0}^{x_1(\delta\omega)} \delta k_1(\delta\omega) dx - \int_{0}^{x_2(\delta\omega)} \delta k_2(\delta\omega) dx \right] \\
&= \int_{0}^{x_1} \frac{dk_1}{d\omega} dx + \left.\delta k_1 \frac{dx_1}{d\omega} \right|_{\delta\omega} - \int_{0}^{x_2} \frac{dk_2}{d\omega} dx - \left.\delta k_2 \frac{dx_2}{d\omega} \right|_{\delta\omega} .
\end{aligned*}
These derivatives, $dx_i/d\omega$, depend on the slope, since
$dx = dk_1 / S_1$.
For simplicity let us consider the case of linear dispersion, where $\delta k_1 = \Delta \beta_1^{(1)} \delta \omega$, $\delta k_2 = \Delta \beta_1^{(2)} \delta \omega$.
Therefore:
\begin{equation}
\frac{dx_1}{d\omega} = \left(\frac{dk_1}{dx_1}\right)^{-1} \frac{dk_1}{d\omega} = \frac{\Delta\beta_1^{(1)}}{S_1} ,~
\frac{dx_2}{d\omega} = \frac{\Delta\beta_1^{(2)}}{S_2} .
\end{equation}
Hence, returning to our expression for the delay:
\begin{aligned*}
-\Delta T(\delta \omega) &= \int_{0}^{x_1} \Delta\beta_1^{(1)} dx + \Delta\beta_1^{(1)} \delta\omega \frac{dx_1}{d\omega} - \int_{0}^{x_2} \Delta\beta_1^{(2)} dx - \Delta\beta_1^{(2)} \delta\omega \frac{dx_2}{d\omega} + O(\delta\omega) \\
&= \int_{0}^{x_1} \Delta\beta_1^{(1)} dx - \int_{0}^{x_2} \Delta\beta_1^{(2)} dx + \left( \frac{{\Delta\beta_1^{(1)}}^2}{S_1} - \frac{{\Delta\beta_1^{(2)}}^2}{S_2} \right) \delta\omega .
\end{aligned*}
We need $\Delta T = 0$.
Set:
\begin{equation}
\frac{S_1}{S_2} = \left( \frac{\Delta\beta_1^{(1)}}{\Delta\beta_2^{(2)}} \right)^2
\end{equation}
and it remains be to be shown that:
\begin{equation}
\int_{0}^{x_1} \Delta\beta_1^{(1)} dx - \int_{0}^{x_2} \Delta\beta_1^{(2)} dx = 0 .
\end{equation}
By integrating the left-hand side:
\begin{aligned*}
\Delta\beta_1^{(1)} x_1 - \Delta\beta_1^{(2)} x_2 &= \Delta\beta_1^{(1)} \frac{dx_1}{dk} \frac{dk}{d\omega} \delta\omega - \Delta\beta_1^{(2)} \frac{dx_2}{dk} \frac{dk}{d\omega} \delta\omega \\
&= \left(\frac{{\Delta\beta_1^{(1)}}^2}{S_1} - \frac{{\Delta\beta_1^{(2)}}^2}{S_2} \right) \delta\omega
= 0 ,
\end{aligned*}
where we used that:
\begin{equation}
x_i = \int \frac{dx_i}{d\omega} d\omega= \int \left(\frac{dk_i}{dx_i}\right)^{-1} \frac{dk_i}{d\omega} d\omega .
\end{equation}
Hence $\Delta T(\delta \omega) = 0$ and $\phi(\delta \omega)$ is constant.

\section{Spectrometer positive operator-valued measure (POVM) and spectral discretization} \label{sec:app:spectrometerpovm}

A natural question for our experiment is how to reconcile the fact that we have a continuous basis (frequency), but a discrete set of measurement modes: the spectrometer pixels.
Indeed, it is common practice in theory and numerics to discretize fields in the manner:
\begin{equation}
\hat a_i = \frac{1}{\omega_i - \omega_{i-1}} \int_{\omega_{i-1}}^{\omega_i} \hat a (\omega) d\omega
\end{equation}
for sufficiently small intervals such that the outcome converges.
However, the experimental implications of this procedure are less obvious, especially if the intervals happen to be too large with respect to the spectral features.
The correct procedure is to consider the field in the continuous limit and classically accumulate the statistics or probabilities.
This coarse-binning effect can therefore be an effective source of decoherence.
The discrete case is recovered in the limit where the field properties are constant within the bins.

To model the possible decoherence effects of binning, we can use the results and notation from Appendix~\ref{sec:app:multimodeQO:expectations}, modified for the continuous limit.
Based on inspection, the major effect will be in the loss of ``interferometric visibility'' for those statistics and probabilities that include interference-like terms.
Let us consider the three-body moment as an example
(the same argument holds for probabilities and other moments):
\begin{aligned*}
\langle n_i n_j n_k \rangle &- \langle n_i \rangle \langle n_j n_k \rangle - \langle n_j \rangle \langle n_i n_k \rangle - \langle n_k \rangle \langle n_i n_j \rangle + 2 \langle n_i \rangle \langle n_j \rangle \langle n_k \rangle \\
&= \frac{1}{|\Delta \omega|^3} \iiint\limits_{\Delta\omega, \Delta\omega', \Delta\omega''}
2 \mathcal{R} \left\{
{\let\scriptstyle\textstyle\substack{U(\omega, \omega')^* [V(\omega, \omega'') U(\omega', \omega'') + V(\omega', \omega'') U(\omega, \omega'')] \\
{} + V(\omega, \omega') [U(\omega, \omega'')^* U(\omega', \omega'') + V(\omega, \omega'')^* V(\omega', \omega'')] }}
\right\} d^3\omega
\end{aligned*}
As we can see, by binning, the interference terms (made up of the continuous, complex-valued, field covariance functions $U, V$) are averaged over the spectral bins, which can only decrease their magnitude relative to the other moments, the more these terms vary over the frequency bin.

Additionally, the point spread function when imaging one wavelength onto the pixelated detector must be much smaller than one pixel.
This ``pixel-arrival'' error is otherwise a source of additional decoherence from a classical process.

\addtocontents{toc}{\protect\nohyperlinkcontentsline{section}{Practice}{}\%}

\section{Photographs of the Experiment}\label{sec:app:photographs}

\begin{figure}[htbp]
\centering
{\hspace{-0.40\textwidth} \large \bf a.} \\
\includegraphics[width=0.35\textheight]{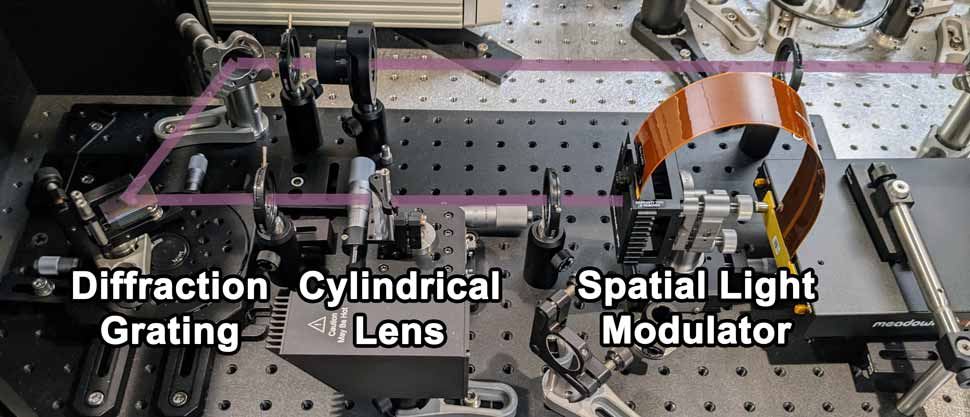} \\
{\hspace{-0.31\textwidth} \large \bf b. \hspace{.31\textwidth} c. \hspace{.29\textwidth} d.}
\makebox[1\textwidth]{
 \includegraphics[height=0.35\textheight]{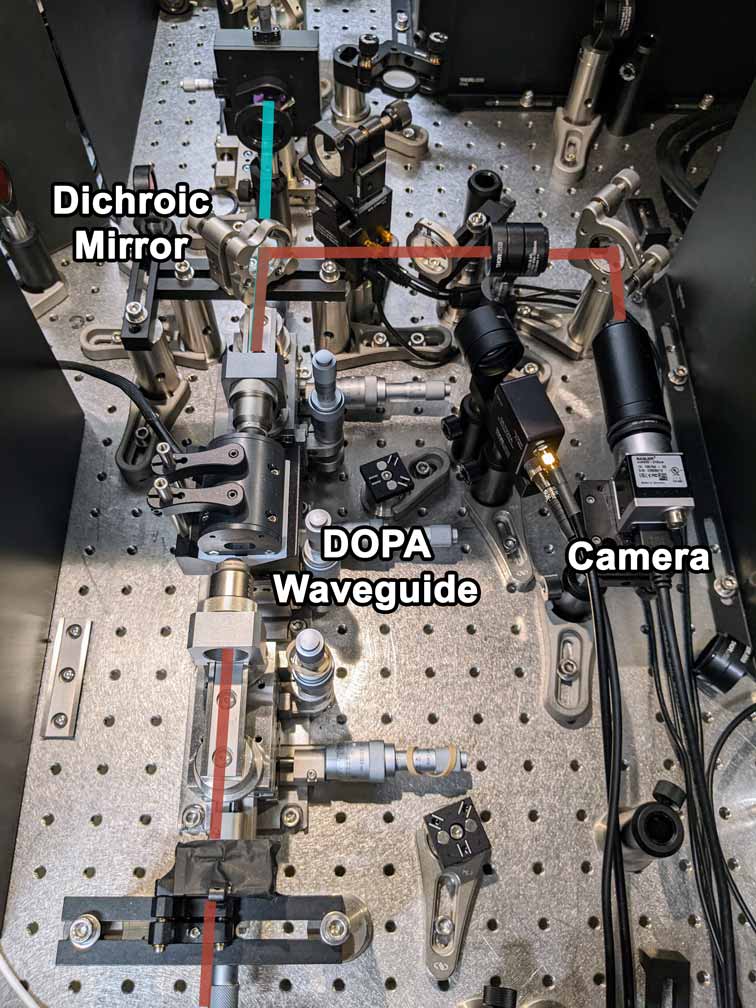}\!
 \includegraphics[height=0.35\textheight]{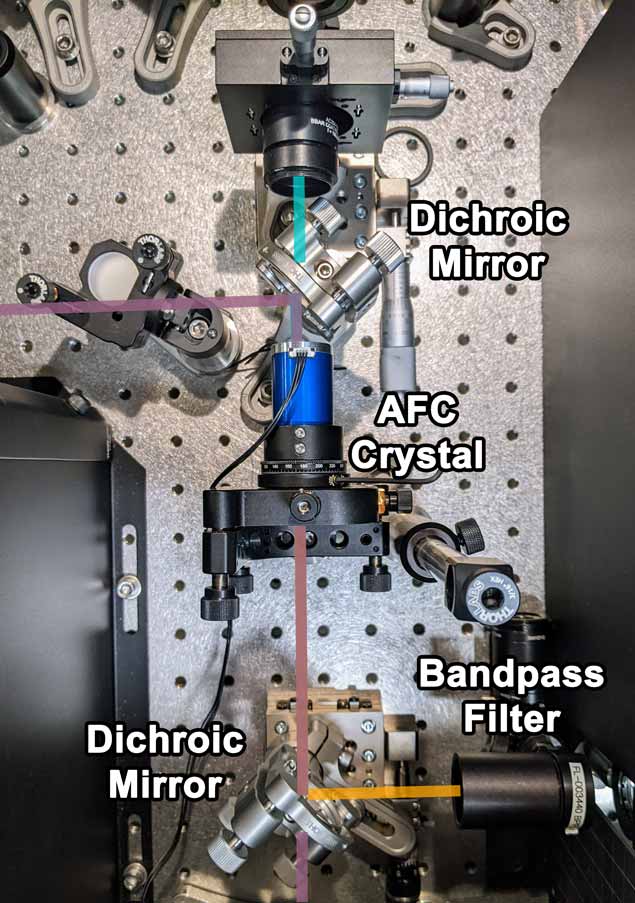}\!
 \includegraphics[height=0.35\textheight]{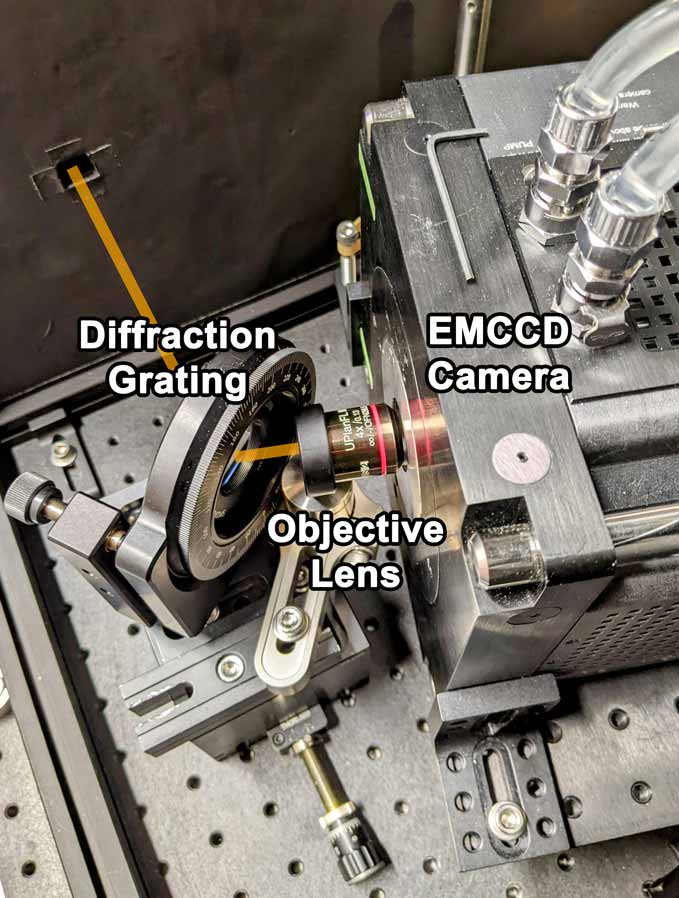}
} \\
{\hspace{-0.08\textwidth} \large \bf e. \hspace{.31\textwidth} f. \hspace{.30\textwidth} g.} \\
\includegraphics[height=0.35\textheight]{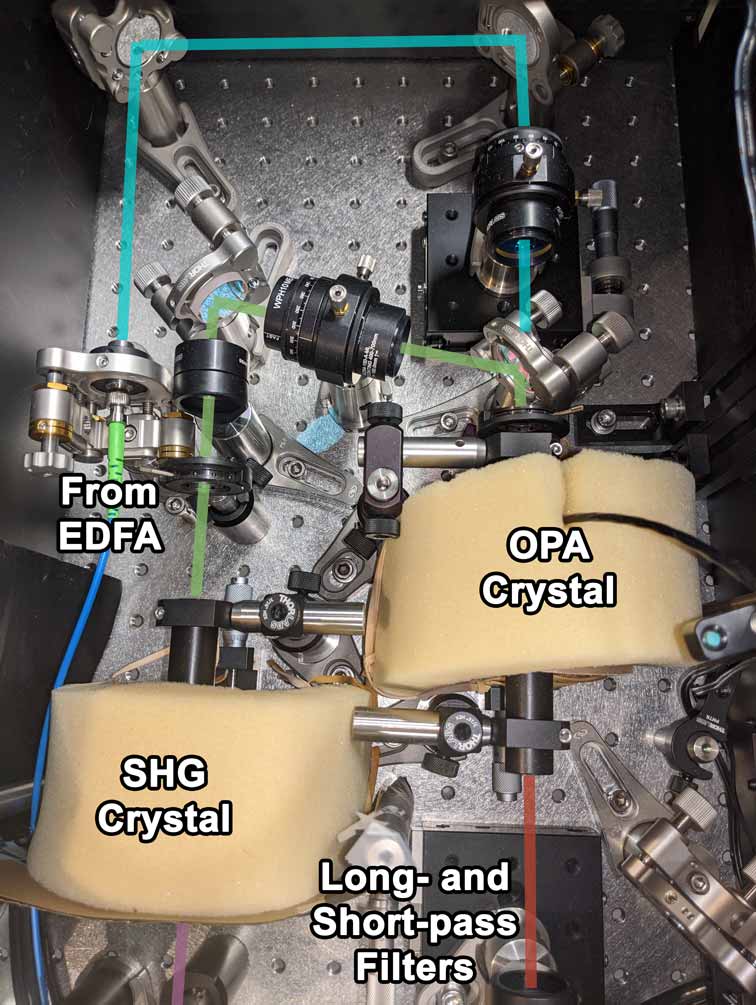}
\includegraphics[height=0.35\textheight]{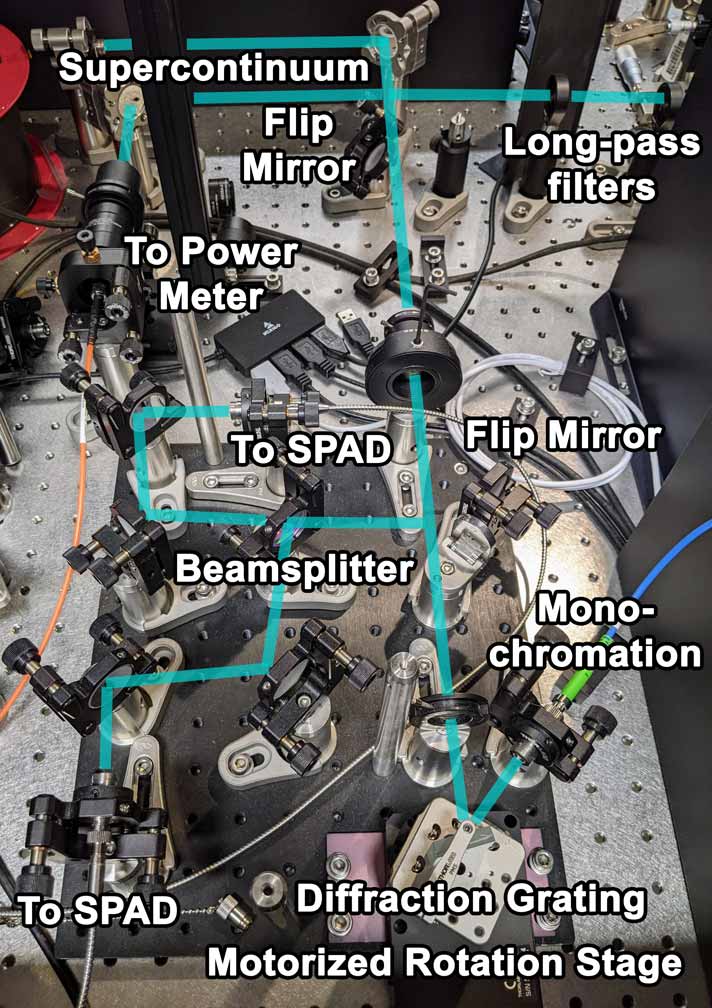}
\includegraphics[height=0.35\textheight]{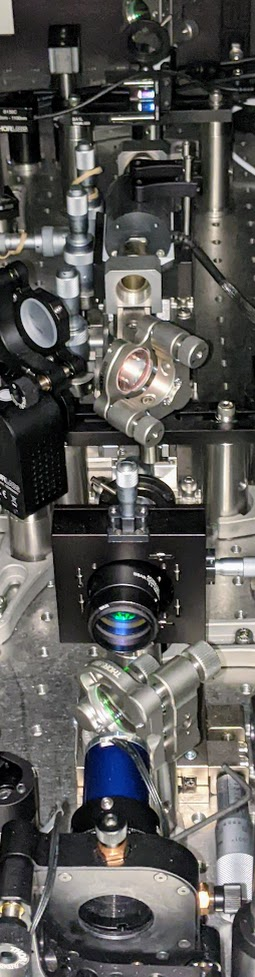}
\caption{
 \textbf{Photographs of the Experiment.}
 The beam paths are overlaid.
 Beam color to wavelength legend:
 \fcolorbox{white}[HTML]{00CCCC}{\textcolor{white}{1550}}
 \fcolorbox{white}[HTML]{996185}{\textcolor{white}{1033}}
 \fcolorbox{white}[HTML]{AE4132}{\textcolor{white}{775}}
 \fcolorbox{white}[HTML]{FFA500}{\textcolor{white}{620}}
 \fcolorbox{white}[HTML]{82B366}{\textcolor{white}{516}}\,nm.
 \textbf{a.} Pulse shaper.
 \textbf{b.} Degenerate optical parametric amplifier (DOPA) waveguide.
 \textbf{c.} AFC.
 \textbf{d.} Spectrometer.
 \textbf{e.} 775\,nm pump generation.
 \textbf{f.} Monochromation and coincidence detection.
 \textbf{g.} DOPA and AFC path.
}
\label{fig:photographs}
\end{figure}

Photographs of notable parts of the experiment are shown in Fig.\,\ref{fig:photographs}.
Important components are labeled and the beam paths are shown as colored overlays.
The apparatus is described in the \nameref{sec:methods}.
Fig.\,\ref{fig:photographs}a shows the pulse shaper, made up of a diffraction grating and spatial light modulator;
b shows the degenerate OPA (DOPA) waveguide and surrounding optics;
c shows the AFC crystal;
d shows the spectrometer, composed of a diffraction grating and EMCCD camera;
e shows the LBO crystals used to generate 775\,nm pulses from the 1033\,nm source;
f shows the characterization setups of 1550\,nm down-converted light, separated by flip mirrors, including the parametric gain (``to power meter''), the coincidence detection (``to SPAD''), and the monochromator, also used with the supercontinuum source;
g shows the DOPA and AFC in succession, the two central components of the setup.

\section{Adiabatic frequency conversion design and operation}\label{sec:app:afc}

\subsection*{AFC Design}\label{sec:app:afc:afcdesign}
\addcontentsline{toc}{subsection}{\nameref{sec:app:afc:afcdesign}}

\begin{figure}[htb]
\centering
\includegraphics[width=1\normaltextwidth]{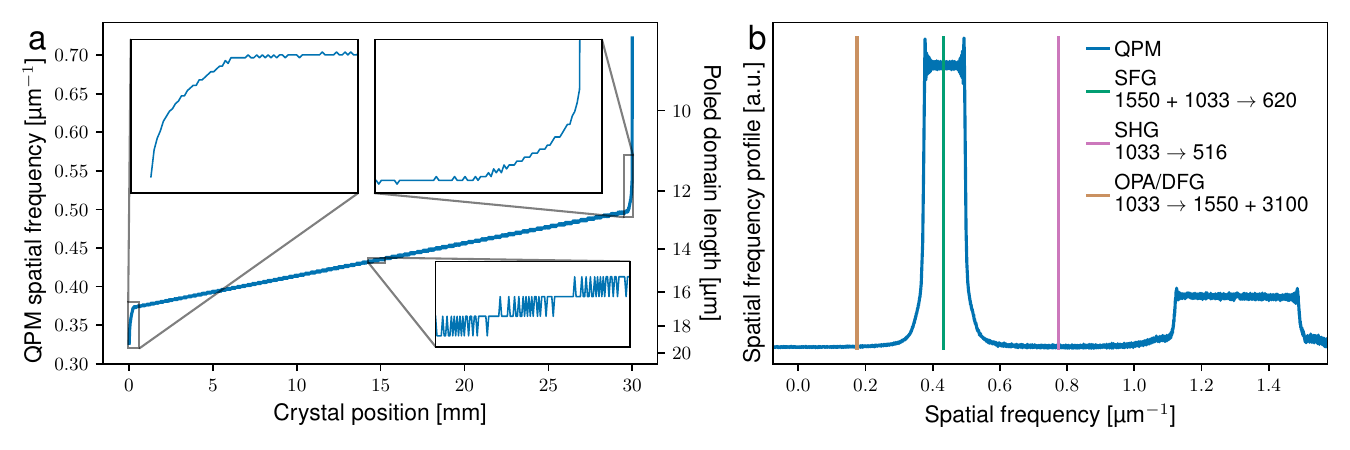}
\caption{
 \textbf{AFC design.}
 \textbf{a.} Crystal domain poling; the spatial frequency and domain length are plotted against position throughout the crystal.
 \textbf{b.} Spatial frequency distribution of the AFC crystal.
 This is compared to the quasi-phase matching frequencies of desired and undesired nonlinear processes.
}
\label{fig:afcdesign}
\end{figure}

\begin{figure}[htb]
\centering
\includegraphics[width=.778\normaltextwidth]{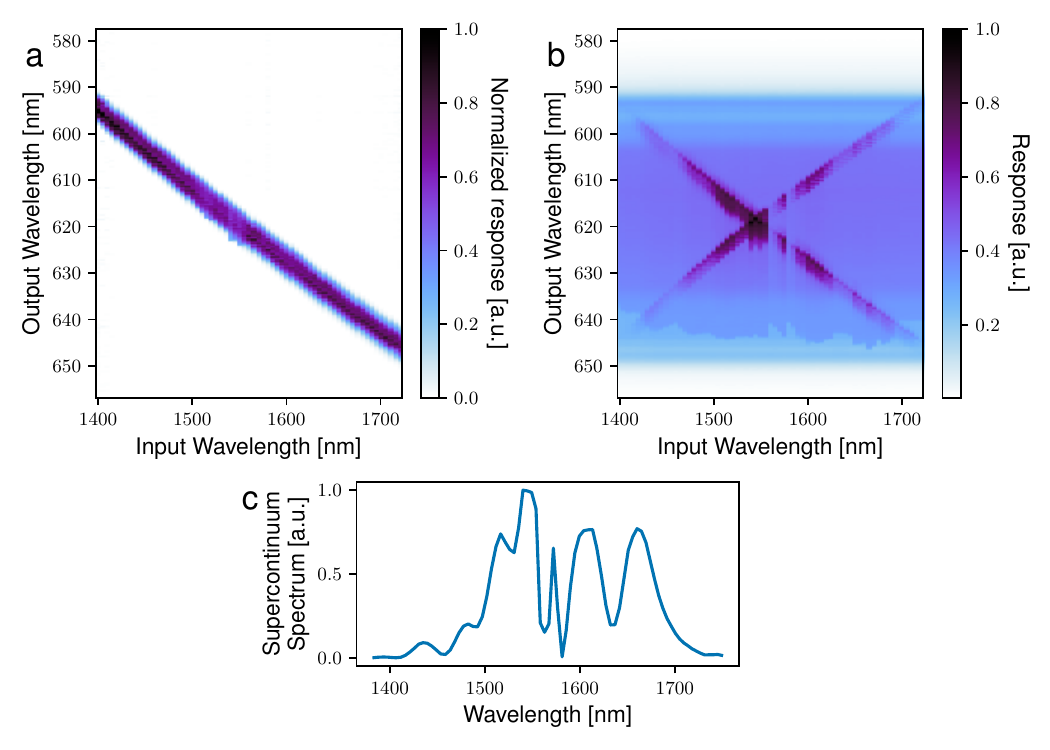}
\caption{
 \textbf{AFC input-output transformation.}
 We measure the intensity (phase-averaged) transformation imparted by the AFC for a given pump, with classical light input.
 \textbf{a.} The visible output transformation matrix of the AFC for single-wavelength inputs.
 The individual output measurements are normalized.
 \textbf{b.} The same measurement while also pumping the OPA (unnormalized).
 The amplified vacuum light constitutes the constant background spectrum.
 \textbf{c.} Spectrum of the supercontinuum source used as an input, prior to monochromation.
}
\label{fig:afcgreens}
\end{figure}

Fig.\,\ref{fig:afcdesign}a shows the AFC poling period design.
The poling spatial frequency is varied linearly throughout the length of the crystal.
As a result, the propagating light experiences a linearly changing phase mismatch (detuning), which allows rapid adiabatic passage.
The front and back of the crystal sweep through poling periods more rapidly, with a tanh profile, to provide the frequencies which convert at the front and back with some propagation distance with detuning.
The domain lengths are quantized to multiples of 0.025\,\textmu m.

Engineering the poling period is a matter of finding the instantaneous phase $\phi(z)$ for a given varying spatial frequency $\Delta k(z)$, and quantizing the sinusoid of the phase into domains $\chi(z): [0, L] \rightarrow \{-1, +1\}$:
\begin{equation}
\chi(z) = \text{sign} (\sin(\phi(z))) , \quad
\frac{d \phi}{dz} = \Delta k(z)
\end{equation}
in our case, for example, we need some linear $\Delta k(z) \in [\beta_i, \beta_f]$, hence we use:
\begin{equation}
\Delta k(z) = \beta_i + \frac{(\beta_f - \beta_i)}{L} z, \quad
\chi(z) = \text{sign} \left( \sin \left(\beta_i z + \frac{(\beta_f - \beta_i)}{2 L} z^2\right)\right)
\end{equation}

The straightforward implementation of this experiment would use the same wavelength to pump both the OPA and the AFC.
Unfortunately, a quick analysis will reveal that the poling spatial frequencies required to quasi-phase match (QPM) the SFG process overlap with those of undesirable nonlinear processes, in both LN and KTP (the two best ferroelectric $\chi^{(2)}$ materials with domain poling).
Specifically, we would expect phase-matched second harmonic generation, which would be highly detrimental to the experiment: the pump could lose a majority of its energy to the second harmonic, possibly also causing damage to the crystal.
This problem is avoided by using the two different pump wavelengths, 775\,nm and 1033\,nm.
Fig.\,\ref{fig:afcdesign}b shows the spatial frequencies of the AFC poling in relation to possible parasitic processes.

To model the dispersion in KTP we used the z-axis Sellmeier and temperature-dependence equations from Refs.~\cite{FradkinRoseman, EmanueliArie}.

As an independent verification that the AFC works correctly, we measure the classical, frequency, intensity input-output relation of the AFC (phase-averaged Green's function), as shown in Fig.\,\ref{fig:afcgreens}a.
Monochromated supercontinuum is sent through the AFC, and the resulting visible spectrum measured on the camera.
(As described in the \nameref{sec:methods} section, this also serves as a calibration procedure for the camera pixel-to-wavelength correspondence, by amplitude-modulating the 1033\,nm laser to monochromate it.)
Fig.\,\ref{fig:afcgreens}b shows the transformation with the inclusion of the OPA.

\subsection*{Influence of pump shape on frequency conversion}\label{sec:app:afc:pumpspectrum}
\addcontentsline{toc}{subsection}{\nameref{sec:app:afc:pumpspectrum}}

\begin{figure}[htbp]
\centering
\includegraphics[width=0.5\normaltextwidth]{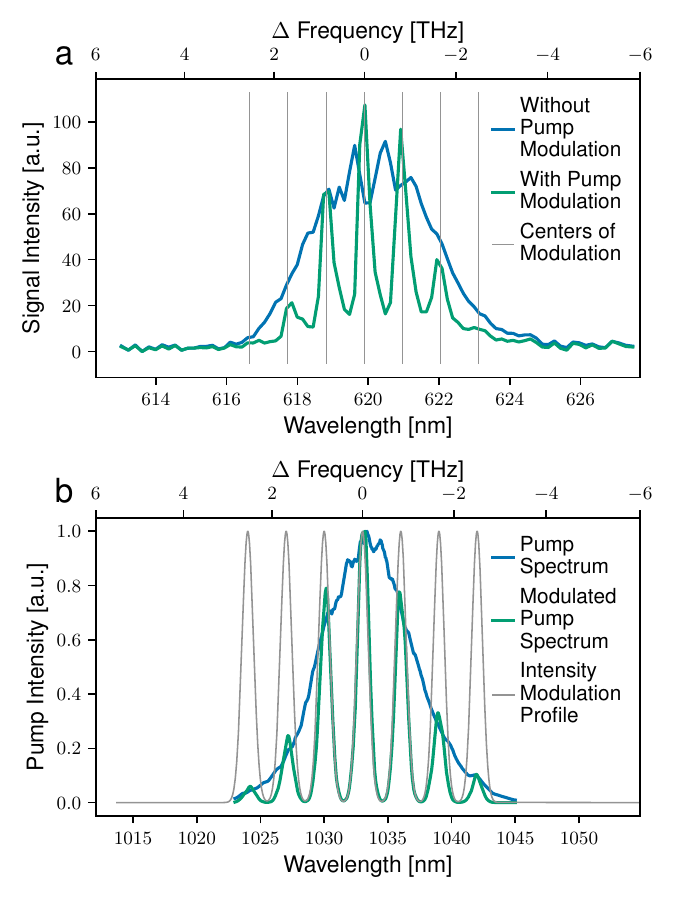}
\includegraphics[width=0.5\normaltextwidth]{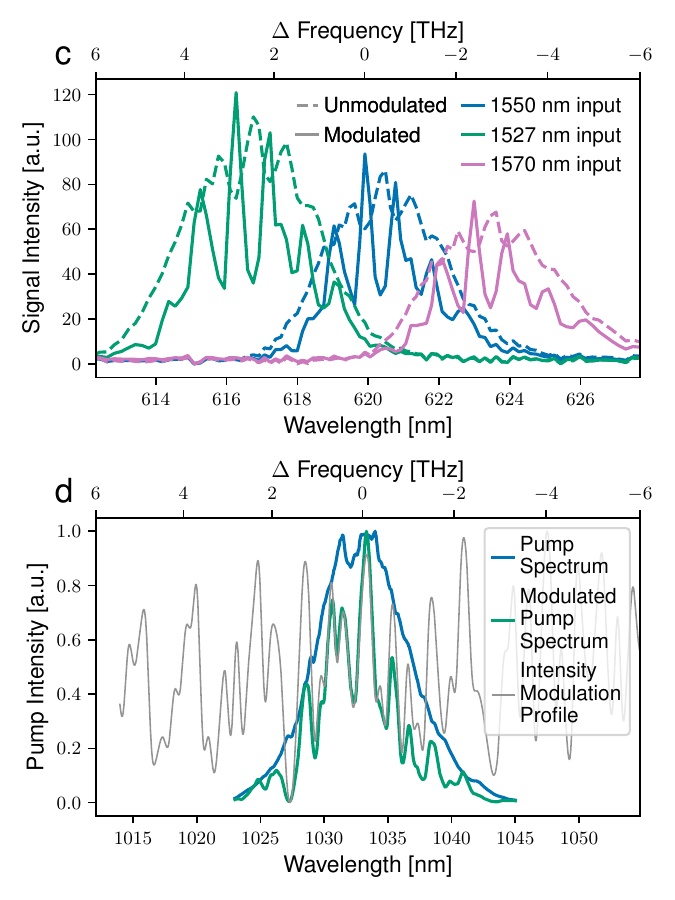} \\
{\vspace{-.5\baselineskip} \includegraphics[scale=0.5]{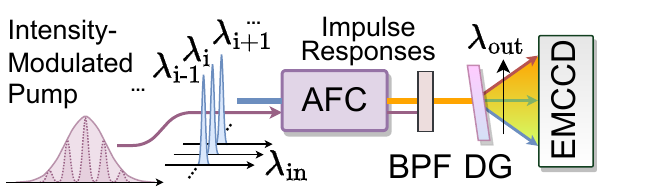}}
\caption{
 \textbf{Programming the unitary transformation by programming the pump shape: the effect of the pump spectrum shape on the transformation.}
 Example of how the pump spectrum affects the frequency conversion profile and linear transformation.
 \textbf{a.} The converted signal spectral profiles, under unmodulated and modulated pumps.
 \textbf{b.} The pump spectra for \textbf{a}, unmodulated and under intensity modulation.
 The conversion profile of the signal closely matches the pump spectrum.
 \textbf{c.} The converted signal spectral profiles, unmodulated and modulated, with three different wavelength inputs.
 The measured signal intensities are essentially translated in frequency.
 \textbf{d.} The modulated and unmodulated pump spectra for \textbf{c}.
 A random intensity modulation pattern is applied.
 $\lambda_\text{in}$ indicates a monochromatic, infrared input light, and $\lambda_\text{out}$ indicates the wavelengths measured by the spectrometer.
}
\label{fig:afcampmod}
\end{figure}

Fig.\,\ref{fig:afcgreens}a shows the (phase-insensitive) experimental measurement of a linear transformation performed by AFC, where the pump is a simple chirped pulse.
Observe that a slice of this transformation function (i.e.\ for a given input; a column of the matrix) resembles a simple Gaussian: the pump spectrum.
Indeed, in this regime, to good approximation, the pump intensity profile directly determines the magnitude of the transformation function.

Fig.\,\ref{fig:afcampmod} demonstrates this effect: by intensity-modulating the pump spectrum (with a fixed phase), for a fixed monochromatic input, we observe that the output signal closely matches the pump profile.
However, by virtue of reducing the overall pump energy, the efficiency is affected.
In addition, a shift in the input signal ostensibly produces an the same, but shifted, output signal.

While the pump imparts its intensity to the magnitude, the spectral phase of the pump determines the phase in the transformation, which is why we observe that pump phase modulation can generate non-trivial differences in the covariance matrices in Fig.\,\ref{fig:covs}.

\begin{figure}[htbp]
\centering
\includegraphics[width=0.75\normaltextwidth]{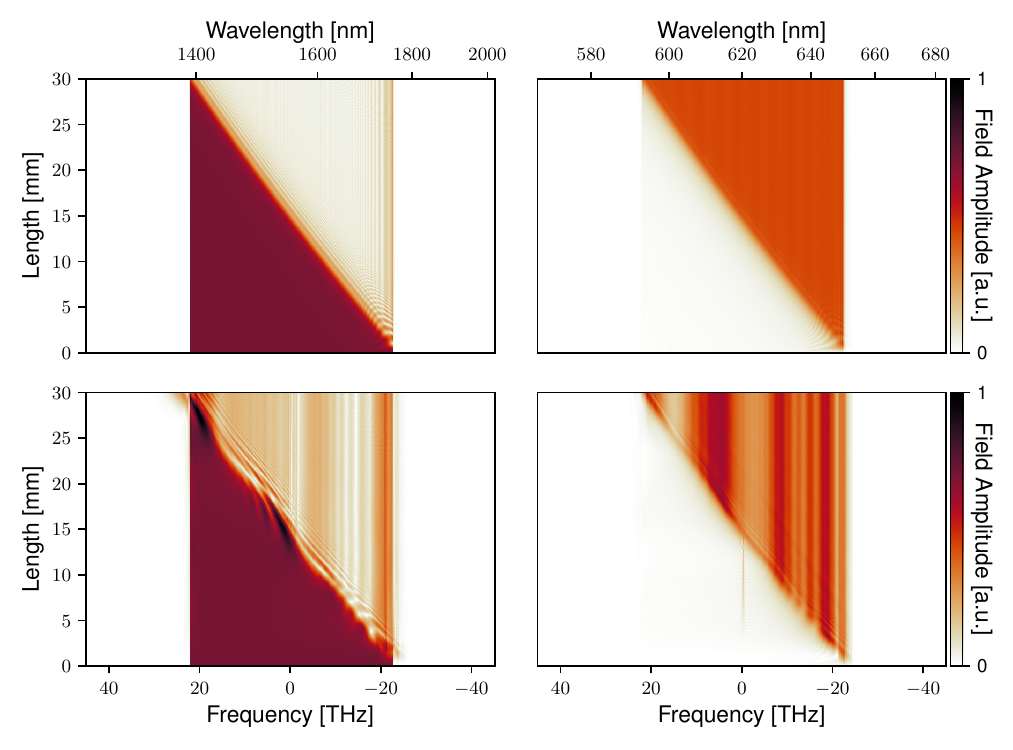}
\includegraphics[width=0.25\normaltextwidth]{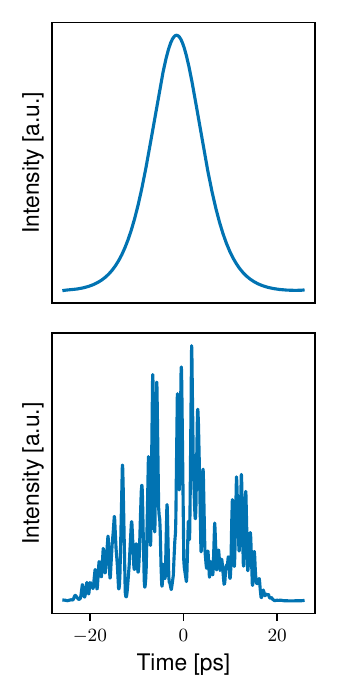}
\caption{
 \textbf{Example conversion dynamics in AFC.}
 The left panel represents the input modes centered around 1550\,nm, the center panel represents the converted modes centered around 620\,nm, and the right panel represents the pump profile used in the simulation.
 \textbf{Top:} conversion of a single broadband input with an unshaped pump.
 \textbf{Bottom:} conversion of a single broadband input with an shaped pump.
 The conversion yields a non-uniform spectrum in the converted wavelength due to interference effects.
}
\label{fig:afcdynamics}
\end{figure}

\begin{figure}[htbp]
\centering
\includegraphics[width=1\normaltextwidth]{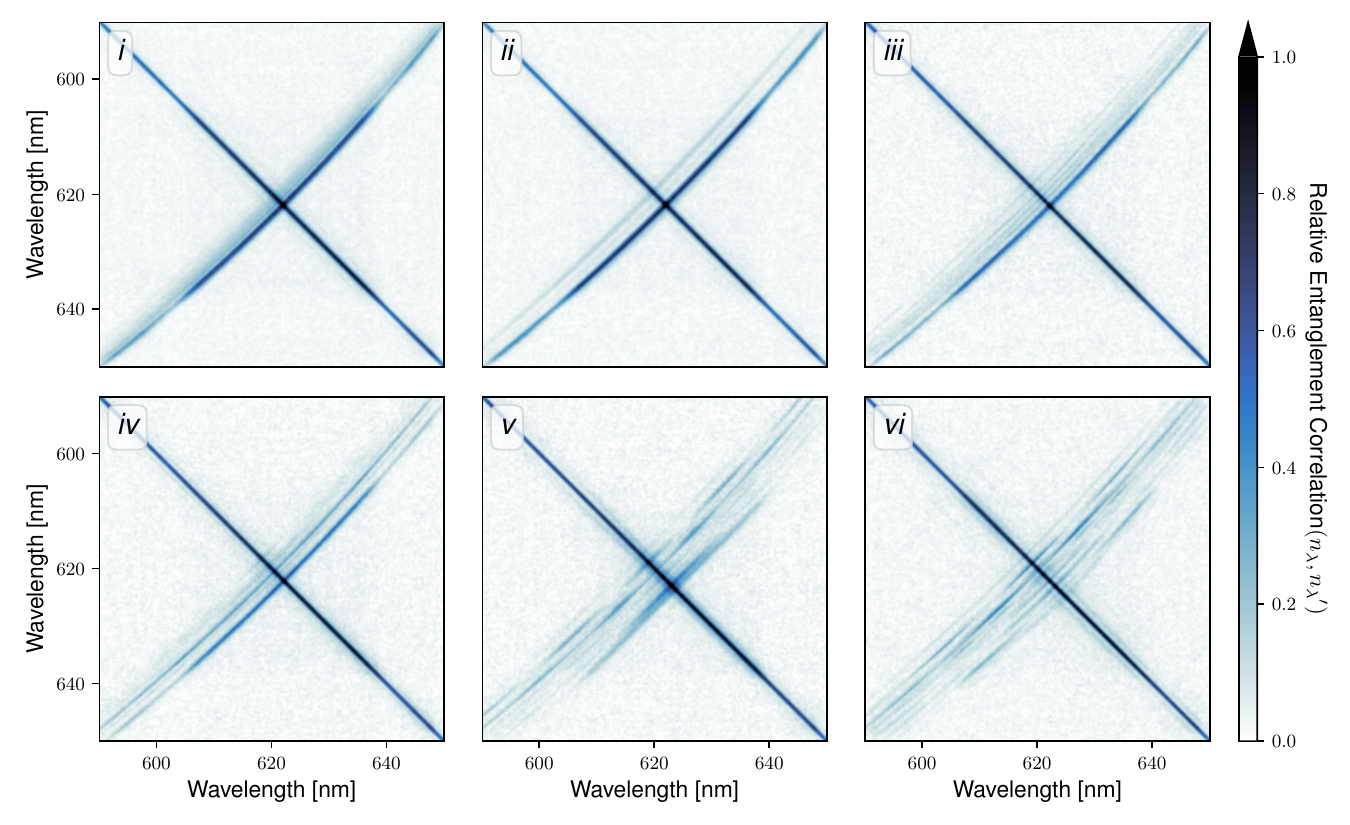}
\includegraphics[width=1\normaltextwidth]{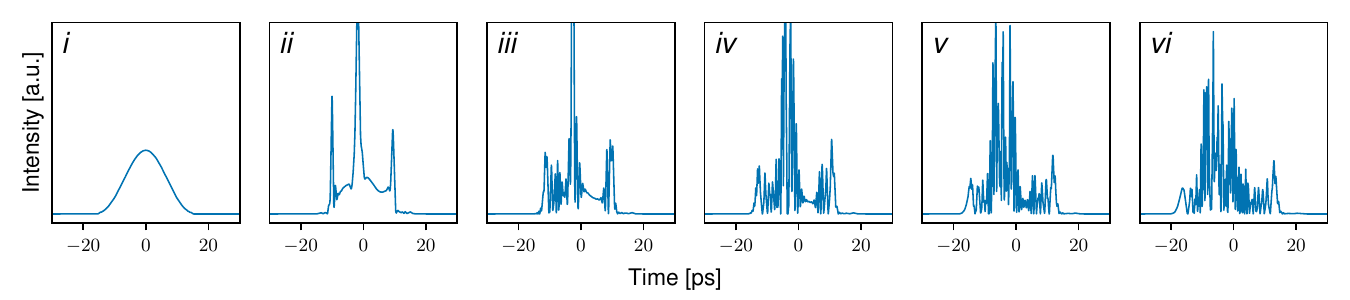}
\caption{
 \textbf{Photon spectral correlations while varying the AFC pump phase modulation.}
 The top panels are photon number correlation plots, where \textit{i} is the state generated by AFC with a chirped pulse, \textit{vi} with the shaped pulse in Fig.\,\ref{fig:covs}\textit{iv}, and the panels in between are with intermediately spaced phase modulations of the AFC pump.
 These six panels show the evolution of the correlation structure as the phase profile varies.
 The corresponding temporal profiles of the pumps, estimated based on the spectral phase modulations, are shown in the bottom panels.
 In contrast to Fig.\,\ref{fig:covs}, the diagonal portion of the correlation matrix has not been subtracted.
}
\label{fig:phasemodscan}
\end{figure}

Fig.\,\ref{fig:afcdynamics} shows the simulated conversion process for a broadband signal.
The conversion process under an unshaped pump converts fairly uniformly,
while the process under a shaped pump is influenced by interference effects, strongly altering the shape of the output.
Fig.\,\ref{fig:phasemodscan} shows a series of experimentally obtained photon spectral correlation matrices as the phase modulation of the AFC pump is gradually varied.

\subsection*{Fluorescence in KTP}\label{sec:app:afc:fluo}
\addcontentsline{toc}{subsection}{\nameref{sec:app:afc:fluo}}

\begin{figure}[htb]
\centering
\includegraphics[width=1\normaltextwidth]{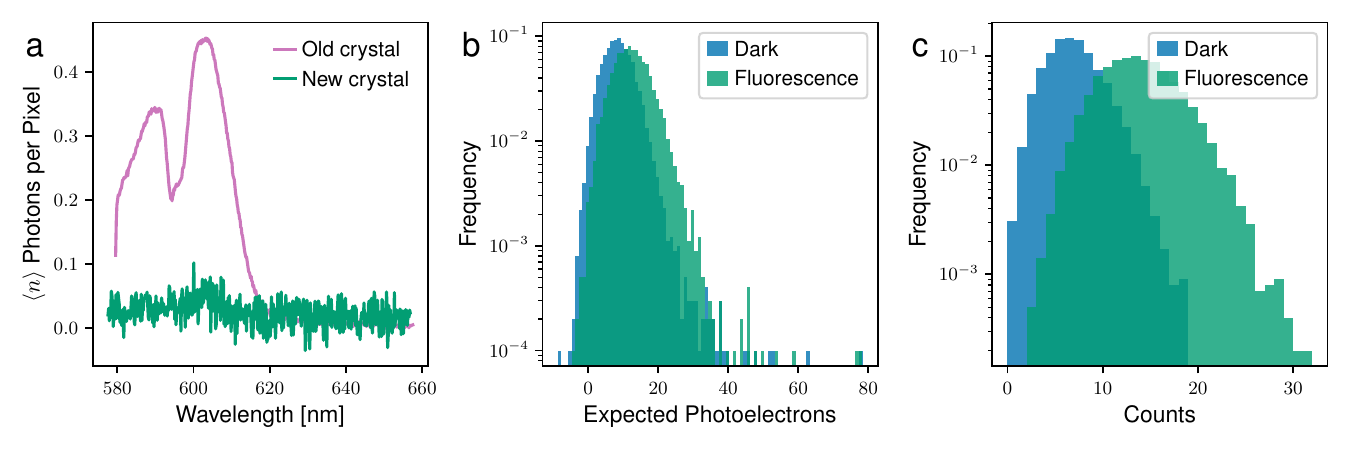}
\caption{
 \textbf{KTP fluorescence.}
 \textbf{a.} Spectrum of the fluorescence background emanating from KTP.
 The old crystal (regularly exposed to a high fluence of green light) emits a distinct spectrum around 600\,nm.
 Fluorescence from the new crystal is low and almost indistinguishable from camera noise.
 \textbf{b.} Cumulative analog signal with dark frames and fluorescence background.
 \textbf{c.} Thresholded counts with dark frames and fluorescence background.
 Experiments \textbf{b--c} were performed with 5$\times$512 pixels.
}
\label{fig:fluorescence}
\end{figure}

We have noticed that the 1033\,nm pumped KTP crystal generates a spectrum centered around 600\,nm.
This has been previously observed (see \cite{HegdeZelmon} and citing articles; \cite{XuJia}).
This fluorescence behavior is likely due to the absorption of green 517\,nm second-harmonic light by crystal defects.
The green light is generated parasitically in the crystal.

We also observe that the fluorescing appears to worsen with crystal use.
The fluorescence in a new crystal is practically undetectable, but becomes gradually brighter as it is subjected to high peak-power green light.

Ref.~\cite{HegdeZelmon} reported that different growth methods can suppress this behavior in KTP, however, a different material may have to be considered for future experiments, as this is a problem inherent to KTP.

Fig.\,\ref{fig:fluorescence}a shows the fluorescence spectrum for a strongly fluorescing crystal, which had been subjected to a high fluence of green light over time through use as an AFC crystal (compare to Fig.\,\ref{fig:afc}a).
Below this curve is the background measured in the experiments in Fig.\,\ref{fig:sampling}, performed with a relatively new crystal.
Fig.\,\ref{fig:fluorescence}b and c compare the fluorescence signal accumulated by the camera, in analog and thresholded mode respectively.
For comparison, dark frames are plotted together (camera shutter closed).
On average, fluorescence and background light account for fewer than 10 additional photons per shot.

Dark counts can be reduced by using fewer rows of pixels, at the cost of pseudo-photon number resolving (if thresholding).
This experiment used 5 rows of pixels (see Fig.\,\ref{fig:sampling}c).

\section{Spectrometer design and validation}\label{sec:app:spectrometer}

\subsection*{Design}\label{sec:app:spectrometer:design}
\addcontentsline{toc}{subsection}{\nameref{sec:app:spectrometer:design}}

In order to verify the correlation structure between frequency modes generated by our quantum light source, we design a spectrometer that has a uniformly high spectral resolution and low loss around $\lambda_0 =$~620\,nm, the central wavelength.
Here we outline the design considerations to achieve the desired spectral resolution.

The number of frequency modes we can resolve is ultimately limited by the number of pixels in each row of the EMCCD camera.
We choose an EMCCD camera with a sensor size of 512 by 512, instead of a larger one (e.g.,\ 1024 by 1024), because the
better signal-to-noise ratio that would likely preserve more information on our light source
(see below for considerations in the choice of camera).
The spectrometer is designed for a 60\,nm bandwidth (a spectrum between 590\,nm and 650\,nm).
The ideal spectral bin width in each EMCCD pixel is thus $\Delta \lambda$ = 60/512 $\approx$ 0.12\,nm.
The design of the optics must satisfy the following two conditions to provide this resolution:
\begin{enumerate}
  \item The spatial resolution of the lens must be finer than the pixel size of the EMCCD camera, which is $x_p =$~16\,\textmu m.
  The focal length of the objective lens we chose (Olympus UPLFLN4x) is 45\,mm.
  To match the focal spot size ($1/\mathrm{e}^2$ diameter) to the pixel size, the $1/\mathrm{e}^2$ beam diameter at the back aperture of the objective lens should be at least $2 f \lambda_0 / (\pi x_p / 2)) =$ 2 $\times$ 45\,mm $\times$ 620\,nm/($\pi \times$ 8 \textmu m) $\approx$ 2.22\,mm, which is smaller than the back aperture diameter (11.7\,mm).

  \item The angular resolution of the grating must exceed $\Delta \lambda$, which requires the beam to cover at least $\lambda_0 / \Delta \lambda =$ 620\,nm/0.12\,nm $\approx$ 5300 grating lines.
  Since the grating (Ibsen PCG-1908/675-972) has a line density of 1908 lines per mm, this means the minimum $1/\mathrm{e}^2$ diameter of the beam on the grating should be at least (5300 lines)/(1908 lines/mm) $\approx$ 2.8\,mm.
  The grating has an area of 20\,mm~$\times$~10\,mm, sufficiently large for a beam of this size.
\end{enumerate}

In conclusion, to achieve both high spatial and spectral resolution, the beam size must be at least 2.8\,mm, which is well within the clear aperture size of the grating and objective lens.
The target beam size was designed to be approximately 8\,mm, and the actual beam size was slightly smaller than that.

To maximally preserve the quantum properties of the light, all the optics, including routing mirrors, grating, and focusing lens, should have uniformly low loss around 620\,nm.
The overall quantum efficiency of the spectrometer, including the EMCCD camera quantum efficiency, is measured to be around 75\% at 633\,nm.

\subsection*{Imaging resolution}\label{sec:app:spectrometer:imaging}
\addcontentsline{toc}{subsection}{\nameref{sec:app:spectrometer:imaging}}

\begin{figure}[htb]
\centering
\includegraphics[width=0.778\normaltextwidth]{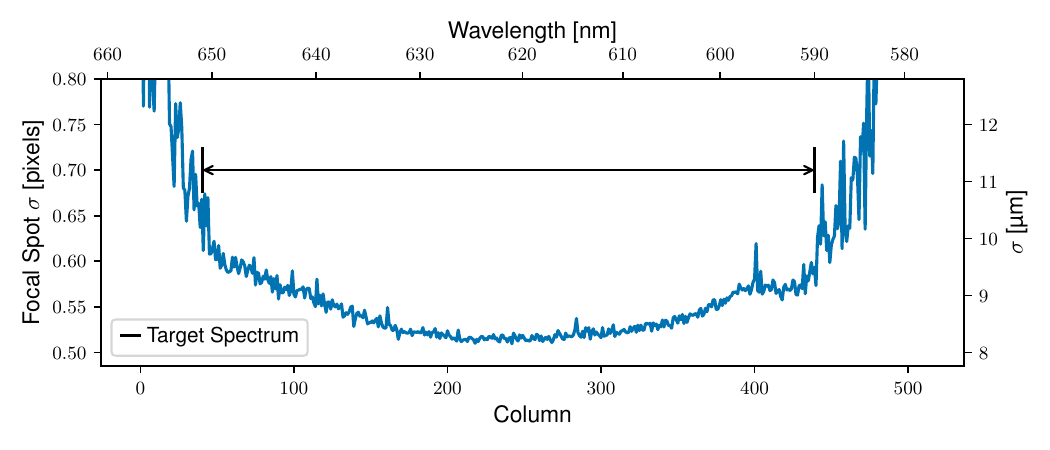}
\caption{
 \textbf{Focal spot size throughout the image plane.}
 Results of a Gaussian fit to the intensity spread at every column; the standard deviation $\sigma$ is plotted.
 The intensity drops outside the target spectrum region, hence the sharp transitions in the fits.
}
\label{fig:pointspread}
\end{figure}

As mentioned in the Discussion, there are imperfections with the detection setup.
As may be seen in Fig.\,\ref{fig:afc}a, we do not make full use of the number of pixels: between 10--20\% of the spectrometer measures wavelengths outside the target bandwidth.
Ideally, customized optics would allow the focal length of the objective lens to match the range of the diffraction angles from the grating to the width of the sensor.
As is, the number of detection modes and squeezed modes are close, a possible cause of decoherence.

Relatedly, the monochromatic focus should be well sub-pixel (16\,\textmu m), however, in
this experiment we achieved a focus spot with standard deviation $\sigma \approx 0.6$ pixels (Gaussian fit) -- on the order of the size of a pixel.
We believe the spatial mode quality of the beam is reduced when it is converted, reducing the tightness of the focus.
This is also a cause of decoherence because it becomes more difficult to measure the frequency of each photon, and correlations become blurred (Appendix~\ref{sec:app:spectrometerpovm}).

The focal spot size of the spectrometer is shown in Fig.\,\ref{fig:pointspread}, throughout the wavelength axis (columns) of the camera.
The average intensity distribution along the 5 vertical pixels of each column is fit to a Gaussian.
The point spread function is a radially symmetric Gaussian in these experiments.
It is fairly constant over the image plane (where illuminated), but larger than expected.
The energy falling outside of a given pixel is therefore about $\text{erfc}(1/\sigma\sqrt{2}) \approx$ 5--10\%, which -- in the worst case -- could amount to an equivalent amount of loss and noise, due to these photons erroneously illuminating neighboring pixels (although this is effectively highly correlated noise and not likely to be as detrimental as thermal noise).

\subsection*{Evaluation of EMCCD camera settings}\label{sec:app:spectrometer:camchoice}
\addcontentsline{toc}{subsection}{\nameref{sec:app:spectrometer:camchoice}}

\begin{figure}[htb]
\centering
\includegraphics[width=0.778\normaltextwidth]{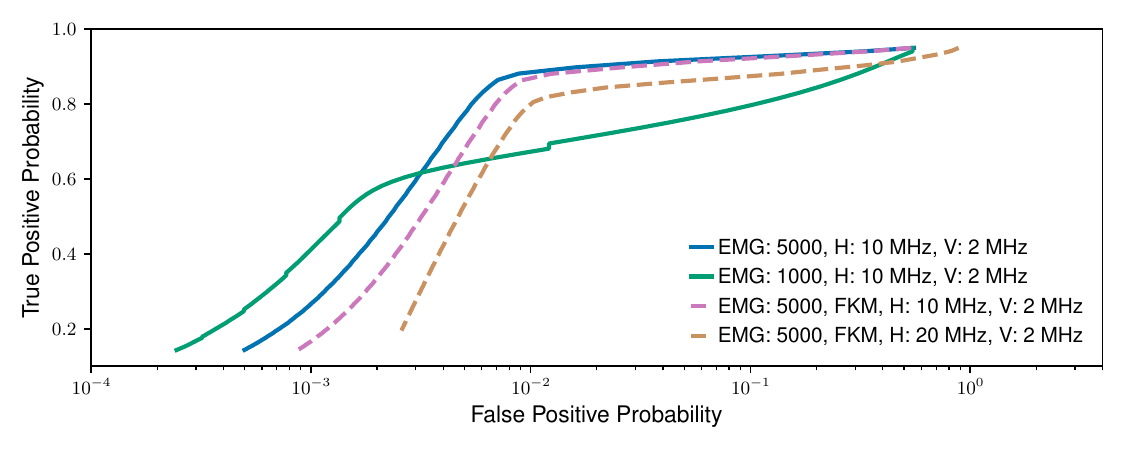}
\caption{
 \textbf{Camera performance.}
 \textbf{a.} ROC curves for different cameras and operational modes.
 EMG, EM gain; H, horizontal line rate; V, vertical line rate; FKM, fast kinetic mode.
}
\label{fig:camevaluation}
\end{figure}

In order to quantify the performance of the different configurations of the EMCCD camera, we consider their ROC curves.
As discussed in Appendix~\ref{sec:app:emccd:thresh}, these represent the trade-off between the photon-detection efficiency (PDE) and the dark count rate: true positive and false positive probabilities.
These are shown in Fig.\,\ref{fig:camevaluation}.

\subsection*{Calibrating the EM gain}\label{sec:app:spectrometer:emgaincal}
\addcontentsline{toc}{subsection}{\nameref{sec:app:spectrometer:emgaincal}}

\begin{figure}[htb]
\centering
\includegraphics[width=\normaltextwidth]{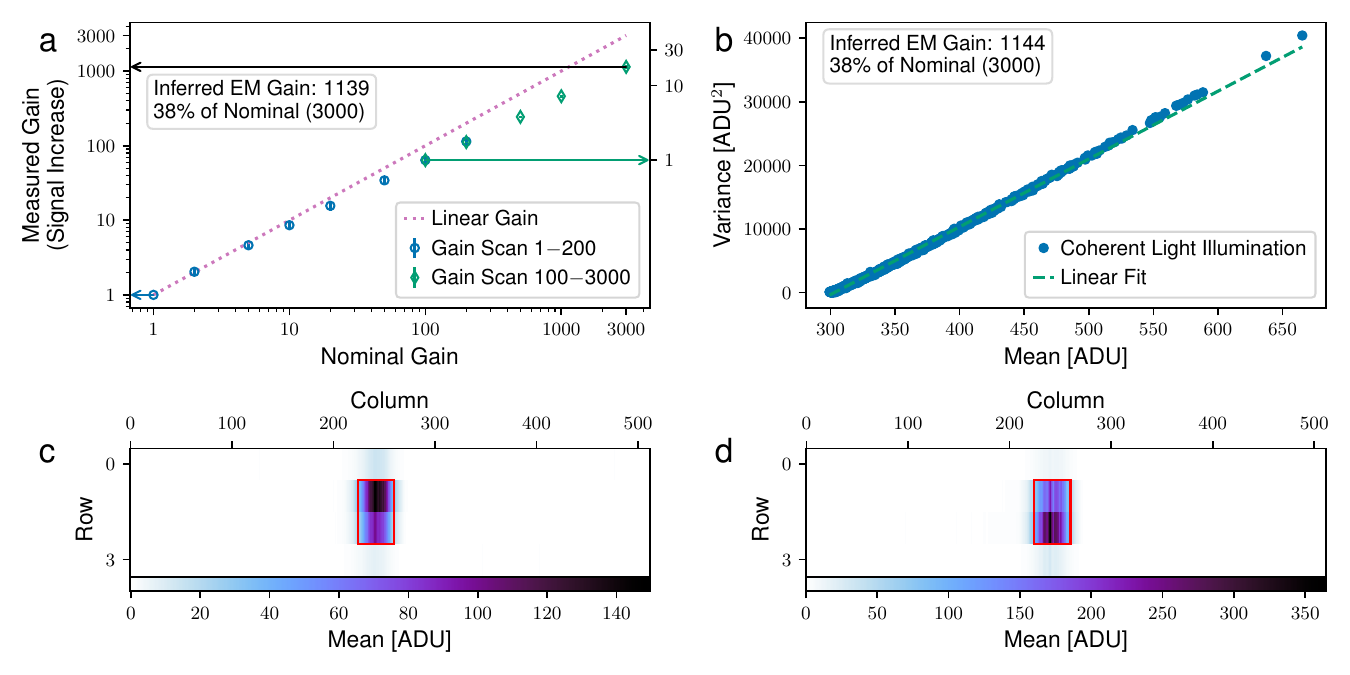}
\caption{
 \textbf{Calibrating the EM gain with scaling and photon statistics.}
 \textbf{a.} Direct method.
 A constant illumination was measured with the camera, as the gain was increased up to a nominal value of 200.
 The relative increase in the signal, as a function of gain and averaged over pixels, is plotted.
 The intensity was reduced and the collection was repeated for gains nominally between 100 and 3000.
 By combining the two datasets we find that a nominal gain of 3000 only yields a 1139-fold increase in signal compared to unit gain (38\% of nominal).
 \textbf{b.} Photon statistics.
 The camera was illuminated with coherent light (converted 1550\,nm laser), which we expect to be Poissonian (equal photon variance and mean).
 The means and variances of the intensities -- as measured by different illuminated pixels -- are plotted.
 This was repeated for different average intensities of the illumination.
 A linear fit is performed on the entirety of these data.
 Its slope is used to infer the EM gain (accounting for a factor of 2 excess gain noise and the analog-to-digital conversion factor).
 The fit implies a gain of approximately 1147 instead of 3000 (38\% of nominal).
 \textbf{c.}
 Illumination profile for the experiment presented in panel~\textbf{a}.
 This image is the mean of those collected for the unit gain datum point.
 The red box indicates the region of pixels included in the analysis.
 The bias was subtracted from all images.
 \textbf{d.}
 Illumination profile for the experiment presented in panel~\textbf{b}.
 This image represents the mean value of those collected with the highest intensity coherent light illumination.
 The red box indicates the region of pixels included in the analysis.
 The bias was subtracted from all images.
}
\label{fig:gaincal}
\end{figure}

An accurate measure of the the camera's EM gain is important for determining the the optimal thresholding efficiencies, as well as for analyzing mean intensities and the results of intensity correlation experiments without thresholding.
We observed that the true value of the gain gradually decreased with respect to the nominal value over time, and that higher gains were more affected by this aging process.

There are different methods to calibrate the EM gain: direct (measuring signal strength with different gain for a fixed illumination), and indirect (photon statistics).
Fig.\,\ref{fig:gaincal} presents data showing that these two methods yield equivalent calibrations.
At the time that this particular data was collected, both measurements implied that the total gain was 38\% of the nominal value, when the EM gain was set a value of 3000.

For the photon statistics case, we expect a factor of 2 excess noise due to the EM gain that must be included in the analysis.
This follows the discussion in Appendix~\ref{sec:app:emccd}.
In summary, for any given pixel, we have intensity moments $\langle x \rangle$ and $\langle x^2 \rangle$ in analog-to-digital units (ADU), which correspond to a gain $g$ and photon number moments $\langle n \rangle$ and $\langle n^2 \rangle$.
Under the assumption of Poisson photon statistics:
\begin{aligned*}
\Var(n) &= \langle n \rangle \\
\langle n^2 \rangle - \langle n \rangle^2 &= \langle n \rangle \\
g^{-2} \langle x^2 \rangle - g^{-1} \langle x \rangle - g^{-2} \langle x \rangle^2 &= g^{-1} \langle x \rangle \\
\Var(x) &= 2 g \langle x \rangle .
\end{aligned*}
Note that the overall gain includes the analog-to-digital conversion factor, converting between units of electron number and ADU.

\section{Quantum light with the Ti:Sapphire pump laser}\label{sec:app:tis}

\begin{figure}[htb]
\centering
\includegraphics[width=0.667\normaltextwidth]{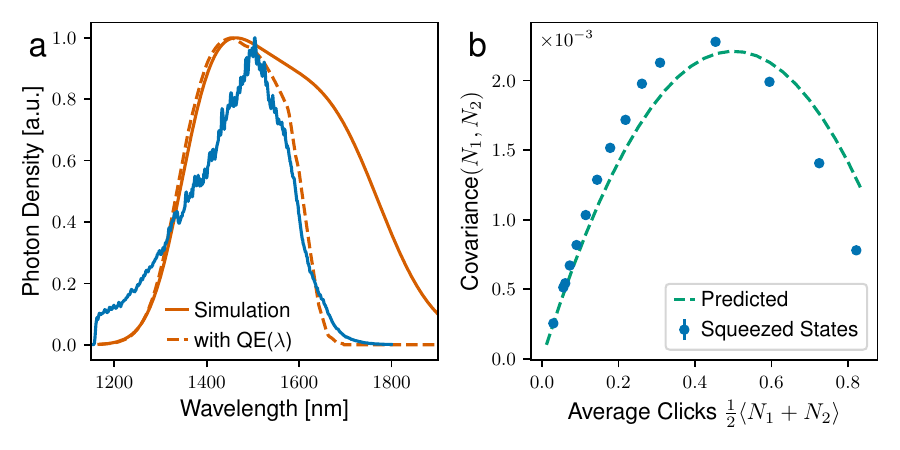}
\caption{
 \textbf{Quantum light from the 100\,fs Ti:S.}
 (See Fig.\,\ref{fig:dopa}d and f for comparison.)
 \textbf{a.} Measured and simulated OPA output spectrum.
 \textbf{b.} Coincidence detection.
}
\label{fig:tisapphire}
\end{figure}

\begin{figure}[htb]
\centering
\includegraphics[width=0.333\normaltextwidth]{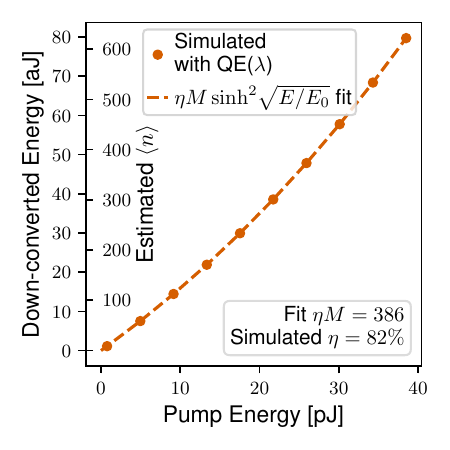}
\caption{
 \textbf{Simulated parametric gain with the 200\,fs pump.}
 (See Fig.\,\ref{fig:dopa}e for comparison.)
}
\label{fig:parametricgain200fs}
\end{figure}

Fig.\,\ref{fig:dopa}e shows the parametric gain measured with the OPA pumped by a 775\,nm, 100\,fs, Ti:S laser, while
although Fig.\,\ref{fig:dopa}d and f are measured with pumping from the 775\,nm converted from the 200\,fs, 1033\,nm laser.
Due to the larger bandwidth of the Ti:S pulses, the OPA output is slightly different in both cases.
Here we show these two other experiments from Fig.\,\ref{fig:dopa} performed with the Ti:S, for comparison.
These results are shown in Fig.\,\ref{fig:tisapphire}.
The Ti:S was necessary to achieve a measurable output for the parametric gain experiment, as the repetition rate of the 775\,nm light generated through the 1030\,nm laser was limited by the optical amplifier (see \hyperref[sec:methods:lbo]{Methods}).
The optical amplifier, used to generate the 1550\,nm OPA seeds, had a maximum average power of 100\,mW, so significantly increasing the repetition rate would proportionally decrease the energy of each 775\,nm pulse.

Similarly, the simulations of the parametric gain we would expect for the 200\,fs pump of the OPA are shown in Fig.\,\ref{fig:parametricgain200fs}.
These correspond to the simulated spectrum in Fig.\,\ref{fig:dopa}c.

\addtocontents{toc}{\protect\nohyperlinkcontentsline{section}{Prospects}{}\%}

\section{Comparison to previously published multimode squeezing systems}\label{sec:app:mmscomparison}

\begin{table}[!htbp]
\centering
\caption{\textbf{Published results in multimode squeezing.}}\label{table:comparisons}
\begin{tabular}{p{0.035\normaltextwidth}p{0.035\normaltextwidth}p{0.09\normaltextwidth}p{0.1\normaltextwidth}p{0.1\normaltextwidth}p{0.09\normaltextwidth}p{0.16\normaltextwidth}p{0.225\normaltextwidth}}
\hline
  \centering Fig. \ref{fig:comparison} Label &
  \centering Ref. &
  \centering Domain &
  \centering \#~Squeezed Modes &
  \centering \#~Photon Detection Modes &
  \centering Average Total \# of Detected Photons &
  \centering Reprogrammability &
  Notes \\
  \hline
  ~ & \cite{PysherPfister} & Frequency & 15 & - & - & Fixed & 4\,dB squeezing \\ 
  1 & \cite{TillmannWalther} & Space & 4 & 4 & 3 & Fixed & Post-selected \\ 
  2 & \cite{CrespiSciarrino} & Space & 2 & 6 & 3 & Fixed & Post-selected \\ 
  3 & \cite{BroomeWhite} & Space & 4 & 4 & 4 & Fixed & Post-selected \\ 
  ~ & \cite{YokoyamaFurusawa} & Time & 10,000 & - & - & Fixed & 6\,dB squeezing \\ 
  ~ & \cite{ChenPfister} & Frequency & 60 & - & - & Fixed & 3.4\,dB squeezing \\ 
  4 & \cite{CarolanLaing} & Space & 4 & 9 & 3 & Fixed & Post-selected \\ 
  5 & \cite{BentivegnaSciarrino} & Space & 4 & 13 & 4 & Fixed & ~ \\ 
  6 & \cite{CarolanLaing2} & Space & 4 & 6 & 6 & Programmable & Post-selected \\ 
  ~ & \cite{XieWong} & Frequency & 9 & - & - & Fixed &  \\ 
  ~ & \cite{ReimerMorandotti} & Frequency & 10 & - & - & Fixed &  \\ 
  7 & \cite{HarderSilberhorn} & NA & 2 & 2 & 80 & NA & Single two-mode squeezer \\ 
  ~ & \cite{YoshikawaFurusawa} & Time & 1,200,000 & - & - & Fixed & 4.3\,dB squeezing \\ 
  ~ & \cite{CaiTreps} & Frequency & 6 & - & - & Fixed & 6.6\,dB squeezing \\ 
  8 & \cite{ZhongPan3} & Space & 24 & 24 & 5 & Fixed & ~ \\ 
  9 & \cite{ZhongPan2} & Space & 6 & 12 & 5 & Fixed & ~ \\ 
  10 & \cite{PaesaniLaing} & Space & 4 & 16 & 4 & Fixed & ~ \\ 
  ~ & \cite{TakedaFurusawa} & Time & 1,000 & - & - & Partial & 1-dimensional connectivity; 5\,dB squeezing \\ 
  ~ & \cite{AsavanantFurusawa} & \raggedright{Time \& Space} & 24,960 & - & - & Fixed & \textgreater4.5\,dB squeezing \\ 
  ~ & \cite{LarsenAndersen} & \raggedright{Time \& Space} & 30,000 & - & - & Fixed & 4.7\,dB squeezing \\ 
  11 & \cite{ZhongPan} & Space & 50 & 100 & 43 & Fixed & ~ \\ 
  ~ & \cite{ChangWong} & Time & 61 & - & - & Fixed & ~ \\ 
  ~ & \cite{YangYi} & Frequency & 40 & - & - & Fixed & 3.1\,dB squeezing \\ 
  ~ & \cite{OmiKannari} & Frequency & 2 & - & - & Fixed \\ 
  12 & \cite{ArrazolaZhang} & Space & 8 & 8 & 11 & Programmable & Bipartite connectivity \\ 
  13 & \cite{ZhongPan2021} & Space & 50 & 144 & 70 & Fixed & Programmable phases \\ 
  14 & \cite{SempereKolthammer} & Time & 10 & 20 & 6 & Partial & 1-dimensional connectivity \\ 
  15 & \cite{MadsenLavoie} & Time & 216 & 216 & 125 & Programmable & 3-dimensional connectivity \\ 
  16 & \cite{DengPan} & Space & 50 & 144 & 100 & Partial & Programmable phases \\ 
  ~ & \cite{RomanParigi} & Frequency & 20 & - & - & Fixed & 2.5\,dB squeezing \\ 
  17 & \cite{YuGuo} & Time & 32 & 32 & 1 & Programmable & Full connectivity \\ 
  18 & \cite{BaoJianwei} & Space & 32 & 32 & 2 & Programmable \\

  ~ & \cite{ZhuLiu} & Space & 16 & 16 & - & Programmable & Fully programmable \\ 
  19 & \cite{YuBianconi} & Time & 40 & 40 & 4 & Programmable & Fully programmable \\ 
  20 & \cite{LiuPan} & Time \& Space & 1024 & 8176 & 2650 & Partial & Full connectivity \\ 
  ~ & \cite{YuPatel} & Time & 100 & 100 & - & Programmable & Fully programmable \\ 

  \hline
  & This work & Frequency & 433 & 512 & 680 & Partial & 500 photons thresholding 
\end{tabular}
\end{table}

\begin{figure}[htb]
\centering
\includegraphics[width=0.5\normaltextwidth]{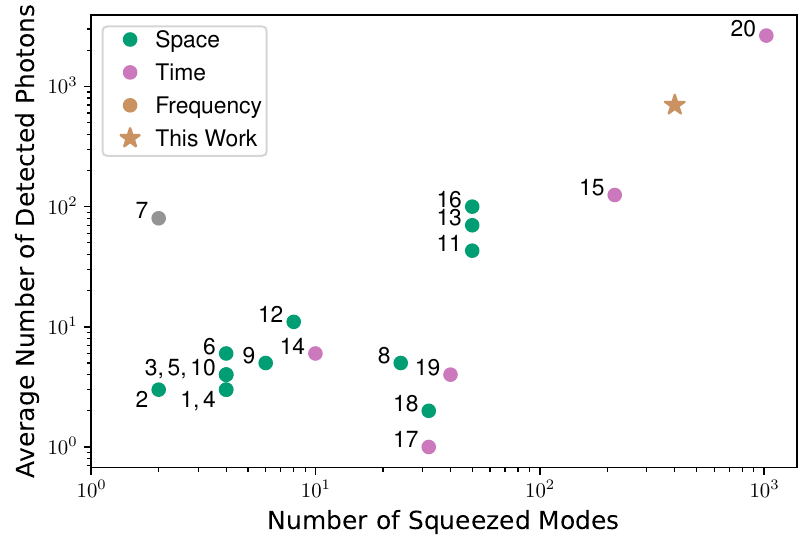}
\caption{
 \textbf{Published results in multimode squeezing.}
 Graphical representation of the experiments collected in Table~\ref{table:comparisons} in which single-photon detection (as opposed to homodyne detection) was performed. The plot axes represent the number of modes that are initially squeezed (before the application of a unitary, if applicable) and the average total number of photons detected (over all modes).
}
\label{fig:comparison}
\end{figure}

Table~\ref{table:comparisons} lists previously published results of experiments demonstrating multimode squeezing, in particular the number of modes and detected photons, where applicable.
A subset of these figures of merit are plotted in Fig.\,\ref{fig:comparison}.
(As we only include references which use some form of squeezed light, this excludes experiments using single-photon emitters.)

\subsection*{Estimating the experimental squeezing from photon statistics}\label{sec:app:squeezingestimate}
\addcontentsline{toc}{subsection}{\nameref{sec:app:squeezingestimate}}

Here we estimate the squeezing in dB that was achieved in the experiment.
As reported in Fig.\,\ref{fig:afc}c--d: $\eta=49\%$ was inferred as the total transmission; $\langle n \rangle = 887$ photons were measured at the maximum pump energy; and $M=433$ modes were effectively present.
We can thus estimate the noise reduction in dB per mode as follows:
\begin{equation}
\text{squeezing in dB} = 10 \log_{10} \left( \eta \mathrm{e}^{-2 r} + 1 - \eta \right), \text{ where } r \approx \sqrt{P/P_0} = \text{arcsinh} \sqrt{\langle n \rangle / \eta M}.
\end{equation}
The average mean photon number per mode is approximately 2.05, or 4.18 prior to loss.
The average squeezing parameter in the pure state (assuming negligible losses during the OPA process) is thus $r=1.46$ per mode (as obtained in the fit in Fig.\,\ref{fig:afc}d, at the largest pump power $P$), corresponding to 12.7\,dB per mode.
After 51\% end-to-end system loss (which includes the losses due to the detectors being imperfect), our experimental results are consistent with squeezing of 2.7\,dB per mode.
Linear-optical losses are currently dominant, and could be improved via customized optics.
We note that our estimate here of squeezing per mode is a somewhat coarse estimate: it has made the assumption that each of the estimated $M = 433$ modes is equally squeezed, but as the simulation results in Fig.\,\ref{fig:dopa}b show, we theoretically expect there to be some variation in the amount of squeezing per mode (especially after the first 200 modes).

\section{Prospects for quantum advantage}\label{sec:app:qadvantage}

\subsection*{Discussion on improvements to the experiment}\label{sec:app:prospects}
\addcontentsline{toc}{subsection}{\nameref{sec:app:prospects}}

Here we expand the discussion on the limitations of our experiment and potential improvements.

As previously mentioned in the \nameref{sec:discussion},
there are currently limitations to our source,
(the combination of our generator of multimode squeezed light, followed by the AFC, which performs a unitary).
The class of unitaries we could experimentally realize were limited to fairly local connectivity (approximately to 10\% of the nearest frequency modes): hence we could programmably entangle each mode to at most $\sim$20\% of the rest (Fig.\,\ref{fig:covs}b--c).
Ideally we would like to be able to realize unitaries with all-to-all connectivity.
This may be possible with broader AFC pump bandwidths and higher pump intensities (as shown in simulations presented here in Appendix~\ref{sec:app:afc:pumpbandwidth}).
In addition, although we have shown control of the joint spectral probability distributions of the state, we currently lack a prescription for what pattern we should program on the pulse shaper for the AFC pump pulse to realize a specific unitary.
In principle, this is achievable if our experimental setup were to have a sufficiently fast means to characterize the realized unitary, allowing us to optimize the pulse-shaping patterns to realize a given unitary.
This can be resolved in the future by building a different characterization setup, for example involving two in-phase, fast-wavelength-sweeping continuous-wave lasers:
this would allows to perform fast, phase-sensitive measurements of our source.

There are also imperfections with the detection setup,
as discussed in Appendix~\ref{sec:app:spectrometer:imaging}.
Most importantly, the frequency binning is suboptimal, which may be a cause of decoherence.
As mentioned in the \nameref{sec:discussion} this could be resolved by: waveguided AFC, to avoid spatio-temporal effects that can prevent diffraction-limited spectrometer resolution; 
using custom imaging optics to optimize the resolution;
and switching to a discrete frequency-bin basis rather than a continuous basis.

A disadvantage of using an EMCCD camera instead of, for example, an array of superconducting nanowire single-photon detectors for measurement is speed: the camera's frame rate of 800\,Hz is slow.
However, the EMCCD camera we used has a linerate of 2\,MHz (up to 3.33\,MHz with higher noise), which is in principle the relevant speed when only using a single row (or few rows) of pixels, as is the case for our measurements.
However, there is currently a practical bottleneck to reading out the data, which is limited to line readouts at up to 15.7\,kHz on our camera model.
Several modern EMCCD cameras, including ours, allow reading out bursts of frames (or regions of whole frames, including lines) at up to MHz rates for short time windows; this could allow fast detection for a limited number of shots in the future.
Finally, the photon-number sampling we demonstrated used threshold detection or pseudo-photon-number-resolving measurements; ideally, future implementations could be capable of true photon-number-resolving measurements up to high photon numbers per mode (e.g., using ultra-low-noise CMOS cameras \cite{HamamatsuCMOS, GigajotCMOS}).

\subsection*{Influence of pump bandwidth on frequency conversion and entanglement structure}\label{sec:app:afc:pumpbandwidth}
\addcontentsline{toc}{subsection}{\nameref{sec:app:afc:pumpbandwidth}}

\begin{figure}[htb]
\centering
\includegraphics[width=1\normaltextwidth]{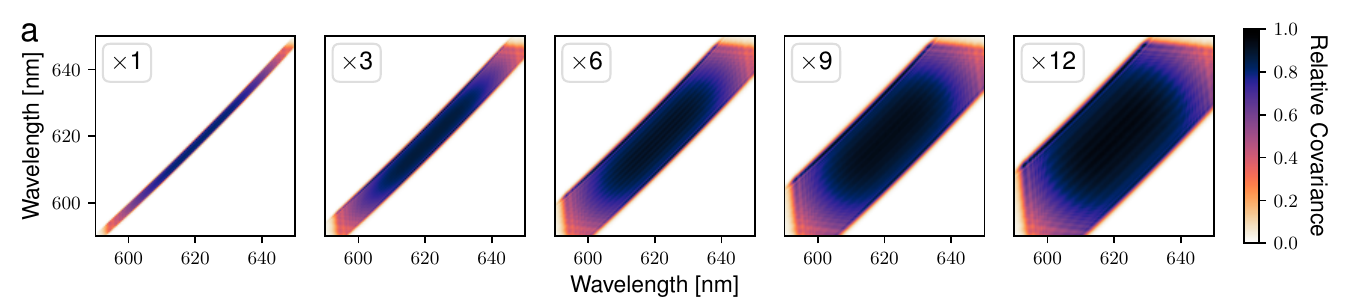}
\includegraphics[width=1\normaltextwidth]{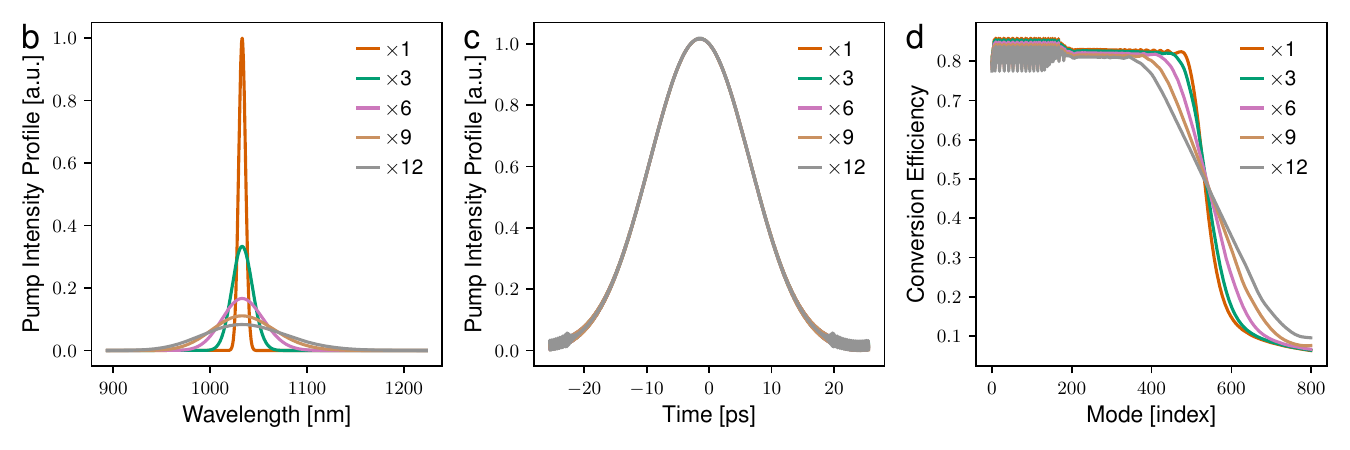}
\caption{
 \textbf{Effect of pump bandwidth on the covariance matrix.}
 These simulations show the AFC driven with pumps of different bandwidths.
 The chirp is inversely proportional to the bandwidth such that the intensity envelope is the same.
 The $\times 1$ indicates the bandwidth for a 200\,fs 1033\,nm pulse, and larger multiples indicate pulses with proportional larger bandwidths.
 \textbf{a.} The covariance matrices.
 \textbf{b.} Corresponding pump spectra.
 \textbf{c.} Corresponding pump temporal pulse profiles, chirped appropriately in order to maintain the same profile throughout.
 \textbf{d.} Conversion efficiency for the major supermodes (see Fig.\,\ref{fig:dopa}).
}
\label{fig:pumpbandwidth}
\end{figure}

As previously explained in Appendix~\ref{sec:app:afclzgrid}, the pump spectrum defines the coupling matrix between the bipartite set of frequency modes.
Specifically, the bandwidth determines the range of output signal frequency modes that one input frequency mode may be converted to.
The pump bandwidth is relatively narrow in our experimental realization -- compared to the signal bandwidth -- which limited the connectivity within the state's entanglement structure.
However, this is only a practical constraint, and a broader bandwidth pump, which can be achieved through spectral broadening methods, can realize far more complex states.

Fig.\,\ref{fig:pumpbandwidth}a shows, in simulation, the influence of the pump bandwidth (Fig.\,\ref{fig:pumpbandwidth}b) on the state's covariance matrix, for the same temporal intensity profile (Fig.\,\ref{fig:pumpbandwidth}c).
The bandwidth directly influences the connectivity.
The supermode conversion profile (Fig.\,\ref{fig:pumpbandwidth}d, with reference to Fig.\,\ref{fig:dopa}a), however, becomes less sharp, due to the wider phase matching.
The AFC phase-matching bandwidth would have to be redesigned appropriately for different pump bandwidths to mitigate this effect.

\subsection*{Validation of photon-counting cameras for quantum advantage}\label{sec:app:qadvantage:cameras}
\addcontentsline{toc}{subsection}{\nameref{sec:app:qadvantage:cameras}}

\begin{figure}[htb]
\centering
\includegraphics[width=0.778\normaltextwidth]{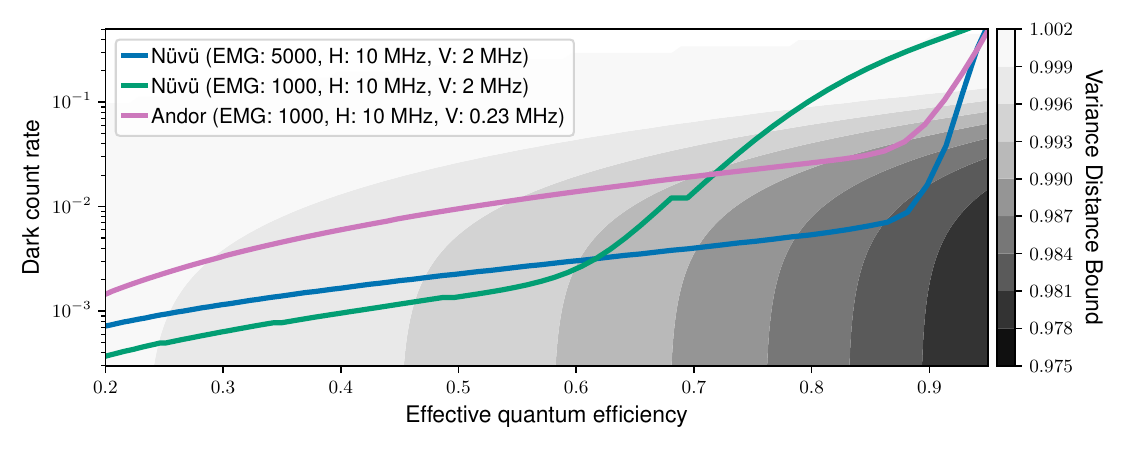}
\caption{
 \textbf{Camera performance vs simulability bound.}
 EMG, EM gain; H, horizontal line rate; V, vertical line rate.
 \textbf{b.} Comparison to the variance distance assuming an overall transmission of $\eta =$ 40\% and squeezing parameter $r=1$.
}
\label{fig:qibound}
\end{figure}

Boson sampling is an intermediate model of linear optical quantum computation \cite{ArkhipovAaronson}.
Realizing boson sampling with a level of post-classical computational complexity requires high performance quantum light sources: a large-scale, low-loss photonic circuit; and high-efficiency single-photon detectors -- all of which are essential building blocks for universal quantum computation using photons.
Gaussian boson sampling (GBS), a variation on this protocol, exploits squeezed vacuum states as input non-classical light sources.
For GBS, two main strategies for classical simulation exist.
The first uses the non-negativity of quasi-probability distributions (QPD) (generalizations of the Wigner function) as a strategy for simulation.
The second uses the fact that in GBS, the marginal distributions of photon numbers (i.e.,\ the probabilities to observe some subset of detection events irrespective of the others) are informative about the complete probability distribution.

For the QPD based simulations, an inequality exists that that demarcates the ``regime of simulability.''
Thus, any finite-sized experiment must pass this inequality to show that it is not simulable by this strategy.
The inequality is given in Ref.~\cite{QiGarciaPatron}:
\begin{equation}
 \text{sech} \left\{ \frac{1}{2} \Theta \left[ \ln \left( \frac{1 - 2 p_D/\eta_D}{\eta e^{-2r} + 1 - \eta} \right) \right] \right\} > e^{-\epsilon^2 / 4K},
\end{equation}
where $r$ is the squeezing parameters, $\eta$ is the overall photon transmission rate, $K$ is the number of squeezed sources, $\epsilon$ is the total variance distance of the experimental GBS samples compared to the ideal cases, $\Theta$ is the Heaviside function, $\eta$ is the transmission, $\eta_D$ is the photon detector efficiency and $p_D$ is the dark count rate.

In order to quantify the simulability of the experiment with different photon-counting cameras, we consider the bound for total variance distance $\epsilon$ as a function of the parameters determined by the cameras: photon detector efficiency ($\eta_D$) and dark count rate ($p_D$) (as discussed in Appendix~\ref{sec:app:emccd:thresh}).
We estimate the total optical transmission rate to be $\eta \approx 40\%$.
In Fig.\,\ref{fig:qibound} we then compare these $(\eta_D, p_D)$-plane trajectories to the left-hand side of the above simulability criterion equation.
We consider two EMCCD cameras: the N\"uV\"u HN\"u 512 IS and the Andor iXon Ultra 888.


\pagebreak

\newgeometry{
    top={26mm},
    headheight={12pt},
    headsep={5.15mm},
    text={\dimexpr8.5in-40mm,\dimexpr11in-50mm},
    marginparsep=5mm,
    marginparwidth=12mm,
    footskip=10.13mm
}

\phantomsection
\addcontentsline{toc}{section}{Appendix References}

\putbib[%
 references%
]
\restoregeometry

\end{bibunit}

\end{appendices}

\end{document}